\documentclass{elsarticle}
\usepackage{epstopdf, epsfig}
\usepackage{amsmath,amssymb,graphicx,float,latexsym,psfrag,subfigure,subfigmat}
\usepackage[usenames, dvipsnames]{color}
\usepackage[colorlinks=true,urlcolor=black,linkcolor=blue,citecolor=blue]{hyperref}

\newcommand{\pd}[2]{\frac{\partial #1 }{\partial #2}}

\newcommand{\Pd}[2]{\frac{\partial^{2} #1 }{\partial #2 ^2}}
\newcommand{\Pdc}[3]{\frac{\partial^{2} #1 }{\partial #2 \partial #3}}
\newcommand{\avg}[1]{\left\langle #1 \right\rangle}
\newcommand{\lpf}[1]{\overline{#1}}
\newcommand{\tlpf}[1]{\widehat{#1}}
\newcommand{\beq}{\begin{equation}}
\newcommand{\eeq}{\end{equation}}
\newcommand{\n}{{\bf n}}
\newcommand{\x}{{\bf x}}
\newcommand{\A}{\mathcal{A}}
\newcommand{\G}{\mathcal{G}}
\newcommand{\F}{\mathcal{F}}

\newcommand{\errprev}{\lpf{e}_{\rm QoI}^{\rm prev}}
\newcommand{\errref}{\lpf{e}_{\rm QoI}^{\rm DNS}}
\newcommand{\fDelta}{{\lpf{\Delta}}}
\newcommand{\etal}{\emph{et al.}}

\newif\ifIncludeFigures
\IncludeFigurestrue
\IncludeFigurestrue

\begin{document}

\begin{frontmatter}

\title{Towards systematic grid selection in LES:\\
iterative identification of the coarse-graining length scale
by minimizing the solution sensitivity}

  \author[UMD]{Siavash Toosi}
  \author[UMD]{Johan Larsson\corref{cor1}}\ead{jola@umd.edu}

  \address[UMD]{Department of Mechanical Engineering,
    University of Maryland, College Park, MD 20742, United States}

  \cortext[cor1]{Corresponding author.}


\begin{abstract}
The accuracy of a large eddy simulation (LES) is determined
  by the accuracy of the model used to describe the effect of unresolved scales, 
  the numerical errors of the resolved scales, and
  the optimality of the length scale that separates resolved from unresolved scales (the
  filter-width, or the coarse-graining length scale).
  This paper is focused entirely on the last of these,
  proposing a systematic algorithm for identifying the ``optimal''
  spatial distribution of the coarse-graining length scale
  and its aspect ratio.
  The core idea is that the ``optimal'' coarse-graining length scale
  for LES is the largest length scale for which the LES solution is
  minimally sensitive to it.
  This idea is formulated based on an error indicator 
  that measures the sensitivity of the solution
  and a criterion that determines how that 
  error indicator should vary in space and direction 
  to minimize the overall sensitivity of the solution.
  The solution to this optimization problem is that
  the cell-integrated error indicator should be equi-distributed;
  a corollary is that one cannot link the accuracy in LES to
  quantities that are not cell-integrated, including the common belief
  that LES is accurate whenever 80-90\% of the energy is resolved.
  The full method is tested 
  on the wall-resolved LES of
  turbulent channel flow and the flow over a backward-facing step,
  with final length scale fields (or filter-width fields, or
  grids) that are close to what is considered ``best practice'' in the
  LES literature.
  Finally, the derivation of the error indicator offers an alternative
  explanation for 
  the success of the dynamic procedure.
\end{abstract}


\end{frontmatter}

\section{Introduction}

Large eddy simulation (LES) is the solution of the Navier-Stokes
equations after they have been coarse-grained through some process,
whether formally by application of a low-pass filter to the
equations or implicitly by discretization on a grid.
In either case, one can define a coarse-graining length scale
(or filter-width)
$\lpf{\Delta}$ that separates the resolved from unresolved scales.
The effect of the unresolved motions is then modeled, either
through an explicitly stated model or implicitly by 
the numerical error
in the discretization.
In either case, this process creates a modeling error which
scales with the filter-width $\lpf{\Delta}$.

The LES equations are solved on a computational grid with 
a grid-spacing $\lpf{h}$,
a length scale which controls both the numerical errors
and the projection errors~\cite{sagaut:06}.
The most common LES approach is to take $\lpf{\Delta}=\lpf{h}$, in which case the
numerical errors are of similar magnitude to the errors caused by the
modeling of the unresolved scales~\cite[cf.][]{ghosal:96,park:07}.
Some studies have sought ``grid-independent LES'' by reducing
$\lpf{h}$ while keeping $\lpf{\Delta}$ fixed~\cite[cf.][]{ghosal:96,gullbrand:03};
while this produces an LES with negligible numerical errors, 
it (of course) does
not remove the modeling errors, and the resulting LES is necessarily
still ``filter-width dependent''.

There are two major ways in which the modeling errors in LES
can be reduced:
(i) by developing models that produce
more accurate predictions 
over a wider range of $\lpf{\Delta}$;
and/or (ii) by choosing more ``optimal" distributions of the filter-width $\lpf{\Delta}$.
Note that in cases where $\lpf{\Delta}=\lpf{h}$,
the correct choice of $\lpf{\Delta}$
would also
depend on
(and lead to a reduction of)
the numerical and projection errors.
In most cases, the choice of $\lpf{\Delta}$ 
is arguably 
at least as important as the 
subgrid/subfilter model and the numerical implementation.
Simply put, provided that the models are consistent 
(for example behaving consistently near solid walls),
most LES codes and subgrid models produce accurate results on
sufficiently ``good'' grids (i.e., filter-width distributions) 
and inaccurate results on sufficiently ``bad'' grids.
This importance of $\lpf{\Delta}$ as a significant modeling parameter
stands in some contrast to the literature
on LES over the last half century, with many papers published on
subgrid modeling (cf. the book by Sagaut~\cite{sagaut:06})
and the influence of numerical errors 
(cf.~\cite[][]{ghosal:96,kravchenko:97,mittal:97,brandt:07,park:07} and many others),
but with few studies devoted to the problem of how to optimally choose
$\lpf{\Delta}$.

The objective of the present work is to develop a systematic algorithm for
finding a nearly ``optimal" $\lpf{\Delta}$ as a function of both space and direction 
(i.e., an anisotropic filter-width).
This ``optimal" distribution,
denoted by
$\Delta_{\rm opt}(\x,\n)$
here
(with $\x$ referring to a spatial location and $\n$ to a direction),
is defined as the coarsest $\lpf{\Delta}(\x,\n)$
for which the LES solution is still sufficiently accurate.
Given the connection between $\lpf{\Delta}$ and $\lpf{h}$
(since the $\lpf{\Delta}/\lpf{h}$ ratio is in practice either unity or a fixed predefined value), 
this work can be viewed equally well
as an algorithm for finding the optimal grid.

The optimal filter-width distribution is (of course) flow-dependent,
model-dependent, and code-dependent.
Since it relies on the LES solution on a given grid, the algorithm is
necessarily iterative in nature:
as the filter-width (grid) improves between iterations, the estimate
of ${\Delta}_{\rm opt}$
becomes more accurate.
This type of process is usually termed ``grid adaptation'', but could
also be called ``filter-width adaptation'' in the context of LES.

The adaptation process is
driven by an ``error indicator''
that, based on an existing LES solution for a given
$\lpf{\Delta}(\x,\n)$,
estimates the characteristics of error generation and
leads to 
a new target
$\check{\Delta}(\x,\n)$
field
that is closer to optimal
(by applying some criteria that determines 
how the error indicator should be distributed).
There have been 
some
attempts to develop such error indicators in
the literature, most of which have been based on relatively heuristic
physics-based arguments.

Some early attempts relied on the
importance of the energy dissipation process
and thus
defined their error indicator as the fraction of
energy dissipation caused by the sub-grid/sub-filter scale (SGS/SFS) model 
to the total~\cite[cf.][]{geurts:02:lesaccuracy}.
This is closely related to using the ratio of the 
eddy viscosity to the molecular viscosity 
as a measure of accuracy~\cite[cf.][]{celik:05:lesquality}.
However,
this general concept is meaningful only at low Reynolds numbers,  
since the whole idea of LES is to avoid having to
resolve the viscous dissipation.

A more successful class of methods 
was inspired by the more 
realistic 
argument that
LES is accurate whenever the contribution 
of the modeled scales to the total kinetic energy
is sufficiently
small~\cite[]{jimenez:00}.
This is a much better assumption,
and more consistent with the purpose and premise of LES.
Pope~\cite[]{pope:04} used this intuitive argument to
suggest that the proportion of resolved to 
total kinetic energy could be used as a local indicator function.
Bose~\cite[]{bose:thesis}
used the kinetic energy in the smallest resolved scales directly
(i.e., without scaling with the resolved or total energy)
as an error indicator.
In both approaches, $\Delta_{\rm opt}(\x,\n)$
was found by requiring a constant and uniform indicator function
everywhere in space
(e.g., that no more than 10\% of the total kinetic energy was in the
unresolved/small scales).
While this general idea of connecting the accuracy of LES to the
amount of small-scale (or unresolved) kinetic energy is quite intuitive and has been
found to work well in several cases~\cite{bose:thesis, toosi:17},
it is important to acknowledge that it is heuristic in nature: there
is no equation showing that error scales with the small-scale or unresolved kinetic
energy.
For example, a perfect SGS/SFS model 
used with $\lpf{\Delta}/\lpf{h}\gg1$
would introduce no modeling or numerical error
regardless of the small-scale kinetic energy.
Similarly, while the unresolved kinetic energy (or its ratio to the total)
may be a good measure of projection or modeling errors in some of the flows
and for some of the variables
it may not be true in all flows or for all variables.

Several researchers tried to modify and improve the
error indicators discussed so far, or
even used the LES solution on more than one grid
(usually combined with Richardson extrapolation)
to define more accurate indicators
\cite[cf.][]{klein:05:lesquality,celik:05:lesquality,freitag:06,klein:08,celik:09},
but still based on the same heuristic ideas about the importance of
energy or dissipation to LES accuracy. 
Some of these modified indicators are:
the modified ``activity parameter''
(ratio of dissipations)
to include
numerical dissipation as well~\cite[]{celik:05:lesquality,celik:09},
the relative SGS viscosity index~\cite[]{celik:09,celik:09:2},
the relative Kolmogorov scale index~\cite[]{celik:09,celik:09:2}, 
combining the energy-based LES error indicator with 
another indicator for numerical errors~\cite[][]{benard:16},
using Richardson extrapolation and the LES
solution on two or three grids to better estimate the total
kinetic energy~\cite[][]{celik:09,celik:09:2}, etc.
Similarly, in the method of Systematic Grid and Model Variation (SGMV),
Richardson extrapolation 
was employed as a way of
deconvolution of the mean velocity~\cite[]{klein:05:lesquality,freitag:06,celik:09,klein:19} 
and Reynolds stresses~\cite[]{klein:19},
where in the process the numerical and modeling errors
(and their effect on the mean fields)
were also approximated.

We should also mention the class of multi-resolution LES 
(MR-LES) methods~\cite[cf.][]{sagaut:06,legrand:aiaa:18}
where two parallel simulations are performed
on two slightly different grids,
with the difference between the two solutions used to infer
the sources of error. 
The chaotic nature of the equations then requires regular
synchronization of the two solutions.

The most sophisticated approach to date
was
developed by
Hoffman and Johnsson~\cite[cf.][]{hoffman:03:chalmers,hoffman:04:siam}
and later Barth~\cite{barth:07}
who defined error indicators 
within a finite-element framework that included 
both the numerical errors and
the estimated modeling
error through a scale-similarity model.
They also solved the adjoint equation
to directly connect the estimated local errors to integrated 
``quantities of interest" (QoIs).
Despite the comprehensive treatment in these papers,
this approach has not been adopted 
extensively in the community.
Part of the reason is probably that the work was focused on the
finite-element approach, 
another is that the adjoint equations diverge exponentially for
long-time integration 
of chaotic flows
\cite[cf.][]{wang:13,wang:14:lss}.

\begin{figure}[t!]
	\centering	
	\includegraphics[width=65mm,clip=true,trim=25mm 0mm 0mm 0mm]{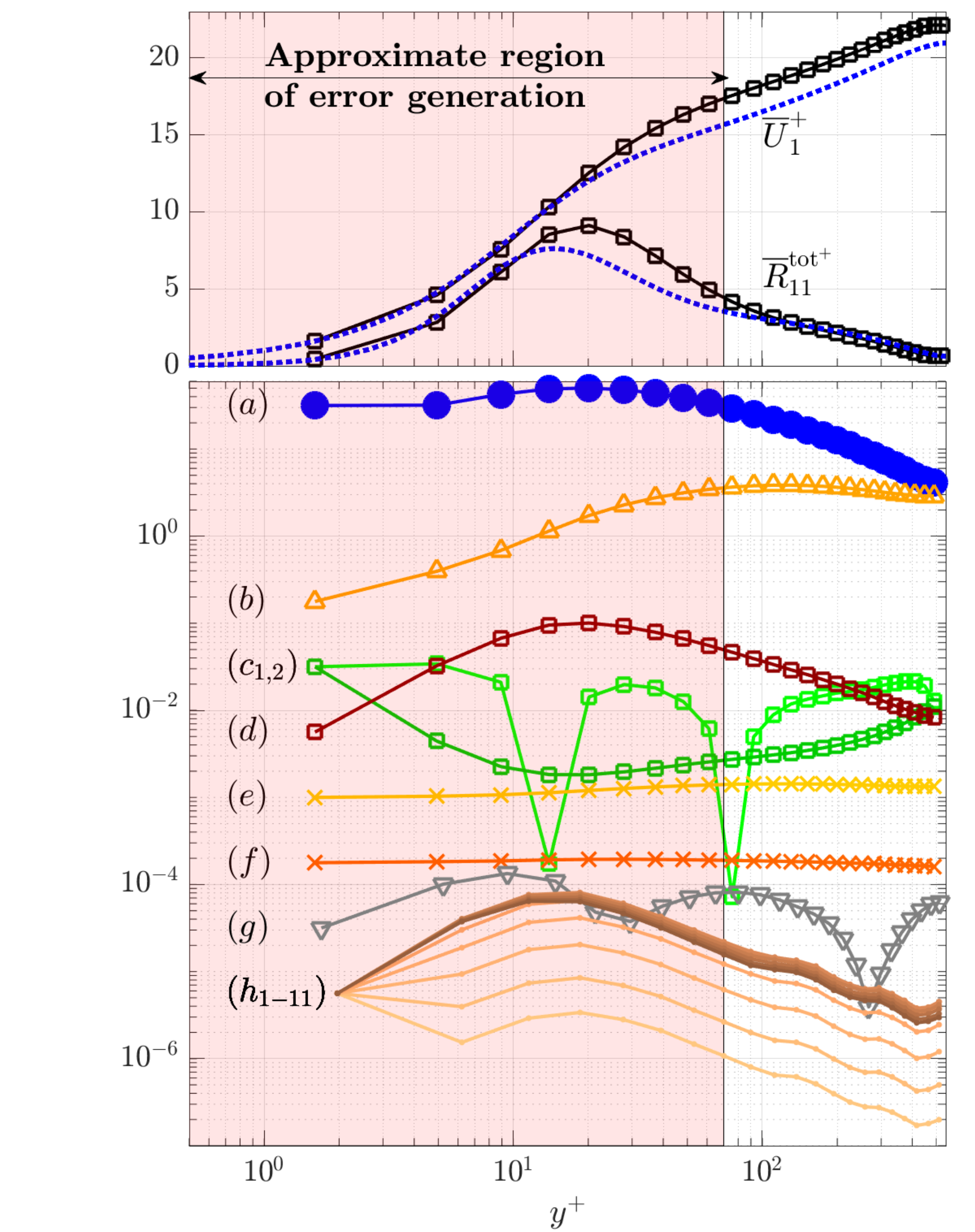}
	\caption{ \label{fig:intro-indicator-all}
	A comparative study of different LES error indicators for a typical 
	wall-resolved LES of the channel flow at $Re_\tau\approx 545$
	using the Vreman model (constant model coefficient of $c_v=0.03$)
	with a near-wall resolution of 
	$(\lpf{\Delta}^+_x,\lpf{\Delta}^+_{y_w}/2,\lpf{\Delta}^+_z)\approx (72,2.7,30)$ 
	in friction units, 
	using a sixth-order accurate numerical method and $\fDelta/\lpf{h}=1.6$ 
	(to minimize the effect of numerical errors and 
	have a fair comparison between different error indicators).
	The region of error generation
	is shaded on both plots.
	The top panel shows the 
	mean velocity and streamwise Reynolds stress,
	while the bottom panel plots different error indicators for:
	$(a)$ the error indicator proposed in this work 
	(defined in Eqn.~\ref{eq:Gn} and
	converted to a single number in the sense of 
	the integrand of Eqn.~\ref{eq:etot}),
	$(b)$ ratio of eddy viscosity to molecular viscosity
	proposed by Geurts and Fr{\"o}hlich~\cite{geurts:02:lesaccuracy},
	$(c_{1,2})$ ratio of unresolved to total kinetic energy
	proposed by Pope~\cite{pope:04} as
	``$1-$modified value'' proposed by Celik~\etal~\cite{celik:09} (light green) 
	and the small scale energy normalized by the local value of kinetic energy (dark green),
	$(d)$ the small-scale energy without local scaling proposed by Bose~\cite{bose:thesis}
	(also equivalent to the previous work of the present authors~\cite{toosi:17}
	applied using a non-directional filter),
	$(e)$ the ratio of effective viscosity index proposed by Celik~\etal~\cite{celik:09,celik:09:2} ($1-$value),
	$(f)$ the ratio of Kolmogorov scale index proposed by Celik~\etal~\cite{celik:09,celik:09:2} 
	for the cube root of the cell volume as length scale ($1-$value),
	$(g)$ sum of the absolute values of the 
	numerical and modeling errors in the mean velocity profile
	from the SGMV method proposed by Klein~\etal~\cite{klein:05:lesquality,freitag:06,celik:09} 
	for the recommended values of $m=2/3$ 
	(scaling exponent for modeling errors)
	and $n=6$ (nominal numerical order of accuracy),
	$(h_{1\mbox{-}11})$ solution error from the MR-LES method proposed by
	Legrand~\etal~\cite{legrand:aiaa:18} 
	(lightest to darkest colors correspond to different times after the last synchronization
	for $(t-t_s) U_b/H=0,\, 0.1,\, 0.2,\, ...,\, 1.0$, respectively).
	The profiles of all error indicators are normalized to make them comparable.
	Note that the error indicators $(a)$, $(d)$, and $(h)$
	are the only successful ones in identifying the region of error generation
	for this specific flow. 
	Refer to the original text of each work for more details.
	}
\end{figure}

A quantitative comparison of most of the aforementioned error indicators
is given in Fig.~\ref{fig:intro-indicator-all} for the case of 
the wall-resolved LES of the channel flow.

One major shortcoming of most existing error indicators
is that they are based on scalar error indicators which are unable to
infer anything about the optimal anisotropy of $\lpf{\Delta}(\x,\n)$.
In other words, if starting from an isotropic grid/filter-width, a
scalar error indicator could never produce an anisotropic final state.
This is inconsistent with, for example, the highly anisotropic grids
required near the wall in LES.
Among the few studies that did address the anisotropy
of the filter
was that of Addad~\etal~\cite{addad:08},
who
defined their $\Delta_{\rm opt}(\x,\n)$ directly
based on an empirical criterion
about the relative size of the filter compared to
the Taylor microscale 
and the RANS dissipation length scale.
Another approach was taken by the present authors in prior work
\cite{toosi:17},
where the optimal anisotropy was inferred from the directionally
high-pass test-filtered solution field
combined with the standard heuristic argument connecting the error
generation to the energy in the small-scale field
(the filter-width selection criterion of that work
was also based on heuristic arguments).
The other works that  
addressed the problem of anisotropy of the grid 
(cf.~\cite[][]{belme:12} as one of the best and most comprehensive examples)
had their error indicators defined 
solely based on the numerical errors
and therefore do not fully address the problem in LES.

In the present work we develop a new error indicator
for LES
that is derived more directly from the governing equation and thus
requires less heuristics.
The key assumption becomes that the LES equations should be minimally
sensitive to a change in the filter-width. 
The final definition of the error indicator
(and in part the reasoning leading to it)
becomes closely related to the
dynamic procedure~\cite[]{germano:91,lilly:92},
and in fact leads to an alternative explanation for
why the dynamic procedure works;
this is discussed briefly in Section~\ref{section:dynamic}.

The present work is focused on statistically stationary problems
for which we seek a stationary grid/filter-width.
In other words,
the grid/filter-width is adjusted only between LES runs, and
the adaptation step becomes solely a
post-processing operation
with no changes needed in the LES solver at all.
We should also emphasize that the
focus here is
entirely on the problem of 
\emph{finding} $\Delta_{\rm opt}(\x,\n)$
and not at all the exact way of \emph{creating} 
this new grid (i.e., on the scientific computing aspect, etc.);
we simply use currently available tools to generate these grids
without worrying about 
parallel performance, data structures, etc.
Other factors like
the grid quality, stretching factor, etc. 
have not been considered either.
Consequently, the results presented in this paper
could presumably be improved 
by imposing some constraints on the grid quality metrics or stretching factors,
or by use of more sophisticated and flexible
grid-generation toolboxes capable of generating a closer grid
to the target $\Delta_{\rm opt}(\x,\n)$.

\section{Methodology}

The governing equation for large eddy simulation (LES)
can be formally derived by applying a low-pass filter with
characteristic filter-width $\lpf{\Delta}$
to the Navier-Stokes equation~\cite[cf.][]{pope:00,sagaut:06}.
In practice, this filtering (or coarse-graining) is often done
implicitly when constructing the computational mesh,
by assuming that the filter-width is either equal to the grid-spacing $\lpf{h}$
or some fixed fraction larger than it.
After replacing the unclosed residual stress tensor
$\tau_{ij}$
by a model
$\tau_{ij}^{\rm mod}(\lpf{u}_k)$,
the resulting equation (for incompressible flows) is
\beq
\label{eq:fns}
	\pd{ \lpf{u}_i}{t} 
+ 	\pd{\lpf{u}_i \lpf{u}_j}{x_j}
= - 	\frac{1}{\rho} \pd{\lpf{p}}{x_i}
+ 	\nu \Pdc{ \lpf{u}_i}{x_j}{x_j}
- 	\pd{\tau_{ij}^{\rm mod}(\lpf{u}_k)}{x_j}
\,,
\eeq
where $\lpf{u}_i$ and $\lpf{p}$ are the resolved velocity and pressure
fields,
and $\rho$ and $\nu$ are density and kinematic viscosity of the fluid
(both assumed constant).

The developments in this paper are based on Eqn.~\ref{eq:fns}
(i.e., for implicitly filtered LES)
which is arguably the most popular formulation.
Derivations for some alternative forms of this equation are given 
in~\ref{sec:appendix-otherGn}, including: 
(i) when the convective flux is written as
$\lpf{ \lpf{u}_i \lpf{u}_j}$ (used when applying an explicit filter in
the solver, known as explicitly filtered LES);
(ii) when solving LES without an explicit subgrid model
$\tau_{ij}^{\rm mod}$ (implicit LES, or ILES);
and
(iii) for compressible LES.

\subsection{Proposed error indicator\label{section:Gn-formulation}}

The idea of this section is to estimate how sensitive the LES equation~\ref{eq:fns}
is to a change in the filter-width $\lpf{\Delta}$ in any given direction and at any given location,
and to use that to define our error indicator. 
The estimate will be derived and computed using a low-pass test
filter, which must be able to filter only in a single direction
(i.e., filter modes with high wavenumber in that single direction)
in order to infer anything about the anisotropy of the optimal
filter-width.
On structured grids such a uni-directional test-filter along the grid
lines is trivial to implement
and there is much flexibility in the choice of
test-filter~\cite[cf.][]{sagaut:06,vasilyev:98}.
To make this work applicable to general geometries and grid
topologies, we will instead use the directional differential filter
from our previous work~\cite{toosi:17} defined as
\beq \label{eq:filter}
\tlpf{\lpf{\phi}}^{(\n_0)} \approx 
\left( 
I + \frac{ \lpf{ \Delta}_{\n_0}^2}{4} \n_0^T \nabla \nabla^T \n_0
\right)
\lpf{\phi}
\, ,
\eeq
where $\lpf{\phi}=\lpf{\phi}(\x)$ is the original resolved LES field,
$\tlpf{\lpf{\phi}}^{(\n_0)} = \tlpf{\lpf{\phi}}^{(\n_0)} \!\!\!\!\!\! (\x)$ 
is the directionally low-pass test-filtered (in direction $\n_0$) field,
$\lpf{\Delta}_{\n_0} = \lpf{\Delta}(\x,\n_0)$ is the filter-width in direction $\n_0$
(where $\n_0$ is the unit direction vector),
and $I$ is the identity tensor
(see~\cite[][]{germano:86:elliptic,germano:86:lesfilter} for the definition of the filter kernel).
The doubly contracted Hessian matrix
$\n_0^T \nabla \nabla^T \n_0$ 
can also be expressed in index notation as 
$n_{0,i} n_{0,j} \partial^2 / \partial x_i \partial x_j= \partial^2 / \partial x_{(\n_0)}^2 $.
For a structured grid with uniform grid-spacing
and using second-order central differencing 
the filter of Eqn.~\ref{eq:filter} simplifies
to a uni-directional box filter of width $2 \lpf{\Delta}(\x,\n_0)$
applied using the trapezoidal rule.
More details about this filter is given in~\ref{sec:appendix-filter}. 
Also, note that the chosen test-filter of Eqn.~\ref{eq:filter} 
is not unique,
but part of a more general class of differential filters
\cite[][]{vasilyev:98,sagaut:06}
that can be modified to 
contain directional information and 
potentially replace Eqn.~\ref{eq:filter}.

Applying the directional test-filter 
to Eqn.~\ref{eq:fns}
yields
(assuming that filtering and 
differentiation commute,
see~\ref{sec:appendix-commute} for how
to include the effect of commutation errors,
and~\ref{sec:appendix-filter} for filters with
low commutation errors)
an evolution equation for the
filtered instantaneous fields at the test-filter level as
\beq
\label{eq:ffns-u}
	\pd{ \tlpf{ \lpf{u}}^{(\n_0)}_i}{t} 
+ 	\pd{\tlpf{ \lpf{u}_i \lpf{u}_j}^{(\n_0)}}{x_j}
= - 	\frac{1}{\rho} \pd{ \tlpf{ \lpf{p}}^{(\n_0)}}{x_i}
+ 	\nu \Pdc{ \tlpf{ \lpf{u}}^{(\n_0)}_i}{x_j}{x_j}
- 	\pd{ \tlpf{ \tau_{ij}^{\rm mod}(\lpf{u}_k) }^{(\n_0)} }{x_j}
\,.
\eeq

An alternative way to obtain an evolution equation for the solution
at the test-filter level is to write the filtered Navier-Stokes equations (Eqn.~\ref{eq:fns})
at the test-filter level instead:
\beq
\label{eq:ffns-v}
	\pd{ \tlpf{ \lpf{v}}^{(\n_0)}_i}{t} 
+ 	\pd{\tlpf{ \lpf{v} }^{(\n_0)}_i  \tlpf{ \lpf{v}}^{(\n_0)}_j}{x_j}
= - 	\frac{1}{\rho} \pd{ \tlpf{ \lpf{q}}^{(\n_0)}}{x_i}
+ 	\nu \Pdc{ \tlpf{ \lpf{v}}^{(\n_0)}_i}{x_j}{x_j}
- 	\pd{ \tau_{ij}^{\rm mod}\left(\tlpf{\lpf{v}}^{(\n_0)}_k \right) }{x_j}
\,,
\eeq
where $\tlpf{ \lpf{v}}^{(\n_0)}_i$ and $\tlpf{ \lpf{q}}^{(\n_0)}$ 
are the resolved velocity and pressure fields at the test-filter level 
$\tlpf{ \lpf{\Delta} }^{(\n_0)} \!\!\!\!\!\! = \tlpf{ \lpf{\Delta} }^{(\n_0)}  \!\!\!\!\!\!\!\! (\x,\n)$.

Defining the difference between the two solutions as
\beq
\nonumber 
\tlpf{ \lpf{e}}^{(\n_0)}_i = \tlpf{ \lpf{v}}^{(\n_0)}_i - \tlpf{ \lpf{u}}^{(\n_0)}_i
\ \ , \ \ 
\tlpf{ \lpf{\Pi}}^{(\n_0)} = \tlpf{ \lpf{q}}^{(\n_0)} - \tlpf{ \lpf{p}}^{(\n_0)}
\,,
\eeq
and subtracting
Eqn.~\ref{eq:ffns-u} from Eqn.~\ref{eq:ffns-v}
yields
an evolution equation for the difference as
\beq
\label{eq:eeq}
\underbrace{
\pd{ \tlpf{ \lpf{e}}^{(\n_0)}_i }{t}
+ \tlpf{\lpf{u}}^{(\n_0)}_j \pd{\tlpf{\lpf{e}}^{(\n_0)}_i}{x_j}
}_{T_1}
- \underbrace{ 
\nu \Pdc{ \tlpf{ \lpf{e}}^{(\n_0)}_i}{x_j}{x_j}
}_{T_2}
+ \underbrace{
\pd{\tlpf{ \lpf{e}}^{(\n_0)}_i \tlpf{ \lpf{e}}^{(\n_0)}_j}{x_j} 
}_{T_3}
+ \underbrace{ 
\tlpf{\lpf{e}}^{(\n_0)}_j \pd{\tlpf{\lpf{u}}^{(\n_0)}_i}{x_j}
}_{T_4}
+ \underbrace{ 
\frac{1}{\rho} \pd{ \tlpf{ \lpf{\Pi}}^{(\n_0)}}{x_i}
}_{T_5}
= \tlpf{ \lpf{ \F }}^{(\n_0)}_i
\,,
\eeq
where
\beq
\label{eq:F}
\tlpf{ \lpf{ \F }}^{(\n_0)}_i \!\!\!\!\!\!\!  (\x)
=
\pd{}{x_j} 
\left(
 	\tlpf{ \lpf{u}_i \lpf{u}_j}^{(\n_0)} 
 - 	 \tlpf{ \lpf{u}}^{(\n_0)}_i  \tlpf{ \lpf{u}}^{(\n_0)}_j
 \right)
 +
 \pd{}{x_j} 
 \left(
\tlpf{ \tau_{ij}^{\rm mod}(\lpf{u}_k) }^{(\n_0)}
-
\tau_{ij}^{\rm mod}\left(\tlpf{\lpf{v}}^{(\n_0)}_k \right)
  \right)
  \,.
\eeq

Terms $T_1$ and $T_2$ in the error evolution equation~\ref{eq:eeq}
describe convective and viscous transport,
term $T_3$ is a nonlinear transport term,
term $T_4$ becomes a production term 
in the governing equation of $\tlpf{ \lpf{e}}^{(\n_0)}_i \tlpf{ \lpf{e}}^{(\n_0)}_j$,
and term $T_5$ is a pressure-like term that keeps $\tlpf{ \lpf{e}}^{(\n_0)}_i$
divergence-free.
The terms in Eqn.~\ref{eq:eeq}
are grouped such that all terms involving $\tlpf{ \lpf{e}}^{(\n_0)}_i$
are on the left while the terms not involving the error are grouped in
$\tlpf{ \lpf{ \F }}^{(\n_0)}_i\!\!\!\!\!\!\!  (\x)$.

The difference $\tlpf{ \lpf{e}}^{(\n_0)}_i$ can be interpreted as 
a measure of sensitivity of the solution
to the filter level used in its computation.
In a chaotic system (like LES), this difference will of course diverge
exponentially (at early times)
and thus rapidly become meaningless 
when $\tlpf{ \lpf{u}}^{(\n_0)}_i$ 
and
$\tlpf{ \lpf{v}}^{(\n_0)}_i$
become uncorrelated.
Having said that, over short time scales, when starting from identical
solutions ($\tlpf{ \lpf{e}}^{(\n_0)}_i=0$),
Eqn.~\ref{eq:eeq} shows that 
$\tlpf{ \lpf{ \F }}^{(\n_0)}_i\!\!\!\!\!\!\! (\x)$
is the source of initial divergence between the two solutions
(since, with $\tlpf{ \lpf{e}}^{(\n_0)}_i=0$, all terms of the left
side of Eqn.~\ref{eq:eeq} are zero).
We can then hypothesize that 
the magnitude of
$\tlpf{ \lpf{ \F }}^{(\n_0)}_i\!\!\!\!\!\!\! (\x)$
remains a meaningful estimate of the error generation in an LES, even
beyond the short time horizon.

The proposed error indicator is then defined as
\beq
\label{eq:Gn}
\lpf{\G}(\x,\n) = \sqrt{ \avg{ \tlpf{ \lpf{ \F }}^{(\n)}_i \!\!\!\!\! (\x)
    \, \tlpf{ \lpf{ \F }}^{(\n)}_i  \!\!\!\!\! (\x) } }
\,,
\eeq
where $\avg{\cdot}$ denotes a suitable averaging operator,
and $\lpf{\cdot}$ signifies the filter-level on which 
the test-filtering is applied (i.e., $\lpf{\Delta}$).
In the present work we are interested in finding the optimal static
grids for statistically stationary problems,
and hence averaging is performed over time and 
any homogeneous spatial directions.
For more general settings the averaging operator could be adjusted accordingly. 
For example, in flows with strong unsteady effects at a slow time-scale
(e.g., vortex shedding) one could use a low-pass time filter,
and for temporally periodic flows
(e.g., pulsating flows) one could use a phase average.

The final definition of the error indicator
given by Eqns.~\ref{eq:F} and~\ref{eq:Gn} 
essentially measures the error in the divergence of the Germano identity tensor.
However, rather than heuristically taking this quantity
to define an error indicator, 
the derivations of this section show that
this is the relevant quantity to minimize in order to
minimize errors related to the filter-width. 
The connections between the error indicator, the 
Germano identity, and the dynamic procedure
are discussed in more details in Section~\ref{section:dynamic}.

The first term (the Leonard-like stress)
in $\tlpf{ \lpf{ \F }}^{(\n)}_i  \!\!\!\!\! (\x)$
can be directly computed from the LES solution $\lpf{u}_i$,
and $\tau_{ij}^{\rm mod}(\lpf{u}_k)$ is also known from the LES.
On the other hand, the SGS stress tensor in the imagined evolution equation at the test
filter level (Eqn.~\ref{eq:ffns-v})
is defined based on the imagined velocity field $\tlpf{ \lpf{v}}^{(\n)}_i$.
One option would be to actually run an additional LES solving
Eqn.~\ref{eq:ffns-v} but in a synchronized way
(similar to the MR-LES methods of Legrand~\etal~\cite[][]{legrand:aiaa:18},
but with the error indicator of this work).
The alternative, which is applied here,
is to use the test-filtered velocity field from the original
LES solution
to expand the SGS tensor as well, i.e.,
\beq \nonumber
\tau^{\rm mod}_{ij} \left( \tlpf{ \lpf{v}}^{(\n)}_k \right)
=
\tau^{\rm mod}_{ij} \left( \tlpf{ \lpf{u}}^{(\n)}_k \right)
+
\mathcal{T}_{ij} \left( \tlpf{ \lpf{e}}^{(\n)}_k \right)
\,,
\eeq
where the dependence of $\mathcal{T}_{ij}$ 
on $\tlpf{ \lpf{u}}^{(\n)}_k $ is purposefully suppressed 
to emphasize that all of its terms contain $\tlpf{ \lpf{e}}^{(\n)}_k $ and
must vanish when
$\tlpf{ \lpf{e}}^{(\n)}_k = 0$
for consistent SGS models
(see the supplementary materials for an example
of $\mathcal{T}_{ij}$ in the case of Smagorinsky eddy viscosity model).
As a result, $T_6 = \partial \mathcal{T}_{ij} / \partial x_j$
can be moved to the left-hand side of Eqn.~\ref{eq:eeq}
where it becomes excluded from the imagined error source
(based on the same reasoning used before).
Note that expanding $\tau^{\rm mod}_{ij} ( \tlpf{ \lpf{v}}^{(\n)}_k )$
using the test-filtered field $\tlpf{ \lpf{u}}^{(\n)}_k$
is not only simpler (and cheaper), but also more consistent with
our current formulation.

The Leonard-like stress 
in the definition of 
$\tlpf{ \lpf{ \F }}^{(\n)}_i  \!\!\!\!\! (\x)$
can be further simplified for a known filter kernel.
For the example of the filter of Eqn.~\ref{eq:filter}
this term takes the form
\beq
\label{eq:Lij}
 	\tlpf{ \lpf{u}_i \lpf{u}_j}^{(\n)} 
 - 	 \tlpf{ \lpf{u}}^{(\n)}_i  \tlpf{ \lpf{u}}^{(\n)}_j
 =
 \frac{\fDelta^2_\n}{2}
 \pd{\lpf{u}_i}{x_{(\n)}}
  \pd{\lpf{u}_j}{x_{(\n)}}
  -
   \frac{\fDelta^4_\n}{16}
 \Pd{\lpf{u}_i}{x_{(\n)}}
  \Pd{\lpf{u}_j}{x_{(\n)}}
  \, ,
\eeq
with no summation over $\n$.
Therefore, the divergence of this term 
has a second derivative in its leading term and somewhat
resembles the truncation or interpolation error of a 
low-order numerical scheme
\cite[cf.][]{toosi:17,benard:16,habashi:00,dompierre:02}.
Note that depending on the test-filter 
(i.e., first, second or higher derivatives in the definition of the differential filter)
the Leonard-like stress can generally resemble 
the truncation error of 
different numerical schemes. 
Also note that the leading term in the
expansion of Eqn.~\ref{eq:Lij} has a similar form
to the Clark model~\cite[cf.][]{sagaut:06,vreman:96,trias:17},
but with a different coefficient (i.e., $1/2$ instead of $1/12$)
due to the difference in filtering.

\subsection{Connection to the dynamic procedure \label{section:dynamic}}

The dynamic procedure~\cite[]{germano:91,lilly:92} 
is a way to compute model constant(s) through
test-filtering,
which has received a lot of attention in the LES community.
It finds the model coefficient that minimizes
\beq \nonumber
\tlpf{ \lpf{e}}_{\rm dyn} = \avg{ (  \tlpf{ \lpf{\mathcal{L}}}_{ij} +  \tlpf{ \lpf{\mathcal{M}}}_{ij} ) (  \tlpf{ \lpf{\mathcal{L}}}_{ij} +  \tlpf{ \lpf{\mathcal{M}}}_{ij} )}
\,,
\eeq
where $\tlpf{ \cdot }$ is a regular test-filter (i.e., not directional), and
\beq \nonumber
\tlpf{ \lpf{\mathcal{L}}}_{ij} 
= \tlpf{ \lpf{u}_i \lpf{u}_j } - \tlpf{ \lpf{u}}_i  \tlpf{\lpf{u}}_j
\ \ , \ \ 
\tlpf{ \lpf{ \mathcal{M}}}_{ij} 
=\tlpf{ \tau_{ij}^{\rm mod}(\lpf{u}_k) } -
\tau_{ij}^{\rm mod}\left(\tlpf{\lpf{u}}_k \right)
\,.
\eeq
There have been multiple explanations for how/why the
dynamic procedure works.
The original explanation
appealed to scale similarity in the inertial 
range (cf.~\cite[][]{meneveau:00,park:09} and references therein),
but as pointed out by others~\cite[cf.][]{jimenez:00,pope:04} this fails to explain
why the dynamic procedure works during transition to turbulence
or in the near-wall region of turbulent
boundary layers (arguably its greatest success).
The lack of any scale similarity
at the test-filter level
in those scenarios 
(the filter is close to the dissipative range in wall-resolved LES)
therefore makes the original explanation unlikely.

Jimenez \& Moser~\cite[]{jimenez:00} suggested that the explanation has
to do (among other things) with dissipation, 
that the dynamic procedure makes the
dissipation by the LES model equal to the production of the
Leonard stresses.
An alternative explanation was put forth by
Pope~\cite[]{pope:04},
who showed that the dynamic procedure can be derived
by requiring that the total Reynolds stress 
(i.e., resolved plus modeled)
should be minimally sensitive to the filtering level, i.e., that
the model coefficient should be chosen to minimize (in magnitude)
\beq
\nonumber
\left(
\tlpf{ \lpf{u}_i \lpf{u}_j }  + \tlpf{ \tau_{ij}^{\rm mod}(\lpf{u}_k) }
\right)
-
\left(
\tlpf{ \lpf{u}}_i  \tlpf{\lpf{u}}_j  + \tau_{ij}^{\rm mod} (\tlpf{\lpf{u}}_k)
\right)
\,,
\eeq
which is equal to minimizing $\tlpf{\lpf{\mathcal{L}}}_{ij} + \tlpf{\lpf{\mathcal{M}}}_{ij}$.
Although not directly stated in~\cite[]{pope:04}, 
the choice of the total Reynolds stress as the critical quantity
presumably comes from the importance of stresses in momentum
transport.

The present derivation of the error indicator $\G$
implies 
a somewhat similar but slightly different
explanation for why the
dynamic procedure works,
without any specific assumption about turbulence properties like scale-similarity
or about the importance of Reynolds stresses, energy or dissipation 
in the accuracy of the LES solution.

The residual force $\tlpf{\lpf{\F}}_i$ of Eqn.~\ref{eq:F}
is simply the divergence of the total tensor
subject to minimization in the dynamic procedure,
i.e.,
\beq \nonumber
\tlpf{\lpf{\F}}_i = \pd{}{x_j} \left(
\tlpf{\lpf{\mathcal{L}}}_{ij}
+
\tlpf{\lpf{\mathcal{M}}}_{ij}
\right)
\,.
\eeq
Pope arrived at the minimization of this tensor by requiring that the
predicted total stress from an LES should be insensitive to the filter
level 
(with a heuristic step that this should lead to a less sensitive,
and hence more accurate, solution);
here, we instead arrive at the same thing by requiring that the
assumed source term in the evolution equation for the difference
between the two solutions at the same filter levels 
(i.e., the solution sensitivity to the filter-width used in its computation)
be as small as
possible.
In other words, while the error in the Germano identity
is definitely a relevant quantity in minimizing the solution sensitivity,
the significance of the derivation of 
Section~\ref{section:Gn-formulation} 
is to show that it is \emph{the} relevant one
(and not just one of the relevant measures).

We should also note that the present work clearly suggests
that the force $\tlpf{\lpf{\F}}_i$ rather than the
tensor $\tlpf{ \lpf{\mathcal{L}}}_{ij} + \tlpf{ \lpf{\mathcal{M}}}_{ij}$
should be minimized in the dynamic procedure.
This has actually been tested before in the literature,
in the work of Morinishi \& Vasilyev~\cite[]{morinishi:01}.
The downside is that this leads to a nonlinear second-order PDE for
the model coefficient, which is presumably why this version of the
dynamic procedure has not received
the attention and popularity it arguably deserved.

Interestingly, 
our tests on the channel flow suggest that using the full tensor to
drive filter-width adaptation
leads to extremely fine cells in the wall-normal direction
and is therefore strongly discouraged.


\subsection{Finding the optimal filter-width \label{sec:optimal}}

The error indicator estimates the introduction of error into the
evolution equation due to insufficient resolution,
but does not automatically determine how much the resolution
needs to be changed for the error to go down to a certain level.
One approach would be to refine the filter by a fixed factor, say cut the filter in half, 
in any direction $\n$ and location $\x$ 
that the value of the error indicator is above a certain threshold.
This is not optimal however. 
It is much better if we can predict 
the change of the error indicator for
any given change in the filter-width,
and then adjust the filter-width proportionally. 
The latter approach requires a direct link between 
the error indicator and filter-width
(i.e., a model). 
In the present work,
we adopt the simplistic model
\beq
\label{eq:model}
\check{\G}(\x,\n) \approx \lpf{g}(\x,\n) \check{\Delta}^{\! \alpha}\!(\x,\n)
\,,
\eeq
where 
$\check{\G}(\x,\n)$ is the predicted value of the error indicator
on the filter-level
$\check{\Delta}(\x,\n)$
and
the ``error source density'' $\lpf{g}(\x,\n)$
is computed from the
existing LES solution as
\beq \nonumber
\lpf{g}(\x,\n) = \frac{\lpf{\G}(\x,\n)}{ \lpf{\Delta}^{\alpha}\!(\x,\n)}
\,.
\eeq
The exponent $\alpha$ should be different in different flow regimes
(free-shear turbulence, near-wall turbulence, etc.; cf.~\cite[][]{klein:07})
and in different directions,
but is simply taken as $\alpha\!=\!2$ in 
our assessments on turbulent channel flow
(Section~\ref{sec:channel})
and the flow over a backward-facing step
(Section~\ref{section:BFS})
without any attempts at finding the best value.
The derivations in the rest of this Section are based on
a constant value of $\alpha$, 
with generalization to the spatially and directionally varying scaling exponent
given in~\ref{sec:appendix-optimal}.

The optimal filter-width distribution is the one that leads
to the highest accuracy
among all possible
filter distributions with the same computational cost.
The ``highest accuracy'' is considered here to be equivalent to the lowest
introduction of ``error'' in the sense of Eqn.~\ref{eq:eeq}.
In the following we assume a grid with hexahedral cells, though
possibly with ``hanging nodes'' and not necessarily with a structured topology.

Assuming that the error source is proportional to the magnitude of
$\check{\G}$,
the total error to be minimized is, for the special case of a grid
with only hexahedral cells,
\beq \label{eq:etot}
\check{e}_{\rm tot} \propto \int_{\Omega}{
\sqrt{ \check{\G}^2(\x,\n_1) +  \check{\G}^2(\x,\n_2) +
  \check{\G}^2(\x,\n_3) } 
\ d\x
}
\,,
\eeq
where the $\n_1$, $\n_2$ and $\n_3$ directions are the three
directions of the hexahedral cells (or in computational space, for a
structured grid).
The computational cost is assumed proportional to the number of
cells in this work, which can be estimated~\cite[cf.][]{belme:12} as
\beq \nonumber
N_{\rm tot} \approx \int_{\Omega}{ \frac{d\x}{\check{{V}}_c (\x) } }
\eeq
where $\check{{V}}_c$ is the volume of a cell.
Assuming a fixed ratio of $\lpf{\Delta }/ \lpf{h}$, we then have
(again assuming a grid with hexahedral cells)
\beq \label{eq:ntot}
N_{\rm tot} \propto \int_{\Omega}{
\frac{d\x}{
\check{\Delta}(\x,\n_1)
\check{\Delta}(\x,\n_2)
\check{\Delta}(\x,\n_3)} }
\,.
\eeq

\begin{figure}[t!]
  \centering	
  \includegraphics[width=60mm,clip=true,trim=0mm 0mm 15mm 5mm]{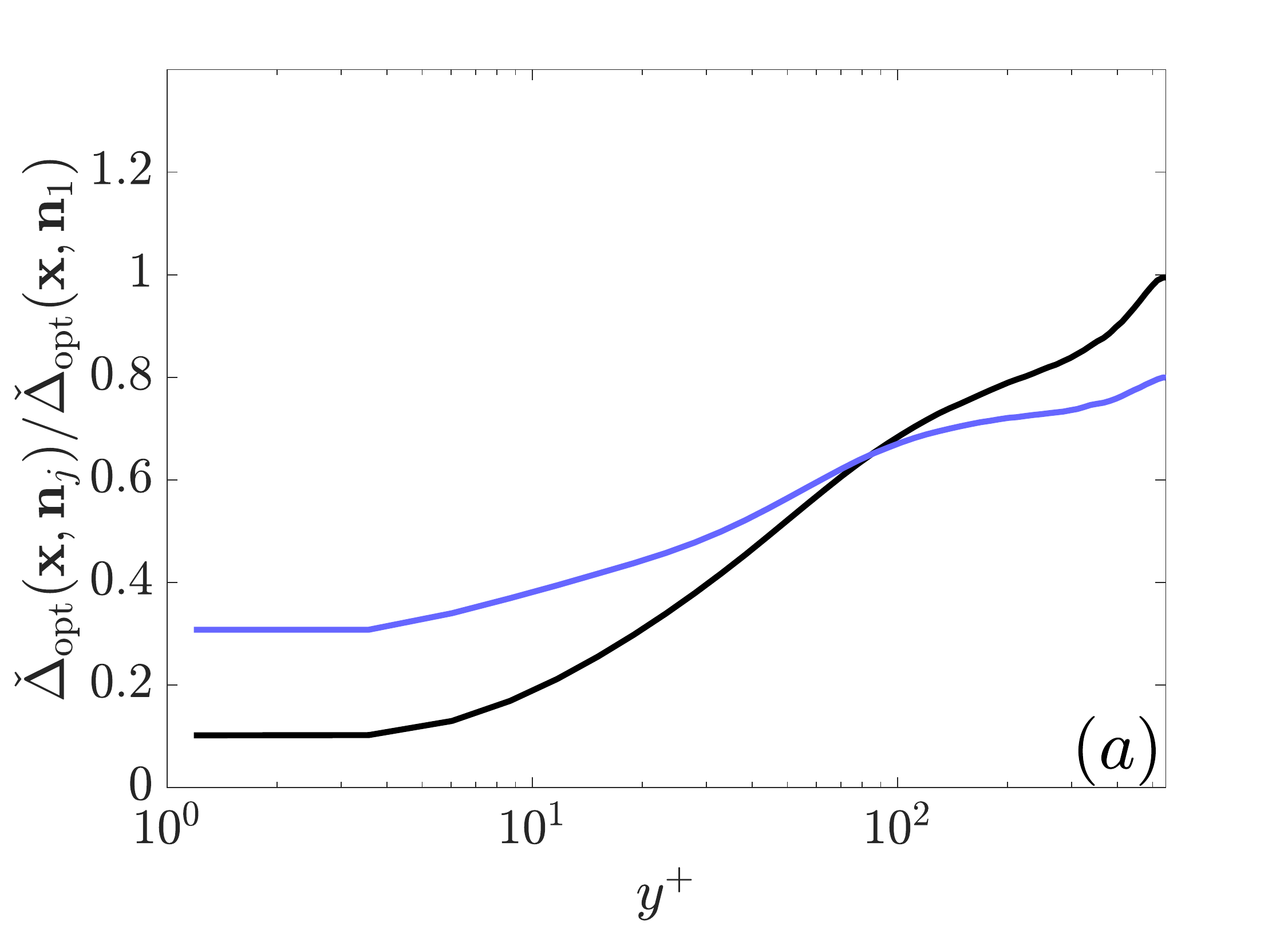}
  \includegraphics[width=60mm,clip=true,trim=0mm 0mm 15mm 5mm]{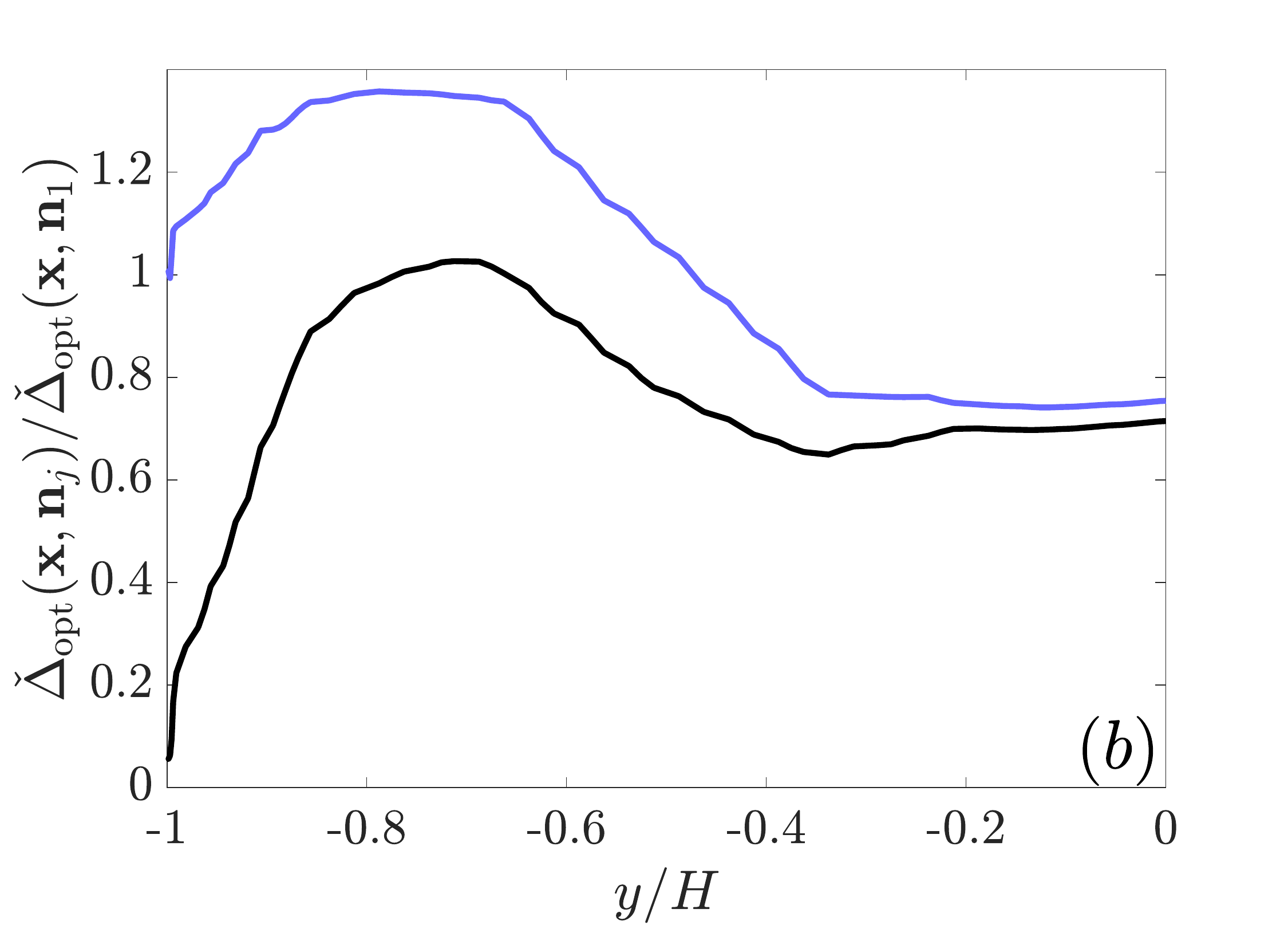}
  \caption{\label{fig:sample-ARs}
    Examples of the predicted optimal cell aspect ratios
    $\check{\Delta}_{\rm opt}(\x,\n_2)/\check{\Delta}_{\rm opt}(\x,\n_1)$ (black lines)
    and
    $\check{\Delta}_{\rm opt}(\x,\n_3)/\check{\Delta}_{\rm opt}(\x,\n_1)$ (blue lines)
    for 
    (\textit{a}) turbulent channel flow
    and 
    (\textit{b}) an $x$-normal plane inside the recirculation region
    of the flow over a backward-facing step
    ($x/H=2$, see Section~\ref{section:BFS}).
  }
\end{figure}

These expressions for the total error and the computational cost are
simplistic and could of course be made more realistic. 
For example,
the computational cost could include the size of the time step,
especially for compressible solvers.
One advantage of these simple estimates is that the optimal solution
can be found analytically.
First, it is quite easy to show that the optimal solution has the
error indicator
$\check{\G}(\x,\n)$
equally distributed among the different directions, i.e., that
it is equal in all directions $\n$ for every fixed location $\x$
(assuming that $\alpha$ is the same in all directions; 
see~\ref{sec:appendix-optimal} for more details,
including directionally and spatially varying $\alpha$).
This can be solved (for the special case of hexahedral cells treated
here) 
to yield
\beq \label{eq:optimal-direction}
\check{\G}_{\rm opt}(\x, \n_i) = \overline{g}(\x,\n_i) \check{\Delta}_{\rm opt}^\alpha(\x, \n_i)
= \overline{g}_{\rm vol} (\x) \check{\Delta}_{\rm vol, opt}^\alpha(\x)
\ , \ \ 
i=1,2,3
\,,
\eeq
where
\beq \nonumber
\overline{g}_{\rm vol} =
\left(
\overline{g}(\x,\n_1)
\overline{g}(\x,\n_2)
\overline{g}(\x,\n_3)
\right)^{1/3}
\ , \ \ 
\check{\Delta}_{\rm vol,opt} = \left(
\check{\Delta}_{\rm opt}(\x,\n_1)
\check{\Delta}_{\rm opt}(\x,\n_2)
\check{\Delta}_{\rm opt}(\x,\n_3)
\right)^{1/3}
\,.
\eeq
This implies that the predicted optimal cell aspect ratio is, for example,
\beq \nonumber
\frac{\check{\Delta}_{\rm opt}(\x,\n_j)}{\check{\Delta}_{\rm opt}(\x,\n_1)} = 
\left(
\frac{\overline{g}(\x,\n_1)}{\overline{g}(\x,\n_j)}
\right)^{1/\alpha}
\ , \ j=2,3
\,.
\eeq
Examples of the predicted optimal cell aspect ratios
are given in Fig.~\ref{fig:sample-ARs}
for a turbulent channel flow and the region
inside the recirculation bubble
in the flow over a backward-facing step.

The distribution of $ \check{\Delta}_{\rm vol,opt}(\x)$ 
as a function of location $\x$
can be found by solving an optimization problem that minimizes
$\check{e}_{\rm tot}$ of Eqn.~\ref{eq:etot}
with an equality constraint on $N_{\rm tot}$.
This reduces to minimizing the Lagrangian
$\mathcal{L}=\check{e}_{\rm tot}+\lambda N_{\rm tot}$ 
(where $\lambda$ is a Lagrange multiplier)
with respect to $ \check{\Delta}_{\rm vol}(\x)$,
which for the simplified versions of $\check{e}_{\rm tot}$ and $N_{\rm tot}$
for hexahedral cells takes the solution
\beq \label{eq:optimal-space}
\overline{g}_{\rm vol} (\x) \check{\Delta}_{\rm vol,opt}^{\alpha+3}(\x)
=
\Lambda 
= 
{\rm const.}
\eeq

Note that Eqn.~\ref{eq:optimal-space} clearly suggests that
the cell integrated error indicator 
(i.e., the error indicator, $\overline{g}_{\rm vol} (\x) \check{\Delta}_{\rm vol,opt}^{\alpha}(\x)$,
multiplied by the cell volume, ${\Delta}_{\rm vol,opt}^{3}(\x)$)
is the quantity that must be uniformly distributed in 
order to achieve the optimal state. 
A very important implication of this equation
is that the notion
of setting quantitative guidelines
on the error indicators 
(e.g., that a good LES is the 
one that resolves 90\% of the 
total turbulent kinetic energy)
is necessarily suboptimal
(if not meaningless) 
and should always be avoided.

Equations~\ref{eq:optimal-space} and~\ref{eq:optimal-direction}
define our optimal filter-width $\check{\Delta}_{\rm opt}(\x,\n)$
on any given grid with a specified $N_{tot}$ number of cells.

\subsection{The stopping criterion}

The question of convergence under grid refinement (or filter-width
refinement) can be a bit philosophical in the context
of LES.
Since LES is by definition under-resolved, the solution 
will necessarily change and develop smaller scales as one refines the
filter-width.
So in a point-wise (in space and time) sense, 
the LES solution does not converge
for finer filter-widths,
at least not until the DNS limit is reached.
While true, this is not a very
practical definition of convergence for LES.
The National Research Council~\cite{nrc:12:vvuq} suggests that the
best-practice is to identify important simulation outputs
(``quantities of interest'', or QoIs),
defined as 
deterministic functionals
of the solution, 
and assess the convergence of these specific outputs only.
This makes sense: 
if the QoIs we are interested in
did \emph{not} converge well before the
DNS limit (in parts of the domain where LES is used), 
LES would be a pointless tool.
In other words, LES makes sense for QoIs
that are generally functions of
the larger
scales of turbulence (e.g., lift, drag, pressure rms, Reynolds stress, etc.,
that do not \emph{directly} depend on the fine scales of the solution)
but not if the purpose is to predict 
the fine-scale turbulent quantities
(e.g., some short-distance structure function, molecular dissipation or similar,
that are functions of the small-scale instantaneous solution).
Coming back to the adaptation algorithm,
the main point is that 
(i) the convergence should only be judged for
specific QoIs, 
and not based on whether or not the instantaneous solution is converged,
and 
(ii) this judgement cannot be directly 
based on the estimated local sources of error;
i.e., one cannot assume that the LES is accurate
if a certain portion of the turbulence is resolved 
or if the proposed error indicator is below a certain threshold
(even if that threshold in set on the cell integrated error indicator,
consistent with Eqn.~\ref{eq:optimal-space}), 
as that acceptable threshold depends on 
the flowfield and the specific QoIs.

In principle, the adjoint equation can produce such a link between a
QoI and the local error sources, which is why it is possible to
estimate the convergence of the QoI in the adjoint-weighted residual
method \cite[cf.][]{fidkowski:11}.
However, for a chaotic system like LES the adjoint fields
become singular
for long time integration~\cite[cf.][]{lea:00,wang:13,wang:14:lss}
that is required in statistically stationary flows,
and the possible workarounds~\cite[cf.][]{wang:13,wang:14:lss}
are still orders of magnitude more expensive than
solving the forward LES equation. 
As a result,
we are not considering adjoints
in this work.
This implies that the proposed adaptation
algorithm does not contain a criterion for when to stop the process:
this must instead be judged by a user after computing the QoIs and
assessing their convergence.
Note that 
once the adjoint equations can be solved the computed adjoint fields 
can be included in the definition of $\check{e}_{\rm tot}$ in Eqn.~\ref{eq:etot}
(with some minor modifications)
to enable a direct link between the estimated local source of errors 
($\tlpf{\lpf{\F}}_i^{(\n)}\!\!\!\!\!(\x)$ in that case)
and the error in the QoIs.

Assuming that we have $M$ quantities of interest $\lpf{Q}_m$ in the simulation
allows for the total error in these QoIs to be defined as
\beq \label{eq:eQoI-general}
\lpf{e}_{\rm QoI}^{\rm ref} = \sum_{m=1}^M {w_m \delta \lpf{Q}_m^{\rm ref}}
\,,
\eeq
where $\delta \lpf{Q}_m^{\rm ref}$ is the change in $\lpf{Q}_m$
compared to a reference solution
and
$w_m$ is an appropriate weight with $\sum w_m=1$.

Two different reference solutions are used to compute the total error in this paper:
(i) the LES solution on 
the previous grid 
that was used to compute the error indicator and generate the current grid
(labeled $\errprev$);
and
(ii) a converged DNS solution (labeled $\errref$).
In a practical situation only the former is available, and hence
convergence must be judged based on $\errprev$ alone.
The DNS-based error $\errref$ is computed here solely to judge the
true accuracy of each solution in order to assess the adaptation process.

The first grid that satisfies the criterion on $\lpf{e}_{\rm QoI}^{\rm ref} $ is
taken as the ``optimal'' grid in this work. A more conservative
criterion would be to require multiple sequential grids to satisfy the
convergence criterion.

The proposed method is next assessed on turbulent channel flow
(Section~\ref{sec:channel})
and the flow over a backward-facing step
(Section~\ref{section:BFS}).

\section{Assessment on turbulent channel flow at $Re_\tau \approx 545$\label{sec:channel}}

The filter-width adaptation problem is inherently
an optimization problem:
we should therefore check whether the predicted grids/filter-widths
are ``optimal'', in the sense of leading to the best accuracy at the
lowest cost.
While true optimality is 
extremely difficult
to assess in the context of LES
(probably impossible, due to the presence of modeling errors
and the uncertainties introduced by the projection errors),
the turbulent channel flow is arguably as close as we can get given
the many decades of experience with this flow in the LES community
and the detailed analyses available
\cite[cf.][]{meyers:07,rezaeiravesh:18}.
For the turbulent channel cases, we therefore ask whether the
adaptation algorithm can produce grids close to the 
$\Delta_{\rm opt}^+
\approx
(40,1,20)$
or so that is widely considered a ``good'' grid for wall-resolved LES.

All simulations are started from exceedingly coarse grids that are
essentially ignorant of the flow physics; this is done to test the
robustness with severely underresolved solutions. 
In the same spirit, we push the resolution of the final grids to the DNS limit,
to make sure that the method is still robust when the LES model 
becomes effectively inactive.
The idea here is that, no matter how coarse or fine the grid might be, 
a robust method should always drive the grid towards a distribution
that leads to lower errors in the solution.

To further test the robustness of the method,
we consider three different 
approaches:
(i) LES with a mixture of modeling and numerical errors;
(ii) LES where the modeling errors are dominant; and
(iii) DNS, which is purely affected by numerical errors.

The predicted filter-widths and associated solution accuracy are
compared to those from our previous work
\cite{toosi:17}
in~\ref{section:An}.

\subsection{Code and problem specification}

The code used for this problem is the \emph{Hybrid} code, which
solves the compressible Navier-Stokes
equations for a calorically perfect gas on structured Cartesian grids
using sixth-order accurate central differencing schemes with a split
form of the convective term 
(skew-symmetric in the limit of zero Mach number)
for increased numerical stability.
Time-integration is handled by classic fourth-order Runge-Kutta.
The code solves the implicitly-filtered LES equations
with an explicit eddy viscosity model.

The bulk Reynolds number $Re_b = \rho_b U_b H/ \mu_w$
(where $\rho_b$ is the bulk density, $U_b$ is the bulk velocity, $H$
is the channel half-height and $\mu_w$ is the viscosity at
the wall) is 10,000, which leads to a friction Reynolds number of
about $Re_\tau \approx 545$.
The bulk Mach number is 0.2.
The computational domains are of size
$(L_x,L_y,L_z)=(10H,2H,3H)$.
Since the code is structured, the grid-spacing in the wall-parallel
directions is taken as the smallest predicted value along $y$.
The simulations are integrated for a time of $200 H/U_b$ (around $11 H/u_\tau$)
before collecting statistics over
a period of $600 H/U_b$ (slightly more than $32 H/u_\tau$),
by post-processing 400 snapshots that are
$1.5 H/U_b$ (close to $0.08 H/u_\tau$)
apart from each other.
The convergence error is found to be sufficiently small as to not
affect any of the conclusions in this study.
This long integration time is primarily required for convergence of the mean profiles, 
while the adaptation process can actually be performed with averages
collected over a much shorter time since the error indicator 
is primarily affected by small scales.
A careful study of the statistical convergence of the error indicator 
and its predicted grids
is delayed until Section~\ref{sec:statisticalConvergence}.

To measure convergence, the QoIs are defined based on the mean
velocity and the Reynolds stresses. Specifically,
the errors in the QoIs are defined as
\beq  \label{eqn:err_channel_delQ}
\begin{aligned}
\delta \lpf{Q}_{1}^{{\rm ref}}
&=
\frac{
\int_a^b
  \left| \lpf{U}_{1}^+ - \widetilde{U}^+_{1,{\rm ref}} \right|
  d \left( \ln y^+ \right)
}{
\int_a^b \widetilde{U}^+_{1,{\rm ref}}
 \, d \left( \ln y^+ \right)
}
\\
\delta \lpf{Q}_{2\mbox{-}5}^{{\rm ref}}
&=
\frac{
\int_a^b
  \left| \lpf{R}_{ij}^{{\rm tot}^+} - \widetilde{R}_{ij,{\rm ref}}^{{\rm tot}^+} \right|
  d \left( \ln y^+ \right)
}{
\int_a^b \widetilde{R}_{kk,{\rm ref}}^{{\rm tot}^+}/2
 \, d \left( \ln y^+ \right)
}
\ , \ \ 
(i,j) = (1,1),\,(2,2),\,(3,3),\,(1,2)
\,.
\end{aligned}
\eeq
where 
$\lpf{U}_1 = \avg{\lpf{u}_1}$
and 
$\lpf{R}^{\rm tot}_{ij} = \avg{\lpf{u}_i^\prime \lpf{u}_j^\prime} +
\avg{\tau_{ij}^{\rm mod} ( \lpf{u}_k)}$
are the mean
velocity and the total Reynolds stress on the LES grid 
(with characteristic filter-width $\lpf{\Delta}$). 
The reference quantities
$\widetilde{U}_{1,{\rm ref}}$ and $\widetilde{R}^{\rm tot}_{ij,{\rm ref}}$
are
taken either from the previous LES grid in the sequence of adapted
grids (for $\errprev$)
or from the DNS of del Alamo \& Jimenez~\cite{delalamo:03}
(for $\errref$).
The integration limits ($a$ and $b$) are taken as 
$y^+=2$ to $y^+=Re_\tau/2$ (i.e., $y=H/2$),
where the core of the channel is excluded since it is the most
affected by the domain size. 
Note that the error in all of the Reynolds stresses is normalized by the
(integral of the) kinetic energy $\widetilde{R}^{{\rm tot}^+}_{kk} /2$.
These five $\delta \lpf{Q}_m^{{\rm ref}}$ are then equally weighted 
to form the final error metric
\beq \label{eqn:err_channel}
\lpf{e}_{\rm QoI}^{\rm ref} = 
\frac{1}{5}
\sum_{m=1}^5 \delta \lpf{Q}_{m}^{{\rm ref}}
\,.
\eeq

\subsection{LES with a mixture of modeling and numerical errors \label{section:DSM}}

We first use the dynamic Smagorinsky model~\cite[]{germano:91,lilly:92}
with filtering and averaging in the wall-parallel directions
to compute $\tau_{ij}^{\rm mod}$.
Since the implicitly filtered LES equations are solved in the code
the filter-width is implicitly assumed to be equal to the grid-spacing, i.e.,
$\lpf{\Delta}/\lpf{h}=1$.
Combined with the use of numerics with low numerical dissipation,
this produces solutions
that are contaminated
by both modeling and numerical errors
of about similar magnitudes~\cite[cf.][]{ghosal:96,kravchenko:96}.

This first grid has a uniform resolution of 
$(\lpf{\Delta}_x,\lpf{\Delta}_y,\lpf{\Delta}_z)/H = (0.20,0.10,0.20)$,
corresponding to 
$(\lpf{\Delta}_x^+,\lpf{\Delta}_{y_w}^+/2,\lpf{\Delta}_z^+) \approx (110, 28, 110)$
if one uses the fully converged friction velocity.
Note that $\lpf{\Delta}_{y_w}^+/2$ is the distance between the first grid point
and the wall.

After performing an LES on this grid, we need to compute the error
indicator which requires the computation of the eddy viscosity
at the test-filter level.
Assuming that the model
coefficient is the same at the grid- and test-filter levels,
this can be computed approximately as
\beq \label{eq:DSM-approximate}
\mathcal{V}_{\rm sgs}^{(\n_0)}
\approx 
\left[ 
\frac{\tlpf{\lpf{\Delta}}^{(\n_0)}}{ \lpf{\Delta}} 
\! \right]^{\! 2}
\frac{ \left\vert \tlpf{ \lpf{S} }^{(\n_0)} \! \right\vert }{ \vert
  \lpf{S} \vert}
\,
\nu_{\rm sgs}
\,,
\eeq
where $\nu_{\rm sgs}$ is the eddy viscosity in the underlying LES.
The effect of this approximation is minor, with a full assessment
shown in~\ref{section:incons}.
We assume $\tlpf{\lpf{\Delta}}^{(\n_0)} \!\!\! / \lpf{\Delta} \approx
\sqrt[3]{2}$
for all 3 directions,
since the test-filter of Eqn.~\ref{eq:filter}
is wider by a factor of two in only one direction.
This assumes that the characteristic filter-width is taken as the
cube-root of the cell volume, which is actually not explicitly
enforced here since the filter-width definition can be absorbed into
the model constant in the dynamic procedure.

\begin{figure}[t!]
	\centering	
	\includegraphics[width=38mm,clip=true,trim=0mm 0mm 10mm 3mm]{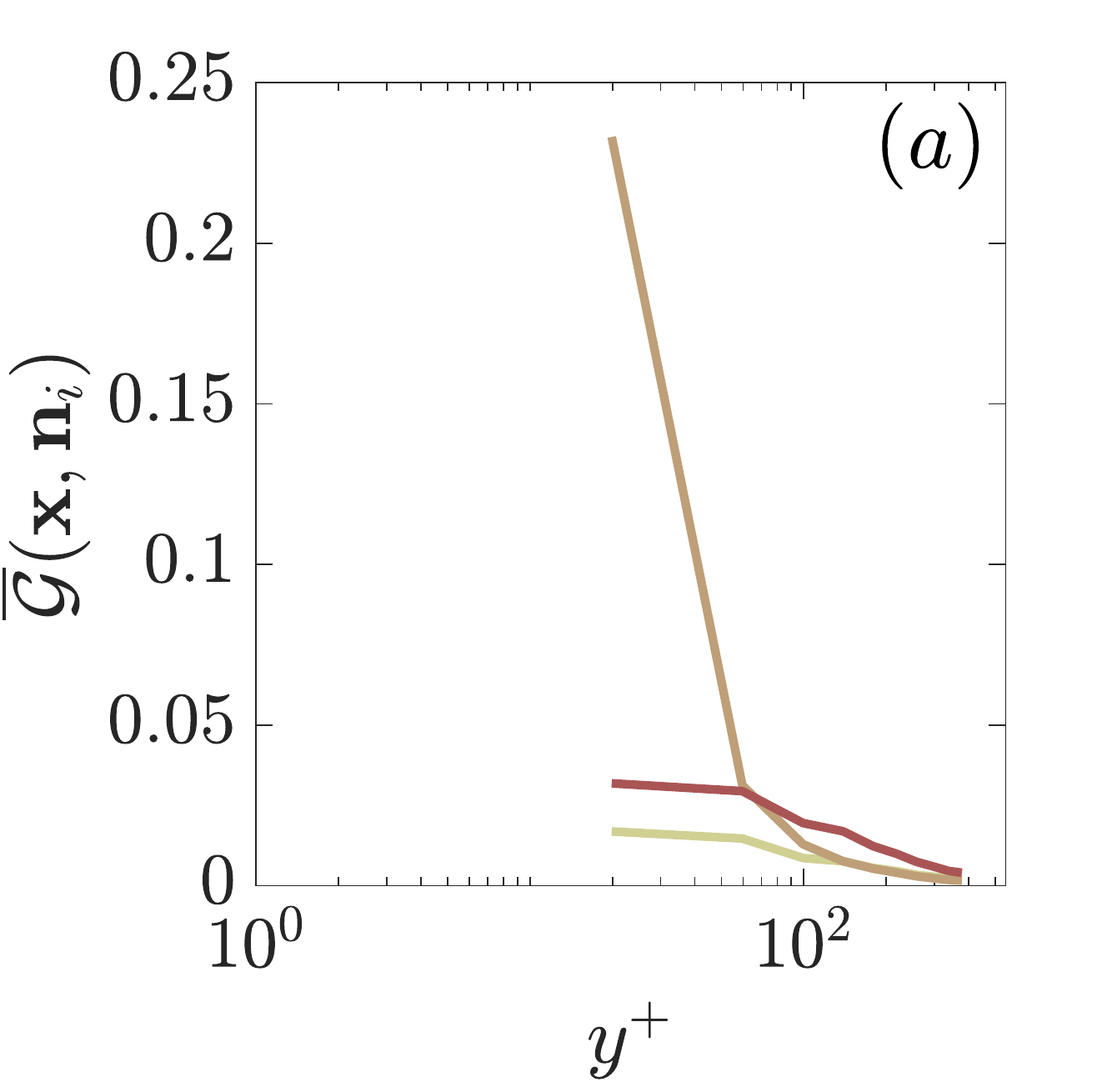}
	\includegraphics[width=38mm,clip=true,trim=0mm 0mm 10mm 3mm]{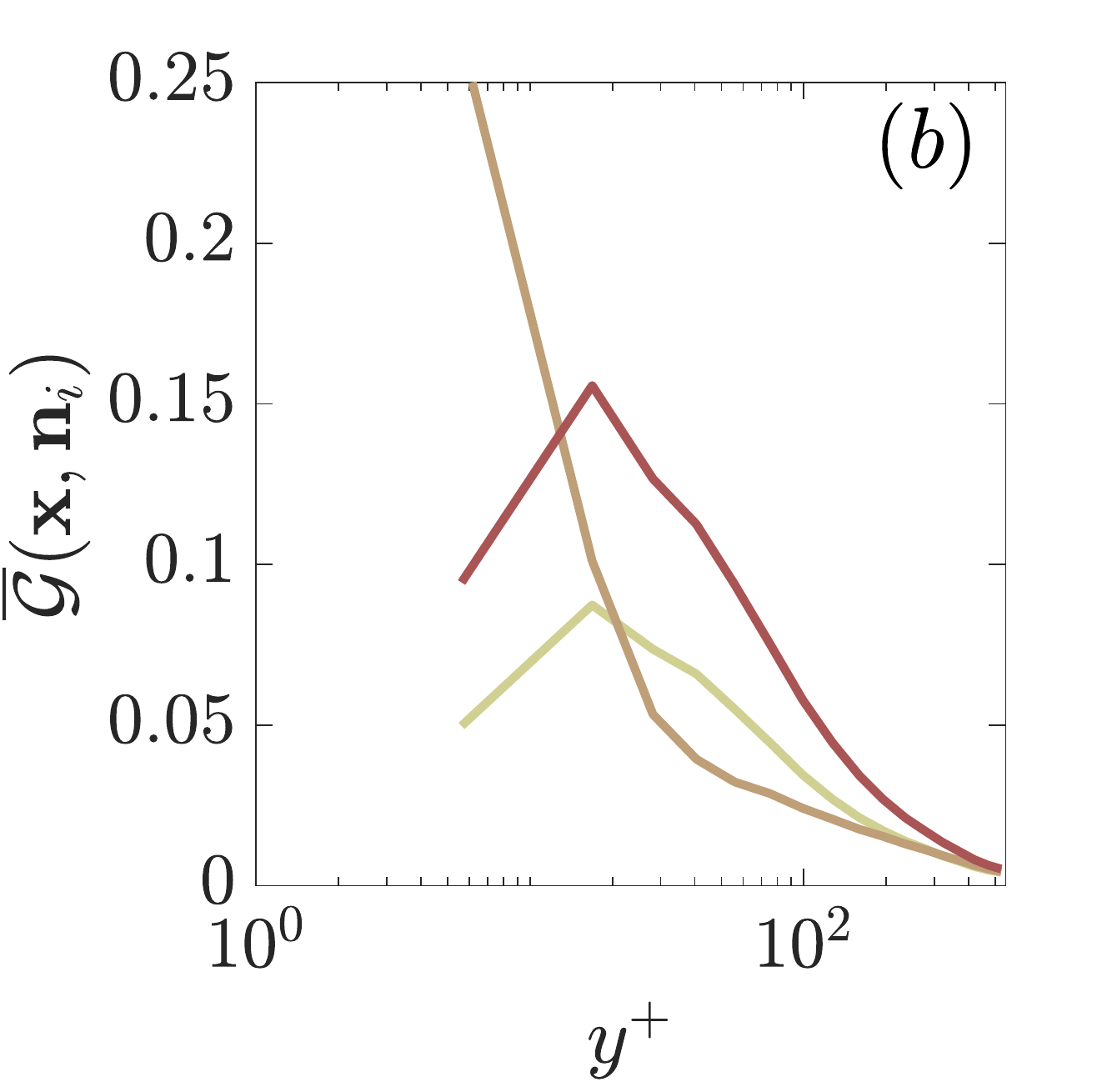}
	\includegraphics[width=38mm,clip=true,trim=0mm 0mm 10mm 3mm]{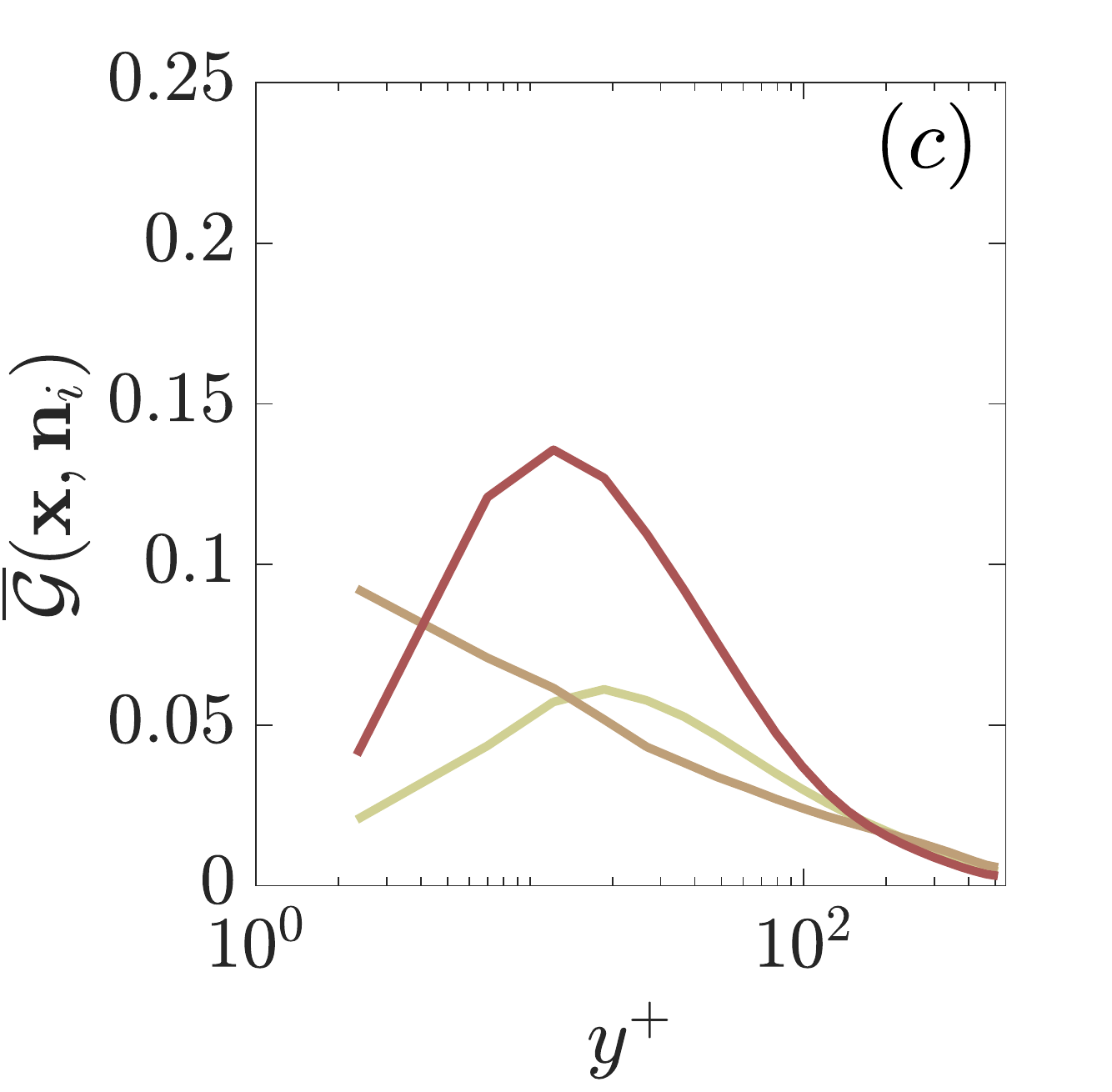}
	\caption{\label{fig:Gn-profiles}
	Computed values of $\lpf{\G}(\x,\n)$
	for grids (\textit{a}) DSM-1, 
	(\textit{b}) DSM-2 and (\textit{c}) DSM-3.
	The streamwise, 
	wall-normal 
	and spanwise 
	directions
	are shown by the lightest to the darkest colors (in that order).
	}
\end{figure}

The computed error indicator on the first grid is plotted in Fig.~\ref{fig:Gn-profiles}.
As expected for this coarse uniform grid, 
the largest error indicator is for the wall-normal direction in the
vicinity of the wall.
The next grid (DSM-2) is then generated 
by enforcing the grid selection criteria of 
Eqns.~\ref{eq:optimal-direction} and~\ref{eq:optimal-space}.
Note that due to the structured nature of the computational grid
we have to take the minimum of the target streamwise and spanwise resolutions
across the channel in order to generate each of the grids.
The constant $\Lambda$ in Eqn.~\ref{eq:optimal-space} is
adjusted (in an iterative process) 
such that the number of grid points in the next grid DSM-2 increases by a factor of 5.

\begin{figure}[t!]
	\centering	
	\includegraphics[width=130mm,clip=true,trim=10mm 10mm 10mm 13mm]{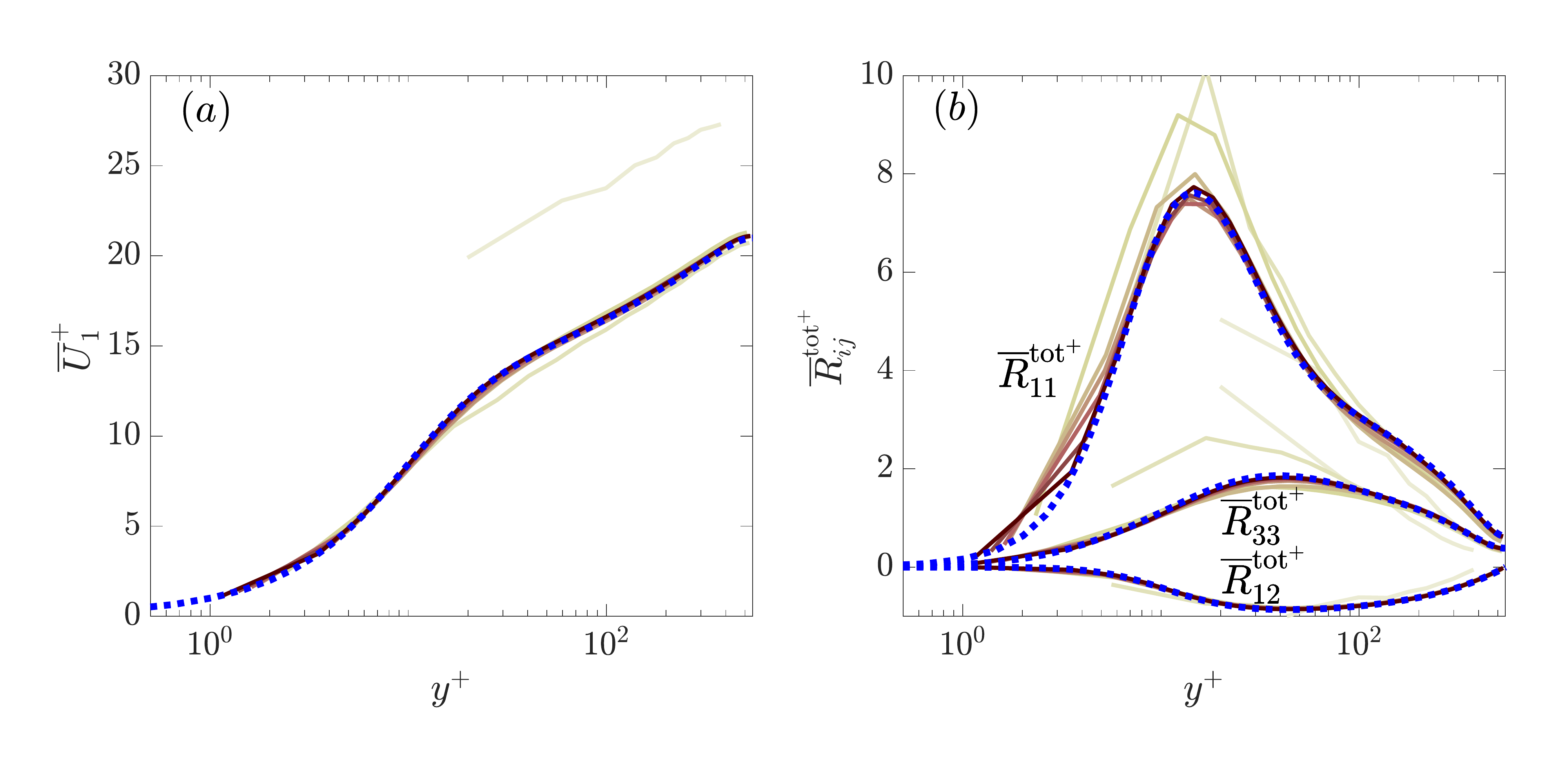}
	\caption{\label{fig:channel-Gn-DSM}
	Convergence of the mean velocity and Reynolds stress profiles
	for the grids in Table~\ref{Table:channel-Gn-DSM}.
	Grids in the sequence are shown by the lightest color for DSM-1
	to the darkest color for DSM-8.
	The dotted blue lines show the DNS solution 
	of del Alamo \& Jimenez~\cite{delalamo:03} at the same $Re_\tau$. 
	}
\end{figure}

The key metrics for all grids are reported in
Table~\ref{Table:channel-Gn-DSM},
and the solutions are shown in Fig.~\ref{fig:channel-Gn-DSM}.

The grid DSM-2 has a grid-spacing of
$(\lpf{\Delta}_x^+,\lpf{\Delta}_{y_w}^+/2,\lpf{\Delta}_z^+)=(77,5.6,55)$.
The solution on this grid is actually not bad
(Fig.~\ref{fig:channel-Gn-DSM}),
but of course not converged.
The error indicator computed from the DSM-2 solution is shown in
Fig.~\ref{fig:Gn-profiles},
and again shows the largest error coming from the wall-normal
resolution near the wall, followed by the spanwise resolution
throughout much of the buffer layer.
The resulting grid DSM-3 produces a solution where the Reynolds
stresses are close to the DNS and where the error indicator values in
the different directions are closer to being balanced, suggesting that
the algorithm has found a nearly ``optimal'' state.

\begin{table}[t!]
\begin{center}
\begin{tabular}{ c  c  c  c  c  c  c  c  }
  Grid
  & $N_{\rm tot}$	
  & $N_y$
  & $(\lpf{\Delta}_x^+, \lpf{\Delta}_{y_w}^+/2 ,\lpf{\Delta}_z^+)$	
  & $(\lpf{\Delta}_x , \lpf{\Delta}_{y_c} , \lpf{\Delta}_z)/H$	
  & $Re_{\tau}$
  &  $\errref$ (\%)
  &  $\errprev$ (\%)
  \\[3pt]
  DSM-1
  & $15k$
  & 20
  & $(80,20,80)$
  & $(0.20,0.10,0.20)$
  & 398 	
  & 32
  & $-$
  \\
  DSM-2
  & $74k$
  & 34
  & $(77,5.6,55)$
  & $(0.14,0.099,0.10)$
  & 553	
  & 11
  & 27
  \\
  DSM-3
  & $251k$
  & 44
  & $(53,2.3,29)$
  & $(0.098,0.091,0.054)$
  & 536
  & 7.3
  & 8.0
  \\
  DSM-4
  & $514k$
  & 50
  & $(45,1.7,19)$
  & $(0.082,0.080,0.035)$
  & 544	
  & 3.3
  & 4.2
  \\
  DSM-5
  & $1.18M$
  & 60
  & $(34,1.4,13)$
  & $(0.063,0.065,0.024)$
  & 544	
  & 1.8
  & 2.1
  \\
  DSM-6
  & $2.53M$
  & 72
  & $(25,1.6,10)$
  & $(0.046,0.052,0.018)$
  & 542
  & 1.1
  & 1.0
    \\
  DSM-7
  & $5.80M$
  & 90
  & $(18,1.4,7.6)$
  & $(0.033,0.041,0.014)$
  & 540	
  & 1.1
  & 0.6
      \\
  DSM-8
  & $11.1M$
  & 108
  & $(14,1.2,6.3)$
  & $(0.025,0.033,0.012)$
  & 541
  & 0.9
  & 0.6
\end{tabular}
\end{center}
\caption{ \label{Table:channel-Gn-DSM}
  Sequence of grids generated for LES of channel flow at $Re_\tau\approx545$
  using the dynamic Smagorinsky model.
  $N_{\rm tot}$ is the total number of grid points,
  while $N_y$ denotes the number of points across the channel.
  $\lpf{\Delta}_\n = \lpf{\Delta}(\x,\n)$ is both the filter-width and
  the grid-resolution.
  Friction resolutions $\lpf{\Delta}_\n^+$ are computed based on grid-specific values.
  $\lpf{\Delta}_{y_w}^+/2$ is distance from the wall of the first grid point.
  $\lpf{\Delta}_{y_c} $ is the wall-normal filter-width at the center of the channel.
  $\errref$ and $\errprev$ are defined by Eqns.~\ref{eqn:err_channel_delQ} 
  and~\ref{eqn:err_channel}.
}
\end{table}

The adaptation algorithm is continued until DSM-8.
After the first two adaptations, the target number of cells is doubled
each time.
The solution is effectively converged on grid DSM-4 or DSM-5 depending
on the desired accuracy.
The grid-spacings on grids DSM-4 and up
are quite close to 
what is considered ``best practice'' in LES and DNS for channel flows,
with
$(\lpf{\Delta}_x^+ , \lpf{\Delta}_{y_w}^+/2 , \lpf{\Delta}_z^+)$
of
$(45,1.7,19)$ on DSM-4
and
$(14,1.2,6.3)$ on DSM-8.

\subsection{LES with dominant modeling errors and small numerical errors \label{section:vreman}}

The next test case tries to assess the performance of the error indicator in a
flow where the numerical errors 
are relatively small and 
the solution is mostly dominated by the effect of modeling errors.
This is achieved here by taking 
$\lpf{\Delta}/\lpf{h}=2$
and using the
eddy viscosity model by Vreman~\cite{vreman:04}
with a constant coefficient of $c_v=0.03$.
The use of a filter-width larger than the grid-spacing causes the eddy
viscosity to increase by a factor of 4, which dissipates most of the
energy before reaching the Nyquist limit of the grid.

\begin{figure}[t!]
	\centering	
	\includegraphics[width=130mm,clip=true,trim=10mm 10mm 10mm 13mm]{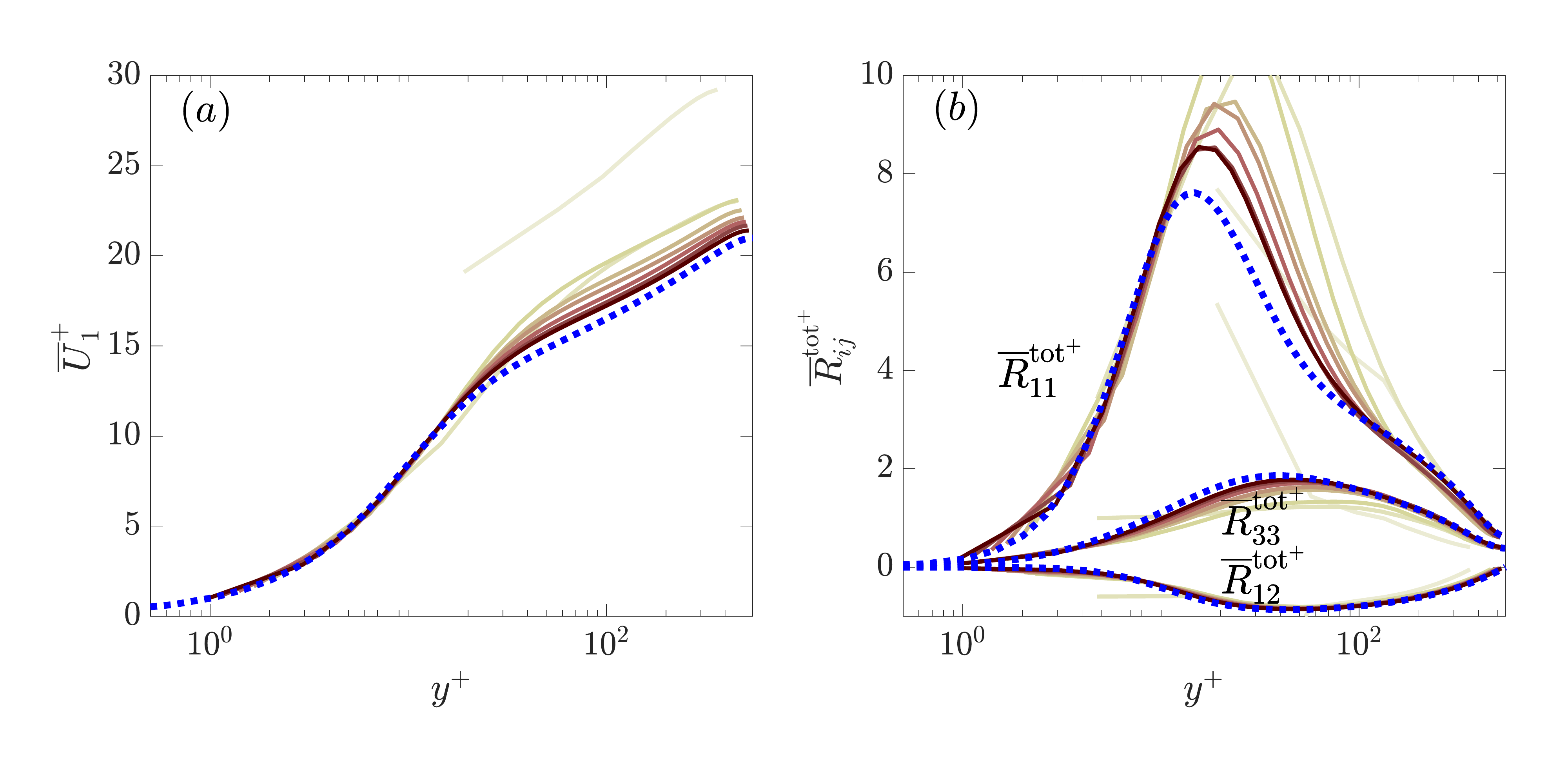}
	\caption{\label{fig:channel-Gn-trueVreman}
	Convergence of the mean velocity and Reynolds stress profiles
	for grids in Table~\ref{Table:channel-Gn-trueVreman}
	generated for LES with the constant coefficient Vreman model
        and $\lpf{\Delta}/\lpf{h}=2$.
	Colors vary from the brightest for the first grid to the darkest for the last one.
	The dotted blue lines are the DNS of del Alamo \& Jimenez~\cite{delalamo:03}.
	}
\end{figure}

The sequence of grids and solutions are summarized in
Table~\ref{Table:channel-Gn-trueVreman}
and Fig.~\ref{fig:channel-Gn-trueVreman}.
The initial grid has the same number of grid points as for the dynamic
Smagorinsky case, but twice the filter-width.
The subsequent grids in the sequence have approximately the same
number of grid points as the corresponding dynamic Smagorinsky cases.

\begin{table}[t!]
\begin{center}
\begin{tabular}{ c c c  c  c  c  c  c  }
  Grid
  & $N_{\rm tot}$	
  & $N_y$
  & $(\fDelta_x^+, \lpf{\Delta}_{y_w}^+/2,\fDelta_z^+)$	
  & $(\fDelta_x , \fDelta_{y_c} , \fDelta_z)/H$	
  & $Re_{\tau}$
  &  $\errref$ (\%)
  &  $\errprev$ (\%)
  \\[3pt]
  Vr-1
  & $15k$
  & 20
  & $(153,38,153)$
  & $(0.40,0.20,0.40)$
  & 382
  & 34
  & $-$
  \\
  Vr-2
  & $73k$
  & 34
  & $(135,9.5,97)$
  & $(0.28,0.22,0.20)$
  & 487
  & 21
  & 25
  \\
  Vr-3
  & $256k$
  & 44
  & $(103,4.7,47)$
  & $(0.21,0.18,0.097$
  & 484
  & 18
  & 7.7
  \\
  Vr-4
  & $517k$
  & 50
  & $(91,4.1,32)$
  & $(0.18,0.16,0.064)$
  & 500
  & 11
  & 6.0
  \\
  Vr-5
  & $1.16M$
  & 62
  & $(68,3.3,24)$
  & $(0.13,0.12,0.048)$
  & 510
  & 9.7
  & 2.3
  \\
  Vr-6
  & $2.51M$
  & 76
  & $(49,2.8,20)$
  & $(0.096,0.096,0.038)$
  & 518
  &7.1
  & 2.8
    \\
  Vr-7
  & $5.83M$
  & 96
  & $(35,2.3,15)$
  & $(0.068,0.075,0.029)$
  & 524
  & 5.1
  & 2.1
    \\
  Vr-8
  & $11.0M$
  & 114
  & $(27,1.9,13)$
  & $(0.052,0.061,0.024)$
  & 530
  & 4.4
  & 1.1
\end{tabular}
\end{center}
\caption{ \label{Table:channel-Gn-trueVreman}
  Sequence of grids generated for LES of 
  turbulent channel flow 
  using the Vreman model with a model constant of $c_v=0.03$ and $\lpf{\Delta}/\lpf{h}=2$.
  Additional details on the notation are given in the caption of Table~\ref{Table:channel-Gn-DSM}.
  All resolutions are based on the filter-width, not the grid-spacing.
}
\end{table}

The solutions converge much more slowly for this case, which is
consistent with the broadly agreed upon notion that
for a given grid-spacing $\lpf{h}$
the choice of $\lpf{\Delta} \approx
\lpf{h}$ leads to the best LES accuracy in most cases.
In other words, that the increase in modeling error for larger
filter-widths is greater than the decrease in numerical error.

More interestingly (in the present context) is that the last few grids
again agree quite closely with the ``best practice'' in LES,
and in fact agree rather well with the grids for the dynamic
Smagorinsky model.
For example, grid Vr-5 in Table~\ref{Table:channel-Gn-trueVreman}
has a grid-spacing (half the filter-width) of $(34,1.6,12)$ in viscous
units,
which is almost identical to the 
resolution of $(34,1.4,13)$
for grid
DSM-5 in Table~\ref{Table:channel-Gn-DSM}.

\subsection{DNS affected solely by numerical errors}

The final channel case is to turn off the LES subgrid model and thus
have only numerical errors.
The adaptation algorithm remains the same except that
${\tau}_{ij}^{\rm mod} = 0$
in both the solver and when computing the error indicator.

\begin{figure}[t!]
	\centering	
	\includegraphics[width=130mm,clip=true,trim=10mm 10mm 10mm 13mm]{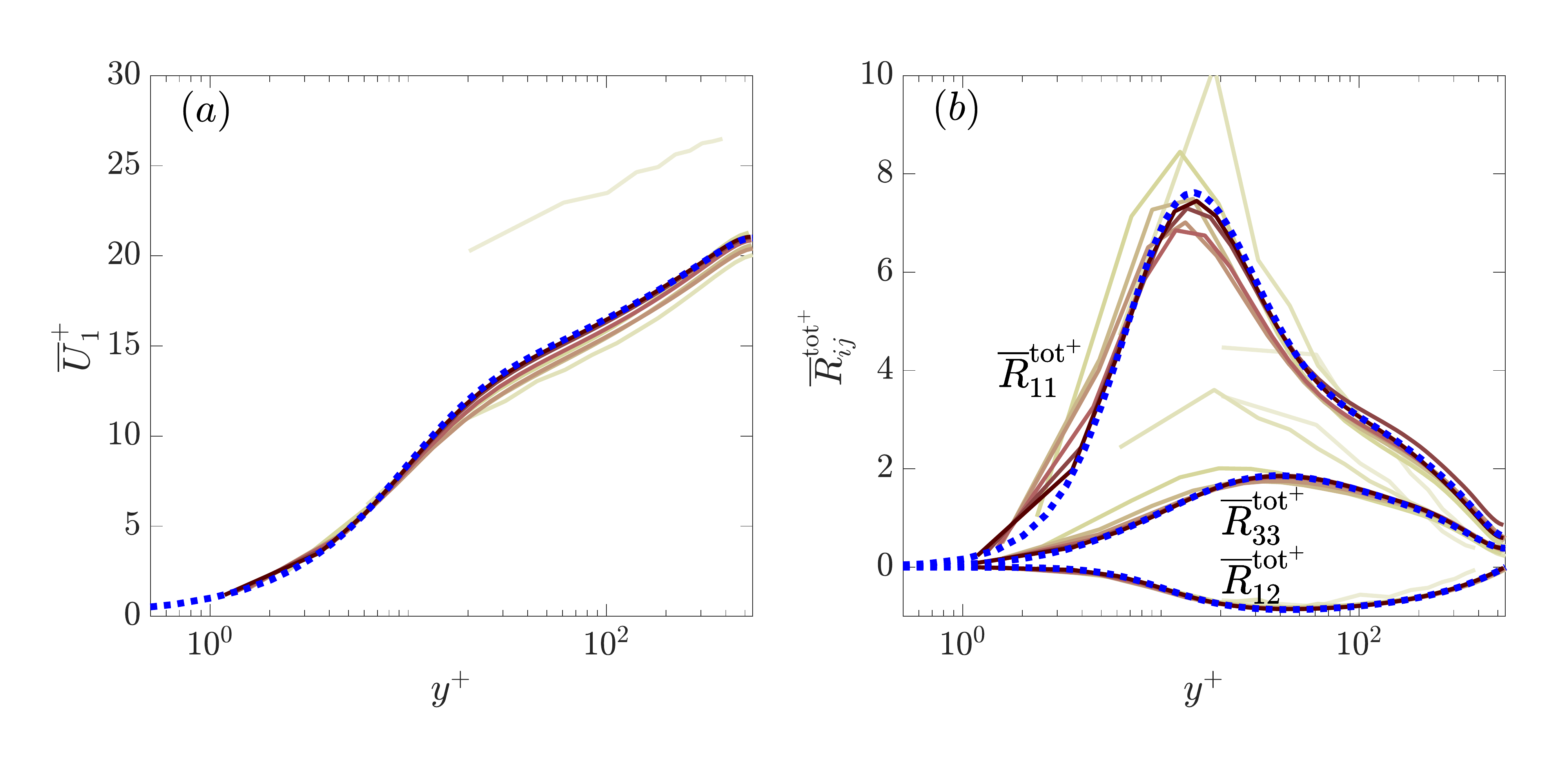}
	\caption{\label{fig:channel-Gn-DNS}
	Convergence of the mean velocity and Reynolds stress profiles
	for grids in Table~\ref{Table:channel-Gn-DNS}.
	Colors vary from the brightest for DNS-1
	to the darkest for DNS-8.
	The dotted blue lines show the DNS of del Alamo \& Jimenez~\cite{delalamo:03}.
	}
\end{figure}

When creating the sequence of grids we target the same number of grid points as in the
previous cases.
Key metrics are
summarized in Table~\ref{Table:channel-Gn-DNS}
with the convergence of the mean velocity and Reynolds stress profiles 
shown in Fig.~\ref{fig:channel-Gn-DNS}.

\begin{table}[t!]
\begin{center}
\begin{tabular}{ c c  c  c  c c  c  c  }
  Grid
  & $N_{\rm tot}$	
  & $N_y$
  & $(\fDelta_x^+, \lpf{\Delta}_{y_w}^+/2,\fDelta_z^+)$	
  & $(\fDelta_x , \fDelta_{y_c} , \fDelta_z)/H$	
  & $Re_{\tau}$
  &  $\errref$ (\%)
  &  $\errprev$ (\%)
  \\[3pt]
  DNS-1
  & $15k$
  & 20
  & $(81,20,81)$
  & $(0.20,0.10,0.20)$
  & 405
  & 30
  & $-$
  \\
  DNS-2
  & $76k$
  & 34
  & $(80,6.2,56)$
  & $(0.14,0.094,0.097)$
  & 578
  & 16
  & 24
  \\
  DNS-3
  & $252k$
  & 44
  & $(51,2.4,31)$
  & $(0.093,0.092,0.057)$
  & 547
  & 6.7
  & 13
  \\
  DNS-4
  & $515k$
  & 50
  & $(43,1.6,22)$
  & $(0.076,0.086,0.039)$
  & 563
  & 5.0
  & 5.3
  \\
  DNS-5
  & $1.14M$
  & 60
  & $(34,1.6,15)$
  & $(0.060,0.067,0.026)$
  & 566
  & 4.4
  & 2.2
  \\
  DNS-6
  & $2.53M$
  & 72
  & $(25,1.5,10)$
  & $(0.046,0.054,0.019)$
  & 553
  & 2.9
  & 2.1
   \\
  DNS-7
  & $5.87M$
  & 90
  & $(18,1.3,7.7)$
  & $(0.033,0.042,0.014)$
  & 545
  & 1.8
  & 2.4
     \\
  DNS-8
  & $11.0M$
  &106
  & $(13,1.2,6.4)$
  & $(0.025,0.034,0.012)$
  & 543
  & 0.9
  & 0.9
\end{tabular}
\end{center}
\caption{ \label{Table:channel-Gn-DNS}
  Sequence of grids generated for DNS of turbulent channel flow at $Re_\tau \approx 545$.
  Additional details on the notation are given in the caption of Table~\ref{Table:channel-Gn-DSM}.
}
\end{table}

The sequence of grids is very similar to those produced for the
dynamic Smagorinsky and constant Vreman models in the previous
sections.
This is partly due to the similarity between wall-resolved LES
and DNS grids in wall-bounded turbulence, 
and partly due to the fact that $\tlpf{\lpf{\F}}_i^{(\n)}$
(Eqn.~\ref{eq:F})
reduces to the divergence of the Leonard-like stress
which based on Eqn.~\ref{eq:Lij}
resembles a truncation error
(although that of a different numerical scheme).

The error in the QoIs is larger for the DNS (no-model)
cases than the dynamic Smagorinsky ones, showing that the model has a
positive effect for this particular flow and code.

\section{Assessment on the flow over a backward facing step at $Re_H = 5100$\label{section:BFS}}

The purpose of this test case is to expose the adaptation algorithm to
a more complex flow, with multiple different canonical flow elements:
an attached boundary layer upstream of the step, 
a free shear layer after the separation, 
an impingement/reattachment region, 
and a large recirculation zone.
This combination of different types of building-block flows is meant
to challenge the adaptation algorithm.

\begin{figure}[t!]
  \centering	
  \includegraphics[width=80mm,clip=true,trim=40mm 5mm 25mm 0mm]{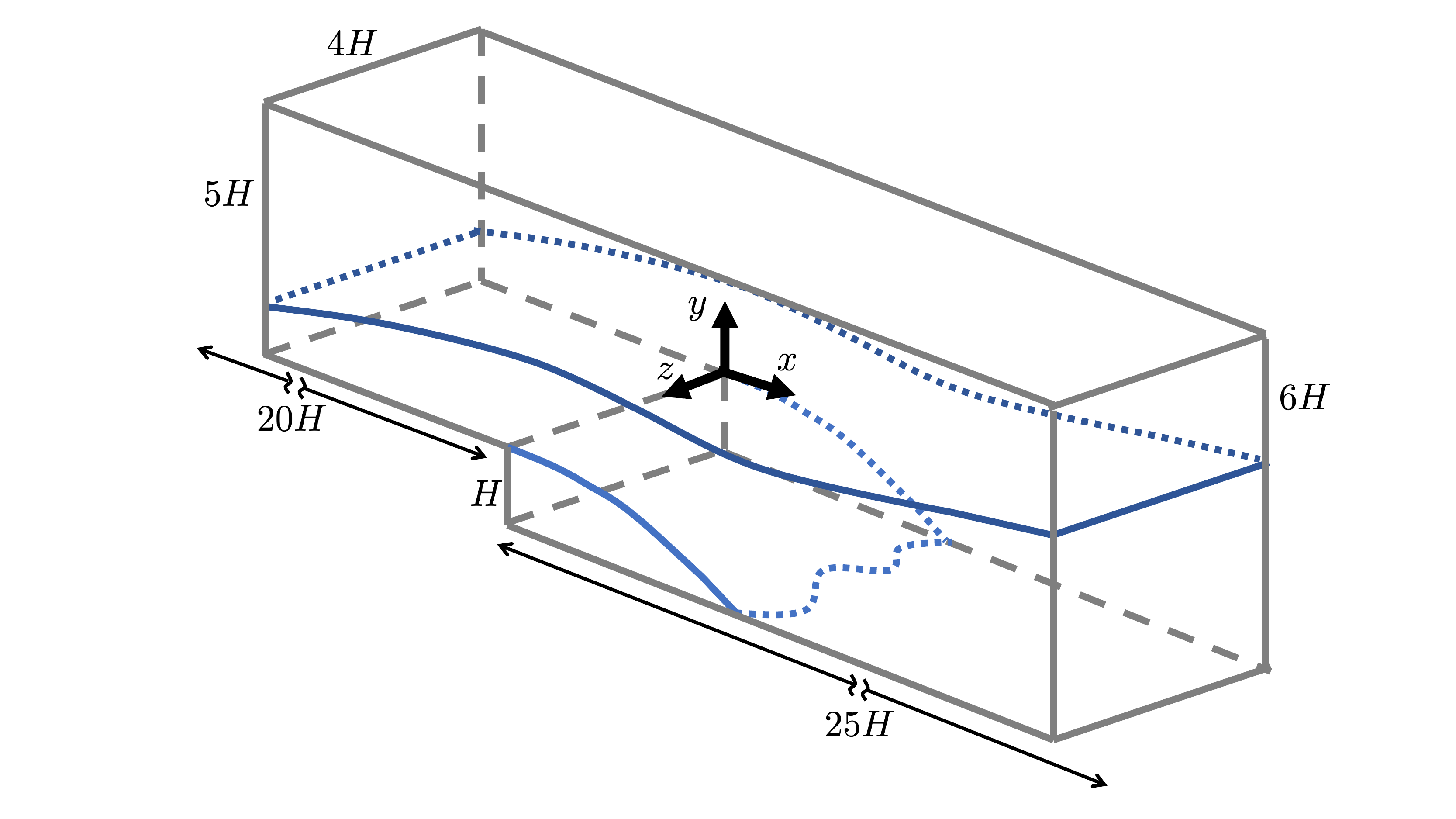}
  \caption{\label{fig:BFS-schematic}
    Schematic of the computational domain 
    for the flow over a backward-facing step.
    The top boundary is a slip wall 
    (modeling the centerline in the experiment)
    while periodic boundary conditions are used in
    the spanwise direction.
    The origin of the coordinate systems is placed
    at the upper corner of the step.
  }
\end{figure}

The flow geometry and conditions are chosen based on
the experiment of Jovic \& Driver~\cite[]{jovic:94:bfs,jovic:95:bfs}
and the DNS of Le~\etal~\cite[]{le:97:bfs}.
The computational domain is shown in Fig.~\ref{fig:BFS-schematic}.
The Reynolds number based on the step height $H$ and inflow 
velocity $U_\infty$ is $Re_H=U_\infty H/\nu =5100$.
This corresponds to a momentum Reynolds number of
$Re_\theta \approx 780$ for the incoming boundary layer (at $x/H=-3$)
and a friction Reynolds number of 
$Re_\tau \approx 208$ based on
the $\delta_{95}$ boundary layer thickness 
(or $Re_\tau \approx 447$ based on $\delta_{99}$)
at that same location.
Note that the flow conditions are close to those of the
experiment and the DNS, but not exactly the same:
the present setup has a thicker boundary layer
compared to that of the experiment 
(which has $Re_\theta \approx 610$).

\subsection{Code and computational details}

The OpenFOAM code version 2.3.1~\cite{openfoam} is used for this test case 
to allow for fully unstructured adapted grids.
Spatial discretization is done using the linear Gauss scheme (second-order accurate),
with second-order backward method for time integration. 
The pressure-velocity coupling is performed using the PISO algorithm
with three iterations of nonorthogonality correction. 
The filter-width is taken as the cube-root of the cell volume.
We use the dynamic $k_{\rm sgs}$-equation 
model~\cite[cf.][]{yoshizawa:93,kim:95,menon:96,chai:12}
that defines the eddy viscosity as
\beq \nonumber
\nu_{\rm sgs} = c_k  \lpf{\Delta} \sqrt{ \lpf{k}_{\rm sgs} }
\eeq
and solves a transport equation for $\lpf{k}_{\rm sgs}$.
This raises the question of how to compute 
$\tlpf{\lpf{k}}^{(\n_0)}_{\rm sgs} $ and thus $\mathcal{V}^{(\n_0)}$
at the test-filter level.
In the present work, we use the simple approach of assuming that 
the eddy viscosity scales as
$\nu_{\rm sgs} \sim \lpf{\Delta}^2  \vert \lpf{S} \vert$
(a consistency requirement for eddy viscosity models),
which then allows us to use the approximate relation
(\ref{eq:DSM-approximate})
to compute the eddy viscosity at the test-filter level.
Similar to the channel case, the effect of this approximation is
assumed to be small (the assessment is in~\ref{section:incons}).
We again take
$\tlpf{\lpf{\Delta}}^{(\n_0)} \!\! / \lpf{\Delta} \approx \sqrt[3]{2} $,
which comes from our definition of the characteristic filter-width
as the cube root of the cell volume.

The quantities of interest for this flow
are taken to be the two non-zero mean velocity components,
the four non-zero Reynolds stress components,
and the friction and pressure coefficient profiles on the horizontal walls.
The errors in the QoIs are defined as
\beq  \nonumber
\begin{aligned}
\delta \lpf{Q}_{1\mbox{-}2}^{\rm ref}
&=
\frac
{
\int \! \! \int_\Omega
 \left| \lpf{U}_{i} - \widetilde{U}_{i,{\rm ref}} \right|
   dxdy
}
{
  0.2 U_\infty \,
  A_\Omega
}
;
\, \, \, \, \, \,
i=1,\,2
\\
\delta \lpf{Q}_{3\mbox{-}6}^{{\rm ref}}
&=
\frac
{
\int \! \! \int_\Omega
  \left|  \lpf{R}^{\rm tot}_{ij} - \widetilde{R}^{\rm tot}_{ij,{\rm ref}} \right|
   dxdy
}
{
  0.015 U_\infty^2 \,
  A_\Omega
}
;
\, \, \, \, \, \,
(i,j) = (1,1),\,(2,2),\,(3,3),\,(1,2)
\\
\delta \lpf{Q}_{7}^{{\rm ref}}
&=
\frac
{
\int_\Psi
  \left| \lpf{c}_{f} - \widetilde{c}_{f,{\rm ref}} \right|
  dx
}
{
  0.002 \,
  L_\Psi
}
;
\\
\delta \lpf{Q}_{8}^{{\rm ref}}
&=
\frac
{
\int_\Psi
  \left|  \lpf{c}_{p} - \widetilde{c}_{p, {\rm ref}} \right|
  dx
}
{
  0.1 \,
  L_\Psi
}
\,.
\end{aligned}
\eeq
where the first two integrals are taken over the region
$\Omega \!: \!(x,y) \in [-10H,20H] \times
[-H,2H]$, with
$A_\Omega = 10H \! \times \! 2H + 20H \! \times \! 3H$ denoting the area of this region.
The remaining two integrals are taken over the horizontal walls in the
region
$\Psi: x \in [-10H,20H]$
with $L_\Psi = 30H$ denoting the normalizing length.
The quantities are scaled by representative values 
to make the $\delta \lpf{Q}_m$ comparable,
and then weighted and added together
to define the convergence metric as
\beq \label{eq:eSol-BFS}
\lpf{e}_{\rm QoI}^{{\rm ref}} 
= 
\frac{1}{3} \sum_{m=1}^2 {
\frac{ \delta \lpf{Q}_{m}^{\rm {ref}} }{2} }
+
\frac{1}{3} \sum_{m=3}^6 {
\frac{\delta \lpf{Q}_{m}^{\rm {ref}}}{4} }
+
\frac{1}{3} \sum_{m=7}^8 {
\frac{\delta \lpf{Q}_{m}^{\rm{ref}}}{2}
}
\,.
\eeq
Similar to the previous section, we consider both $\errprev$ defined with
respect to the previous grid in the sequence (to judge convergence in
a realistic scenario)
and
$\errref$ defined with respect to a converged DNS (to assess the
adaptation method).
The reference DNS is computed on a very fine unstructured grid with about 54M cells.

Each case was run for $500 H/U_\infty$ time units to remove the
initial transients, after which 800 snapshots were collected over a
period of $2000 H/U_\infty$.
The convergence of the averaging was judged by dividing the full
record into four separate batches with 200 snapshots in each,
computing the QoIs for each batch, and then computing the sample
standard deviation between the batch averages.
We then constructed 95\% confidence intervals for each quantity using
the Student's t-distribution
with 3 degrees of freedom~\cite[cf.][]{montgomery:book:11}.
The confidence intervals for the integrated errors in the QoIs are
very small (and thus omitted below), but they are significant for some
of the profiles especially downstream of the step.
This is consistent with the expectation of low frequency unsteadiness
in the separated flow.

We emphasize that the long averaging times are required only for the
solution to converge; the error indicator converges
about an order of magnitude
more quickly
due to its dependence on small scales.
The averaging convergence of the error indicator and the resulting
predicted grids is investigated in
Section~\ref{sec:statisticalConvergence}.

\subsection{Results}

\begin{figure}[t!]
  \centering	
  \includegraphics[width=130mm,clip=true,trim=50mm 10mm 10mm 10mm]{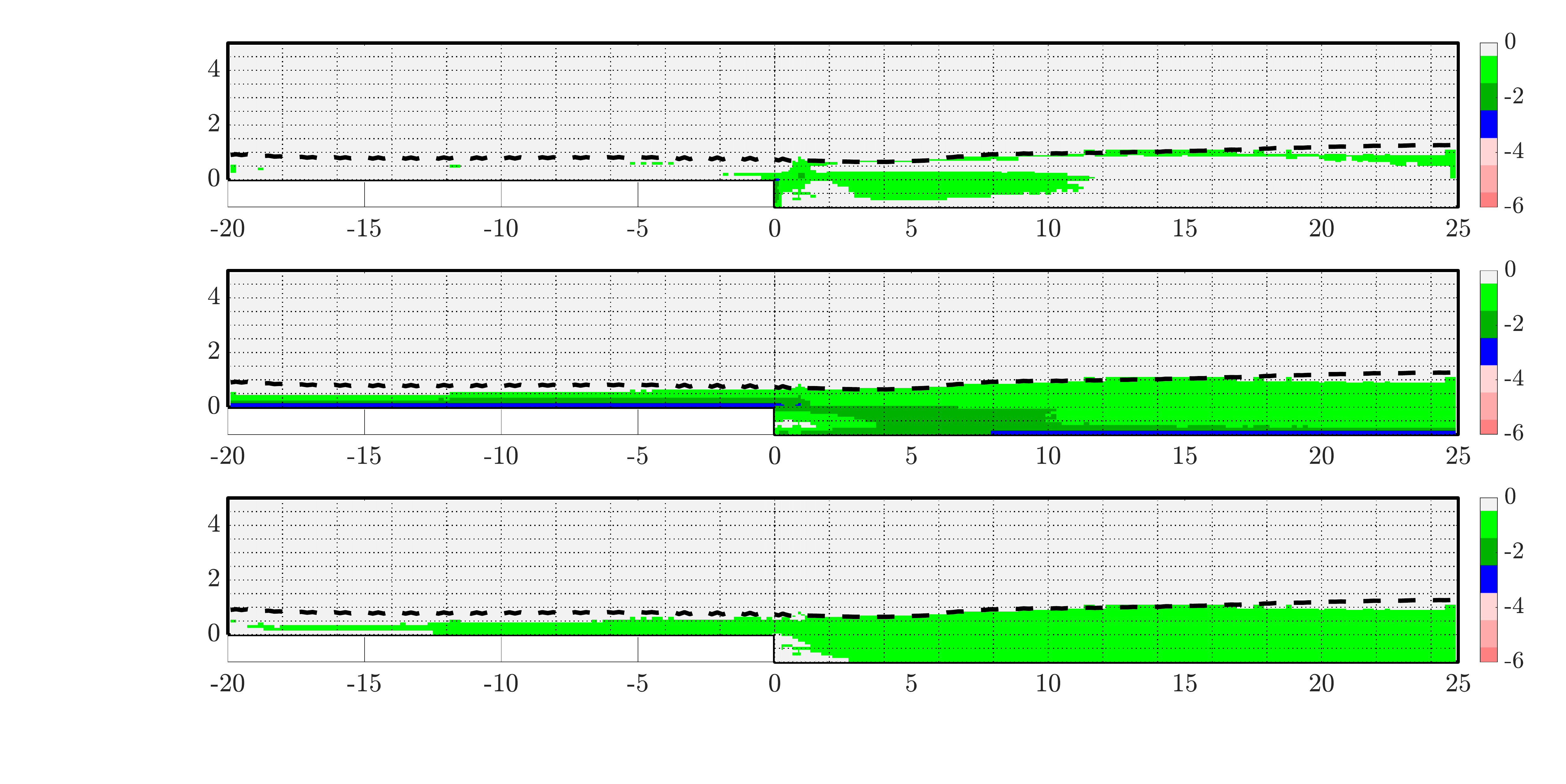}
  \caption{\label{fig:grid-Gn2}
    The grid G-2 from Table~\ref{Table:bfs-Gn}
    illustrated by its refinement levels 
    in $x$ (top),
    $y$ (middle), and $z$ (bottom).
    Refinement levels are computed based on a skeletal grid
    with ${\Delta}_0(\x,\n)=0.2H$ for all $\x$ and $\n$.
    The light green, dark green and blue colors illustrate
    regions with one ($\lpf{\Delta}_\n/H=0.1$), 
    two ($\lpf{\Delta}_\n/H=0.05$), 
    and three ($\lpf{\Delta}_\n/H=0.025$) refinement levels, respectively.
    The white regions 
    correspond to areas of the domain
     that are left untouched 
    (i.e., $\lpf{\Delta}_\n/H=0.2$).
    The dashed line highlights the $\delta_{95}$ boundary layer thickness.
  }
\end{figure}

The initial grid (labeled G-1) has a resolution of
$\lpf{\Delta}(\x,\n)/H=0.2$ everywhere in the domain
except close to the walls where the wall-normal direction is refined
by a factor of two
(we note that the method works equally well without this wall-normal
refinement, it just adds another step in the adaptation sequence).
After computing the LES on this grid, the error indicator is
computed in the three possible directions of refinement/coarsening,
and the target filter-width fields for the second grid (G-2) are
computed.
We then create the actual grid G-2 using the \emph{refineMesh} utility
in OpenFOAM.
Since \emph{refineMesh} can only refine hexahedral cells by factors of
2 in any direction,
the resulting grid is different from the predicted target: e.g., a
predicted target resolution of $0.17H$ in one location/direction will
produce a cell of size $0.10H$ in that location/direction.
The resulting grid G-2 (actually, the target filter-width field before
creating the \emph{refineMesh} input)
is visualized in Fig.~\ref{fig:grid-Gn2}.
Note that the constant $\Lambda$ in Eqn.~\ref{eq:optimal-space} was
adjusted such that the resulting number of cells was approximately
doubled.

Figure~\ref{fig:grid-Gn2} 
illustrates how the adaptation methodology
targets different regions of the domain for refinement.
The algorithm predicts a single level of refinement
in the $y$ direction ($\lpf{\Delta}(\x,\n_y)=\lpf{\Delta}_y=0.1H$) in most of the domain inside the boundary layer,
while the $y$ resolution is predicted to need a second level of refinement ($\lpf{\Delta}_y/H=0.05$)
closer to the horizontal walls and in the shear layer, 
and a third level of refinement ($\lpf{\Delta}_y/H=0.025$) 
in close vicinity of the horizontal walls in both incoming and recovering boundary layers.
The spanwise resolution $\lpf{\Delta}_z$ is targeted for a single level of refinement 
($\lpf{\Delta}_z/H=0.1$) for the most part of the domain 
inside the turbulent boundary layers, 
while the relaminarized region inside the recirculation bubble
is left untouched.
The resolution of the skeletal grid
in the $x$ direction ($\lpf{\Delta}_x/H=0.2$)
is deemed adequate for the most part of the domain,
except near the vertical wall of the step
(where the recirculation bubble causes shear)
and in the shear layer (where the turbulent fluctuations are
significant in all three directions).
We also note that the aspect ratio of the cells in the
boundary layers and the shear layer are quite close
to what we expect from experience for those flows.
The fact that the resulting G-2 grid seems this reasonable from an
``LES experience'' point-of-view is actually quite remarkable, since
it was created entirely by an algorithm from a solution on a highly
underresolved mesh.

\begin{figure}[t!]
	\centering	
	\includegraphics[width=130mm,clip=true,trim=15mm 10mm 10mm 7mm]{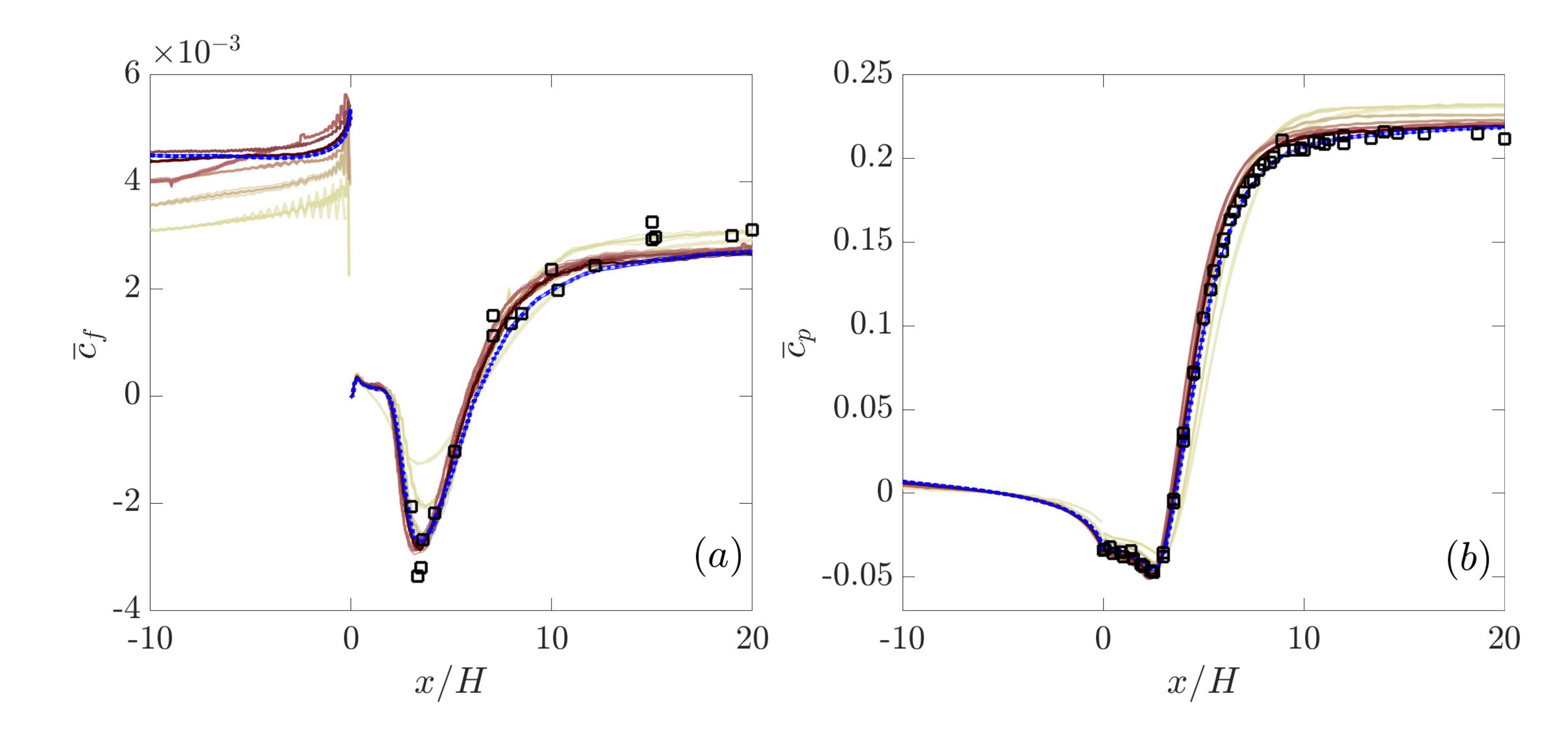} 
	\caption{\label{fig:BFS-Gn-cpcf}
	Convergence of (\textit{a}) friction coefficient $\lpf{c}_f$ 
	and (\textit{b}) pressure coefficient $\lpf{c}_p$
	for LES of flow over a backward-facing step.
	 Grids in Table~\ref{Table:bfs-Gn} 
	 are shown by the lightest color for G-1
	 to the darkest for G-7.
	 Solid lines denote the sample means,
	 while the shaded regions correspond to
	 the approximate confidence intervals
	 (computed locally).
	 The dotted blue lines and their shaded regions 
	 denote our DNS results and their confidence intervals.
	 Symbols correspond to the 
	 experimental data 
	 of Jovic \& Driver~\cite[]{jovic:94:bfs,jovic:95:bfs} with slightly different setup
	 (error bars on the experimental data are not shown).
	 Experimental measurements of $\lpf{c}_f$ and $\lpf{c}_p$
	 are not available upstream of the step.
	}
\end{figure}

\begin{figure}[t!]
	\centering	
	\includegraphics[width=130mm,clip=true,trim=10mm 10mm 10mm 10mm]{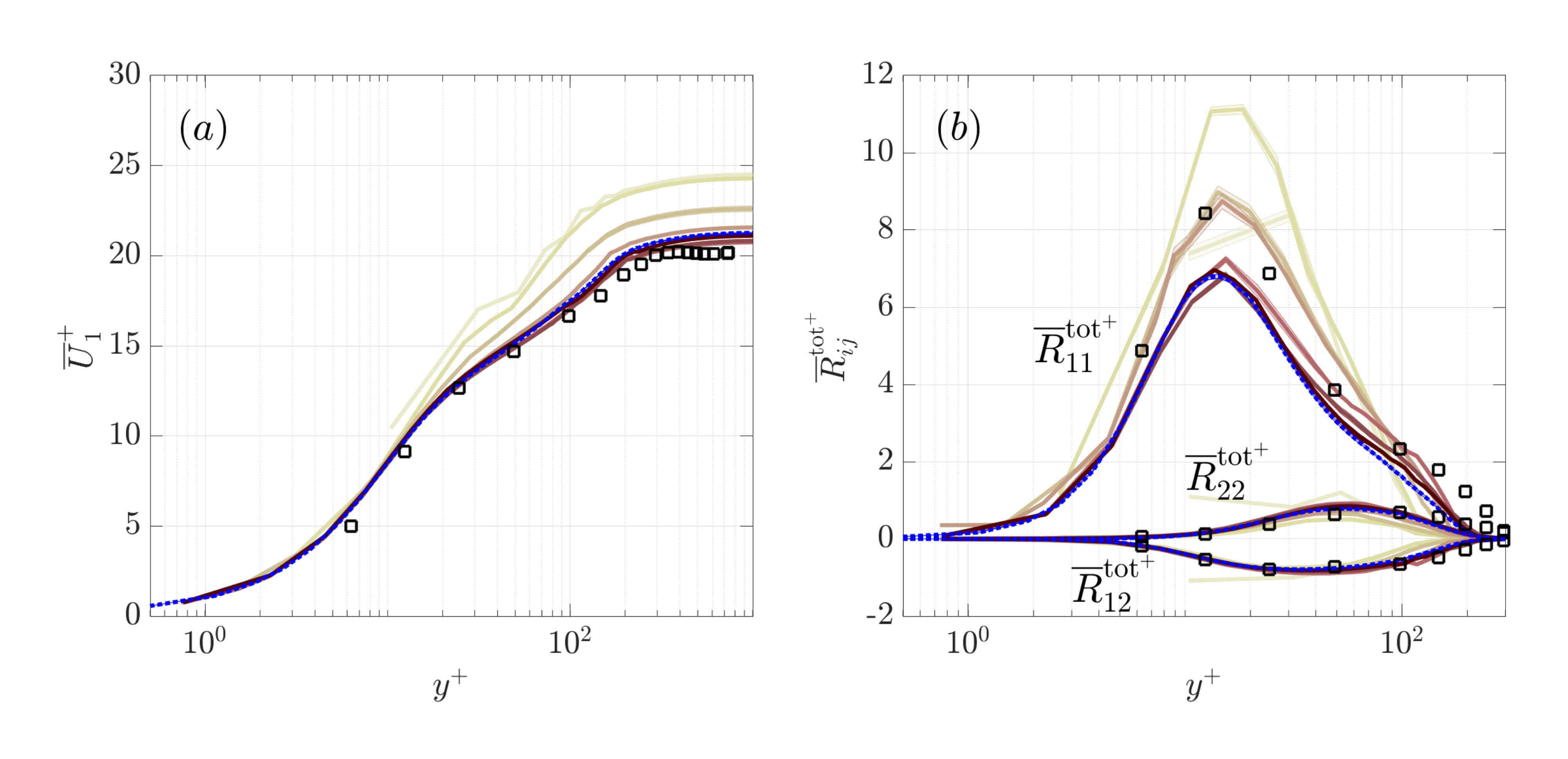} 
	\caption{\label{fig:BFS-Gn-x3}
	Convergence of the mean velocity and Reynolds stress profiles
	for the sequence of grids in Table~\ref{Table:bfs-Gn}
	at the incoming boundary layer at $x/H=-3$.
	 Grids in the sequence are shown by the lightest color for G-1
	 to the darkest for G-7.
	 Solid lines denote the sample means,
	 while the shaded regions correspond to
	 the approximate confidence intervals
	 (computed locally).
	 The dotted blue lines and their shaded regions 
	 denote our DNS results and their confidence intervals.
	 Symbols correspond to the 
	 experimental data 
	 of Jovic \& Driver~\cite[]{jovic:94:bfs,jovic:95:bfs}
	 (error bars on the experimental data are not shown).
	}
\end{figure}

\begin{figure}[t!]
	\centering	
	\includegraphics[width=130mm,clip=true,trim=10mm 10mm 10mm 10mm]{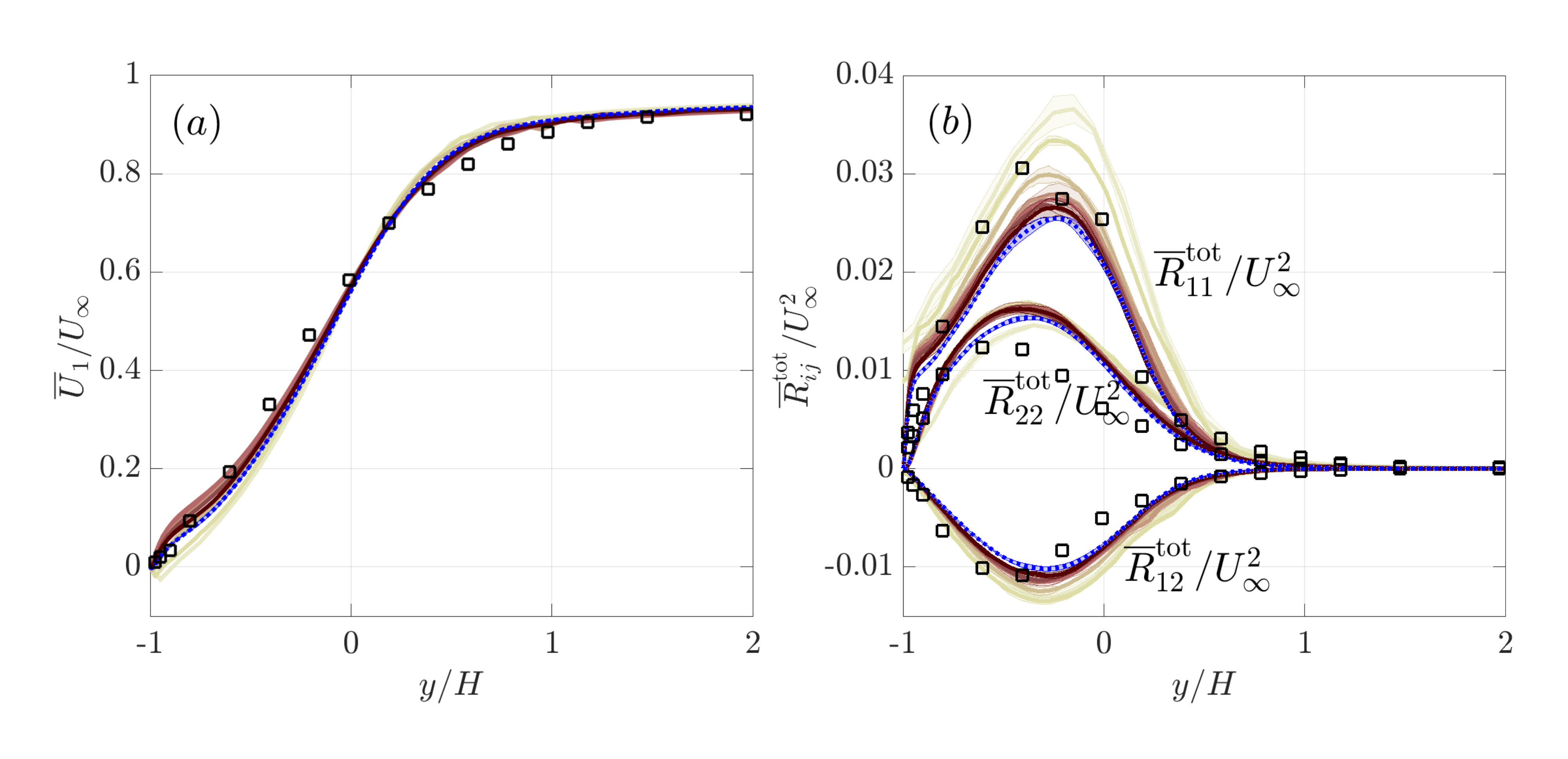} 
	\caption{\label{fig:BFS-Gn-x6}
	Convergence of the mean velocity and Reynolds stress profles
	for grids in Table~\ref{Table:bfs-Gn}
	at $x/H=6$ near the reattachment point.
	See Fig.~\ref{fig:BFS-Gn-x3} for more details.
	}
\end{figure}

\begin{figure}[t!]
	\centering	
	\includegraphics[width=130mm,clip=true,trim=10mm 10mm 10mm 10mm]{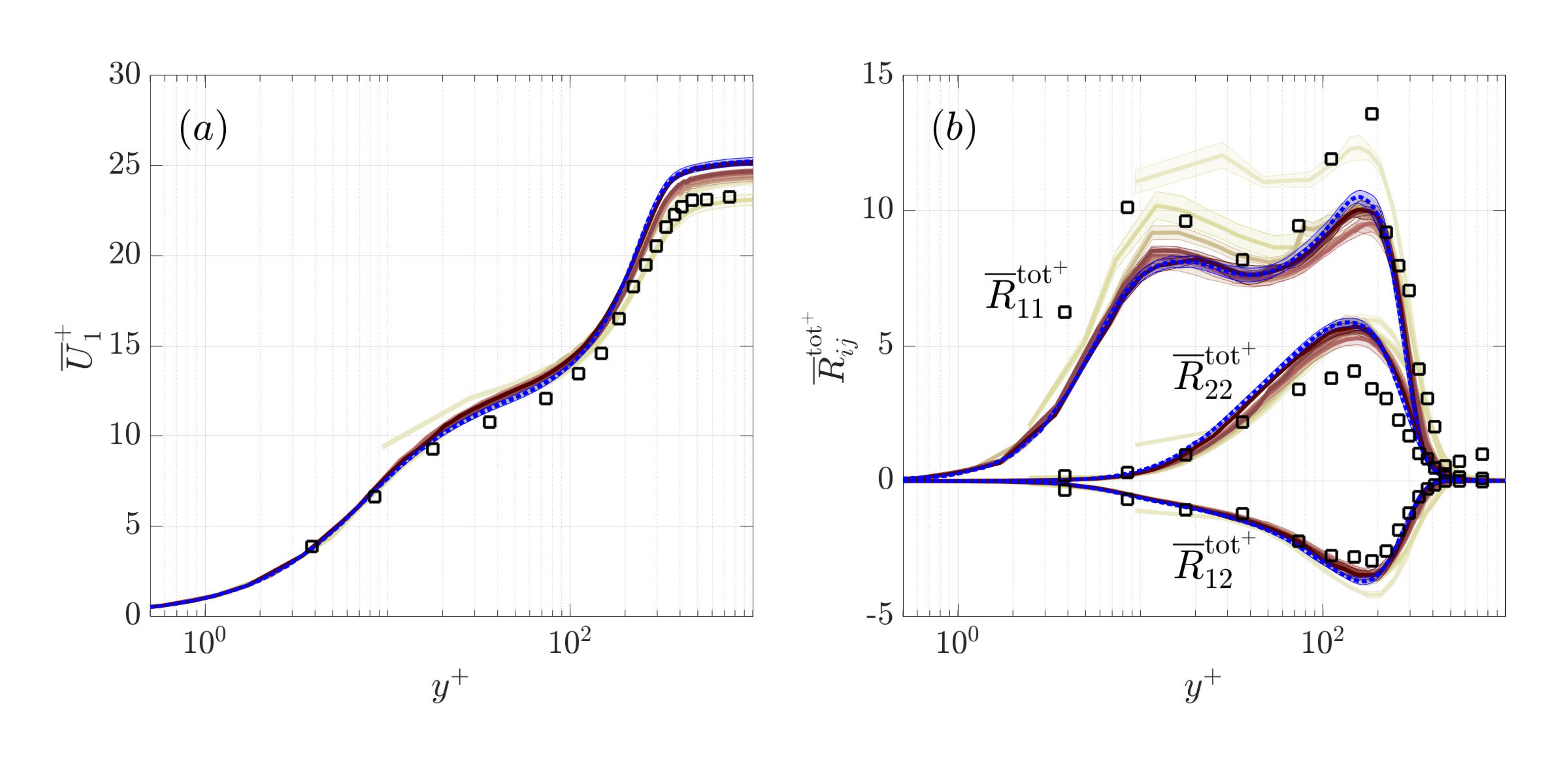} 
	\caption{\label{fig:BFS-Gn-x15}
	Convergence of the mean velocity and Reynolds stress profiles 
	for the recovering boundary layer at $x/H=15$
	for the sequence of grids in Table~\ref{Table:bfs-Gn}.
	See Fig.~\ref{fig:BFS-Gn-x3} for more details
	including interpretation of colors.
	}
\end{figure}

The adaptation process is continued until grid G-7 
where the QoIs are deemed converged.
Each target grid is generated by aiming for approximately doubling the
number of cells, without trying to match this ratio exactly.
The sequence of generated grids is reported in Table~\ref{Table:bfs-Gn}
by their total number of cells $N_{\rm tot}$ 
and QoI errors (both $\errref$ and $\errprev$).
The Table also reports the grid-spacings
in the approaching boundary layer at $x/H=-3$
and shortly after the step at $x/H=1$ (for $y/H=0$)
in the shear layer formed by separation at the step.
The convergence of the QoIs
is shown in Fig.~\ref{fig:BFS-Gn-cpcf} for the pressure and friction coefficients
and Figs.~\ref{fig:BFS-Gn-x3},~\ref{fig:BFS-Gn-x6} and~\ref{fig:BFS-Gn-x15}
for the mean velocity and Reynolds stress profiles at some of the more
interesting locations.

\begin{table}[t!]
\begin{center}
\begin{tabular}{ c c  c  c  c   c }
  Grid
  & $N_{\rm tot}$	
  & $(\lpf{\Delta}_x^+ , \lpf{\Delta}_{y_w}^+/2 , \lpf{\Delta}_z^+)$
  & $(\lpf{\Delta}_x, \lpf{\Delta}_y, \lpf{\Delta}_z)/\delta_{\rm shear}$
  & $\errref$ (\%)
  & $\errprev$ (\%)
  \\[3pt]
  G-1
  & 149k
  & $(42,10,42)$
  & $(0.21, 0.17, 0.33)$
    & 11.1
  & $-$
  \\
  G-2
  & 297k
  & $(42,2.6,21)$
  & $(0.16,0.078,0.16)$
  & 10.5
  & 5.3
  \\
  G-3
  & 611k
  & $(45,1.4,11)$
  & $(0.16,0.049,0.078)$
  & 5.6
  & 6.4
  \\
  G-4
  & 1.32M
  & $(47,1.5,12)$
  & $(0.076,0.038,0.076)$
  & 4.9
  & 3.8
  \\
  G-5
  & 2.13M
  & $(25,0.77,6.2)$
  & $(0.070,0.035,0.035)$
  & 5.4
  & 2.8
    \\
  G-6
  & 3.41M
  & $(25,0.77,6.1)$
  & $(0.068,0.034,0.034)$
  & 3.5
  & 3.4
    \\
  G-7
  & 6.72M
  & $(12,0.76.6.0)$
  & $(0.034,0.017,0.034)$
  & 2.5
  & 2.2
  \\ [2pt]
  DNS
  & 54M
  & $(6.0,0.38.3.0)$
  & $(0.017,0.0086,0.017)$
  & 0
  & $-$
\end{tabular}
\end{center}
\caption{ \label{Table:bfs-Gn}
Sequence of grids generated for LES of flow
over a backward-facing step.
$(\lpf{\Delta}_x^+ , \lpf{\Delta}_{y_w}^+/2 , \lpf{\Delta}_z^+)$
correspond to the boundary layer
resolutions at $x/H=-3$ upstream of the step,
$\delta_{\rm shear}$ is the approximate shear layer thickness
at $(x,y)/H=(1,0)$, and
$(\lpf{\Delta}_x, \lpf{\Delta}_y, \lpf{\Delta}_z)$ 
is the resolution at that location.
See Fig.~\ref{fig:grid-Gn6} for more details.
$\errref$ and $\errprev$ are defined in Eqn.~\ref{eq:eSol-BFS}.
}
\end{table}

The computed error $\errref$ decreases after every adaptation
except for grid G-5.
The relatively large value of the error for this grid
is primarily due to the error in the friction coefficient 
of the incoming boundary layer
(see Fig.~\ref{fig:BFS-Gn-cpcf}),
that happens despite the apparently sufficient 
resolution of the grid,
and affects the entire flowfield 
downstream of the step.

\begin{figure}[t!]
  \centering	
  \ifIncludeFigures
  \includegraphics[width=130mm,clip=true,trim=500mm 0mm 200mm 130mm]{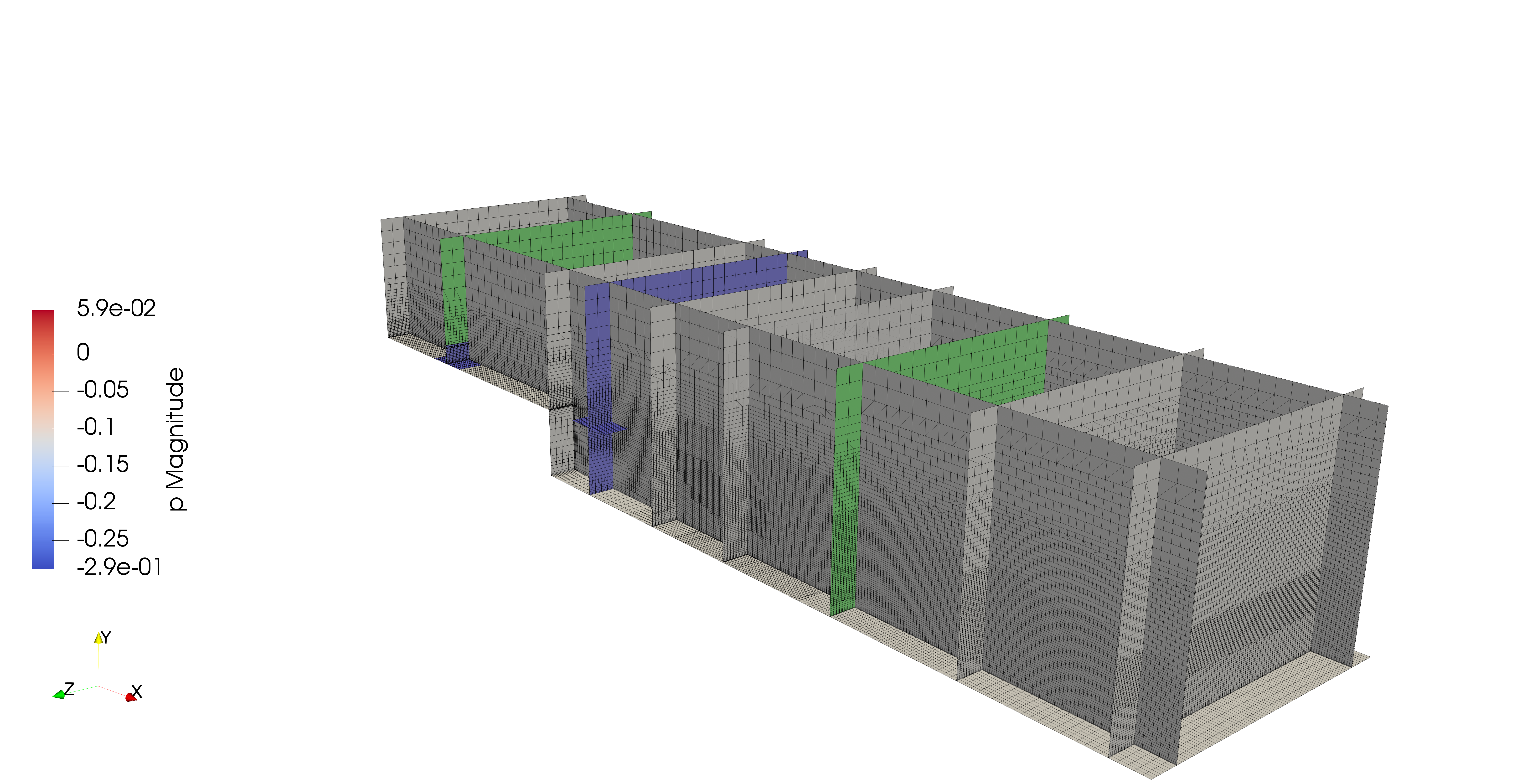}
  \fi
  \caption{\label{fig:grid-Gn6}
    The generated grid G-6 of Table~\ref{Table:bfs-Gn} with 3.41M cells.
    Intersections of the blue planes denote locations 
    whose resolutions are reported in Table~\ref{Table:bfs-Gn},
    while the green planes correspond to $x/H=-3$ and $x/H=6$
    whose velocity and Reynolds stress profiles are plotted in 
    Figs.~\ref{fig:BFS-Gn-x3} and~\ref{fig:BFS-Gn-x6}.
    The grid is resulted from computation of the proposed error indicator 
    (Eqns.~\ref{eq:Gn} and~\ref{eq:F})
    and applying the grid selection criteria of 
    Eqns.~\ref{eq:optimal-direction} and~\ref{eq:optimal-space}
    with no user experience involved.
  }
\end{figure}

Figure~\ref{fig:grid-Gn6} shows the constructed grid G-6 of 
Table~\ref{Table:bfs-Gn} as an example of a converged LES grid
for this specific setup.
Note how complicated this grid has become,
with many transitions between different grid-resolutions
and cells that have completely different aspect ratios
from one region of the domain to another
(e.g., compare the aspect ratios at the locations
reported in Table~\ref{Table:bfs-Gn}).
It is interesting to note how coarse the grid
is in the recirculation bubble,
expect for the wall-normal directions 
that are refined to predict the right level of shear at the wall.
The most important observation is that these predicted
resolutions are very similar to what an experienced user
would use when generating a grid for LES of
the flow over a backward-facing step.

The influence of the initial grid is investigated
in~\ref{section:BFS-initialGrid} and found to be decreasing for every grid in the sequence
with effectively no influence after grid G-4.

\section{Statistical convergence of the error indicator and the resulting grid\label{sec:statisticalConvergence}}

We conclude
by studying the sensitivity
of the error indicator and the predicted grids
to insufficient averaging in time.
Most quantities of interest in LES depend on the large scales of
motion
which then require a relatively long averaging time for adequate
convergence.
In contrast, the error indicator depends on the smallest resolved
scales and should therefore converge more quickly.
The implication is that, in practice, one could reduce the cost of the
adaptation process by running only short simulations on many of the grids.

The statistical convergence assessment is done only for the
backward-facing step flow, for which we have 800 snapshots spaced $2.5
H/U_\infty$ apart in time.
The error indicator computed using all 800 snapshots is labeled
$\lpf{\G}_{\rm ref}(\x,\n)$,
and the refinement level of the resulting grid
is labeled $\lpf{\mathcal{R}}_{\rm ref}(\x,\n)$,
where the refinement level is quantified as
\beq \nonumber
\lpf{\mathcal{R}}(\x,\n)= \log_2{\frac { \lpf{\Delta}(\x,\n) }{ {\Delta}_{0}(\x,\n) }}
\,,
\eeq
where ${\Delta}_{0}(\x,\n)=0.2$ for all $\x$ and $\n$
(this $\lpf{\mathcal{R}}(\x,\n)$ is what was plotted in Fig.~\ref{fig:grid-Gn2}).

We also consider averages over batches of $m$ snapshots,
for which the resulting error indicator and predicted refinement
levels are
labeled
$\lpf{\G}_{m,j}(\x,\n)$
and
$\lpf{\mathcal{R}}_{m,j}(\x,\n)$
where $j$ is the batch number.
The errors due to insufficient averaging are then defined as
\beq \label{eq:eM}
\begin{aligned}
\lpf{\mathcal{E}}_\G(m;j)
&=
\frac{
\sum_i
\int \! \! \int_\Omega
  \left| \lpf{\G}_{m,j}(\x,\n_i) - \lpf{\G}_{\rm ref}(\x,\n_i) \right|
  d\x
}{
\sum_i
\int \! \! \int_\Omega
  \lpf{\G}_{\rm ref}(\x,\n_i)
  d\x
}
\\
\lpf{\mathcal{E}}_\mathcal{R}(m;j)
&=
\frac{
  \sum_i
  \int \! \! \int_\Omega
  \left| \lpf{\mathcal{R}}_{m,j}(\x,\n_i) - \lpf{\mathcal{R}}_{\rm ref}(\x,\n_i) \right|
  d\x
}{
  \sum_i
  \int \! \! \int_\Omega
  \left| \lpf{\mathcal{R}}_{\rm ref}(\x,\n_i)  \right|
  d\x
}
\end{aligned}
\eeq
where $\Omega: \x=(x,y) \in [-20H,25H] \times [-H,5H] $ is the full two-dimensional
domain.
This assessment procedure
is adopted from our previous work~\cite{toosi:17},
where we also found (by visual comparison) that an error threshold of 0.05
is amply low for both of the errors (for this flow problem and this
definition of the error metric).
This threshold is therefore used here as well.

\begin{figure}[t!]
  \centering	
  \includegraphics[width=60mm,clip=true,trim=0mm 0mm 0mm 0mm]{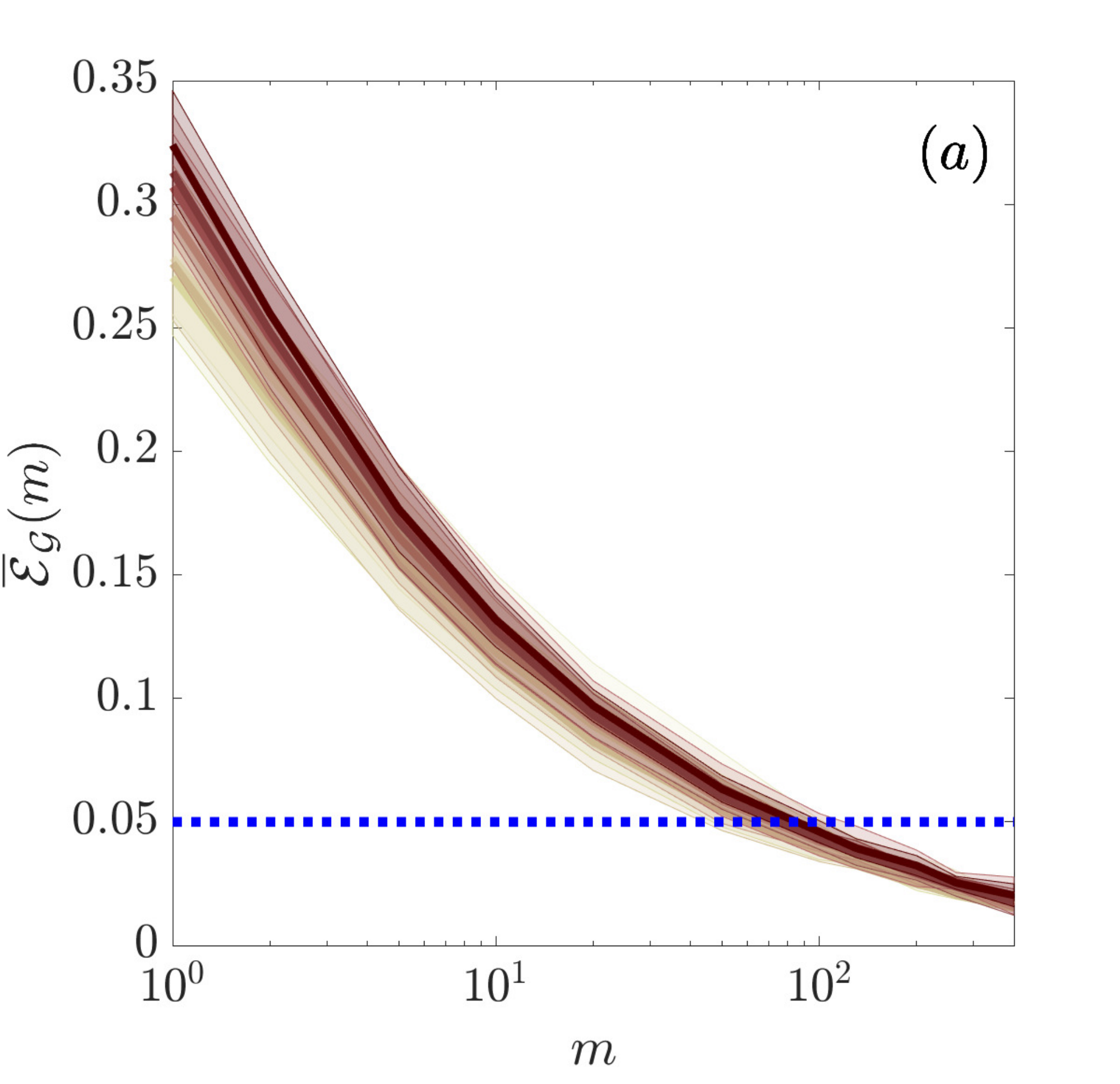}
  \includegraphics[width=60mm,clip=true,trim=0mm 0mm 0mm 0mm]{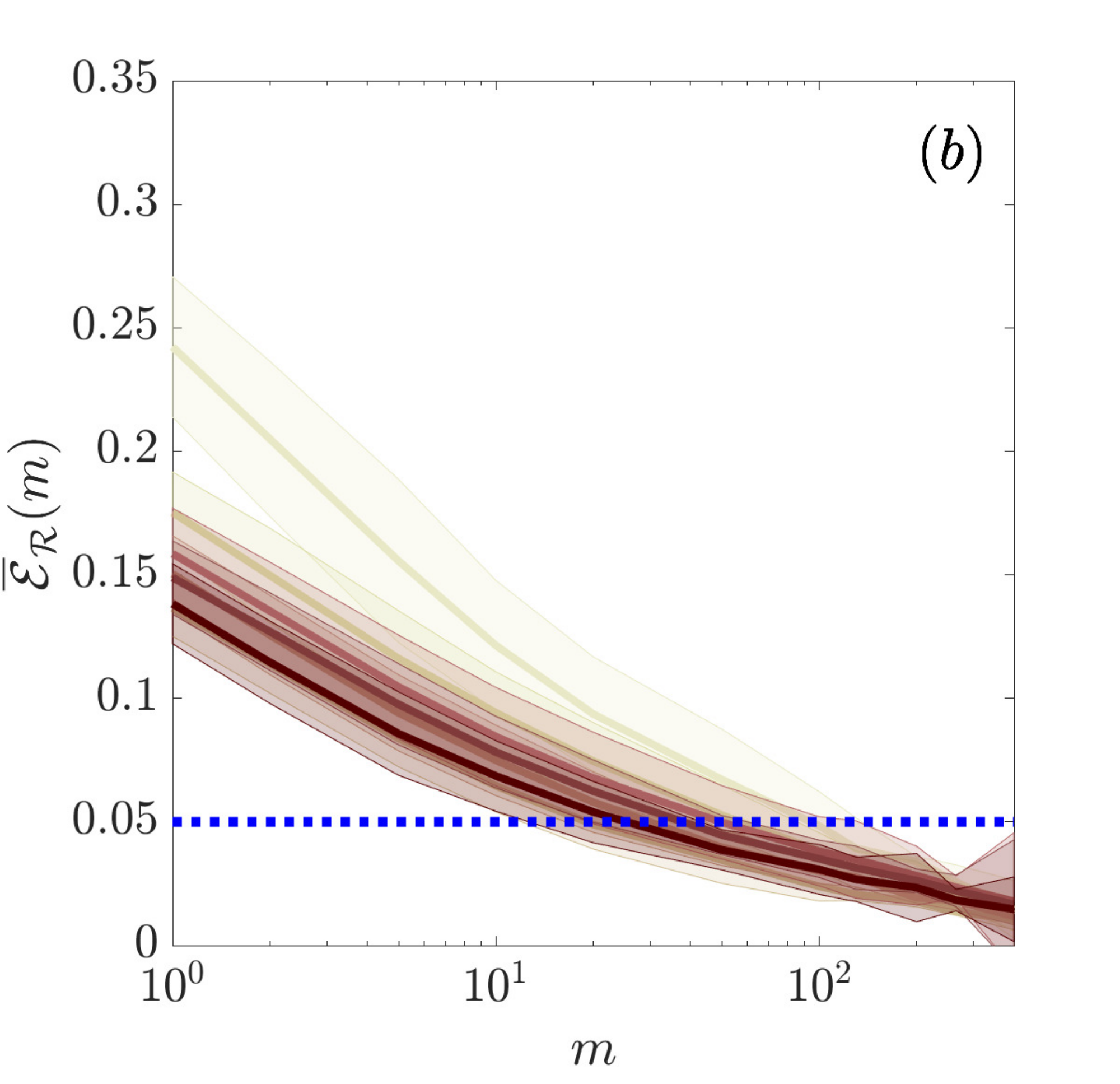}
  \caption{\label{fig:convergence}
    Convergence of 
    (\textit{a}) the error indicator
    and (\textit{b}) the target grids
    with the number of snapshots $m$ used in the averaging.
    The grids in Table~\ref{Table:bfs-Gn} 
    are shown by the lightest color for G-1
    to the darkest for G-7. 
    The solid lines show the sample mean of 
    $\lpf{\mathcal{E}}_\G(m;j)$ and $\lpf{\mathcal{E}}_\mathcal{R}(m;j)$
    (Eqn.~\ref{eq:eM}), 
    while the shaded regions highlight the $90\%$ 
    prediction interval of the computed values.
    When the upper bound of the confidence interval 
    goes below the horizontal dotted line 
    there is a 95\% chance that the error metric for a single realization is
    below 0.05.
  }
\end{figure}

Since both $\lpf{\G}_{m,j}(\x,\n)$ and $\lpf{\mathcal{R}}_{m,j}(\x,\n)$ are random variables,
their errors
are also random variables.
A 90\% two-sided prediction interval 
is computed for each using 
the sample mean and sample standard deviation of
$\lpf{\mathcal{E}}_\G(m;j)$ and $\lpf{\mathcal{E}}_{\mathcal{R}}(m;j)$
and Student's t-distribution
(cf.~\cite[][]{montgomery:book:11} for more details on prediction intervals).
These prediction intervals are shown in Fig.~\ref{fig:convergence}.
When the upper bound of the prediction interval
lies below the acceptable threshold,
there is a 95\% chance that the error metric
in a single realization
is below 0.05.
The approximate integration times required 
for these errors to go below the acceptable threshold are 
summarized in Table~\ref{Table:bfs-time}.

\begin{table}[t!]
\begin{center}
\begin{tabular}{ c c  c  c  c   c }
  Grid
  & $N_{\rm tot}$	
  & $t_{\rm init} U_\infty/H$ 
  & $t_{\rm QoI} U_\infty/H$
  & $t_{\mathcal{G}} U_\infty/H$
  & $t_{\mathcal{R}} U_\infty/H$
  \\[3pt]
  G-1
  & 149k
  & 500 
  & 2000
  & 260
  & 350
  \\
  G-2
  & 297k
  & 500
  & 2000
  & 240
  & 220
  \\
  G-3
  & 611k
  & 500
  & 2000
  & 190
  & 90
  \\
  G-4
  & 1.32M
  & 500
  & 2000
  & 200
  & 120
  \\
  G-5
  & 2.13M
  & 500
  & 2000
  & 300
  & 330
    \\
  G-6
  & 3.41M
  & 500
  & 2000
  & 220
  & 150
    \\
  G-7
  & 6.72M
  & 500
  & 2000
  & 260
  & 110
\end{tabular}
\end{center}
\caption{ \label{Table:bfs-time}
Comparison of the simulation run times
used for removing the initial transients (denoted by $t_{\rm init}$)
and computing accurate mean QoI profiles (denoted by $t_{\rm QoI}$)
with the integration times required for accurate computation of
the error indicator ($t_{\mathcal{G}}$)
and the refinement levels of the next optimal grid ($t_{\mathcal{R}}$).
For regular situations (i.e., where the flowfield does not undergo a bifurcation)
the required integration time to remove the initial transients is
presumably shorter for the error indicator and the target grids compared to
the QoIs; 
however, it has not been investigated in this section.
}
\end{table}

It is quite clear from Fig.~\ref{fig:convergence} and Table~\ref{Table:bfs-time}
that the proposed error indicator and
its predicted grids require
almost one order of magnitude
shorter integration times
compared to what is needed for sufficiently converged QoI profiles
(once the transients are gone).
In other words, we only
need to collect $m \lesssim 120$ snapshots 
(equivalent to an integration time of $\lesssim 300 H/U_\infty$)
to compute sufficiently accurate values of the error indicator,
and the required integration time is similar on all grids.
The error in the target grids show a very similar behavior in general:
grids G-3 to G-7 only require an integration time of
$\lesssim 150 H/U_\infty$
(interestingly, grid G-5 is an outlier here as well, 
requiring an integration time of $330 H/U_\infty$).

Note that the initial transient must of course be removed, 
a time which is highly flow specific; 
for the cases studied here, the initial transient was about 500 time units.

Finally,
we emphasize that although
the exact results presented here
are specific to 
the flow over a backward-facing step,
the sampling frequency,
and other flow parameters,
the conclusion about a much faster convergence of
the error indicator and target grids
compared to the QoIs
is probably more general and
valid for a broader set of 
flows.

\section{Summary, discussion and future directions\label{sec:summary}}

The goal of this paper has been to introduce a systematic approach to
finding the ``optimal'' filter-width distribution
(or, equivalently, the ``optimal'' grid)
for large eddy simulation (LES).
The long-term vision is to (i) reduce the amount of human time spent
on grid-generation (by automating the process),
(ii)
reduce the
computational time (by producing more ``optimal'' grids), and
(iii)
make
the LES simulation process more systematic (e.g., so different users/codes
produce more similar results).

The heart of the proposed adaptation algorithm is the error indicator
$\lpf{\G}(\x,\n)$
defined in Eqn.~\ref{eq:Gn},
which estimates the error introduced into the LES evolution equation
at location $\x$ caused by an insufficient filter-width in direction $\n$.
More specifically, the error indicator measures the initial divergence
between the test-filtered LES solution and an imagined solution to the
test-filtered LES equation.
In other words, it measures how sensitive the LES equation is to small
(directional) changes in the filter-width.
While the error indicator is based on manipulations of the governing
equation, it is also based on the assumption that the source of
initial divergence between these different solutions is a meaningful
measure of the error in the fully nonlinear long-time evolution of the
LES.
This is really the key physical assumption in this work.

The ``optimal'' filter-width is found by equi-distribution of the 
cell integrated error indicator 
(i.e., the value of the error indicator multiplied by the cell volume,
to second order accuracy).
While this was the solution to the optimization problem
for the error indicator of this work
(Section~\ref{sec:optimal}), 
it appears to hold for many of the other error indicators;
generally speaking, if the value of the error indicator is assumed to be proportional to
the local error generation, the overall error is proportional
to the volume integral of the local value,
for which the optimization problem takes the same solution of
equi-distribution of the cell integrated value.
This suggests that the popular approach of
setting a threshold on the error indicator itself,
e.g., judging the accuracy of LES by the ratio
of unresolved to total turbulent kinetic energy,
is at best suboptimal (if not wrong) 
and should be avoided.

The adaptation process is tested on a channel flow and the flow over a
backward-facing step.
For the channel, the algorithm consistently produces
grids/filter-widths that are very close to what is considered
``best practice'' in LES and DNS:
grids with
$(\lpf{\Delta}_x^+, \lpf{\Delta}_{y_w}^+/2, \lpf{\Delta}_z^+)
\approx (45,1.7,19)$
and
$(13,1.2,6.4)$, respectively.
For the backward-facing step, the predicted grids are close to what an
experienced user might produce.
It is essentially impossible to say how ``optimal'' (in
the mathematical sense) the
grids are for this problem, but we note that the error (compared to DNS)
reaches about 5\% with only 600K to 2M cells; it is hard to imagine an
experienced user creating a better grid than that
(note that the code uses a low-order numerical scheme), 
at least without significant trial-and-error.

The subgrid/subfilter model used in the LES solution
enters directly into the definition of the residual forcing term
and thus the error indicator.
This is theoretically advantageous, since a more accurate subfilter
model would produce a lower error indicator for the same resolved
velocity field.

We also note that the proposed error indicator has similarities with
the dynamic procedure but was derived without any appeals to
scale-similarity in the inertial subrange of turbulence.
Its use is not restricted to filter-widths in the inertial subrange,
and the derivation in fact offers an alternative explanation for the
success of the dynamic procedure.

The error indicator was derived for the incompressible
Navier-Stokes equation in this paper, but can be easily extended to other physics.
A derivation for compressible flow is shown in~\ref{sec:appendix-otherGn}, 
which creates a separate error indicator for the energy equation.
Following the same process, one could extend it to chemically reacting
flows, etc.

\subsection{Cost}

One potential criticism of this type of adaptation algorithm is the
additional cost of performing LES on a full sequence of grids.
We make four counter-arguments and observations.

First, assuming that the cell count is doubled at each iteration and
that the time step scales as $N_{\rm tot}^{1/3}$, the total cost of computing
all grids in the sequence (including the final one) is $\lesssim 1.66 N_{\rm final}$.
If the cell count is quadrupled at each iteration, the total cost is
$\lesssim 1.19 N_{\rm final}$ instead.

Second, one could start from a ``best guess'' grid in practice
(based on prior experience with the flow in question),
thus reducing the number of steps of the algorithm.
We only started from exceedingly coarse and ``ignorant'' initial grids
here in order to test the robustness of the method.

Third, as shown in Section~\ref{sec:statisticalConvergence}, the error
indicator converges faster (in terms of integration time) than the LES
itself due to its dependence on the smallest resolved scales.
Therefore, one could run the simulations on the first several grids
for shorter times.

Finally, the minor added computational cost must of course be balanced
against the larger cost saving of having a more ``optimal'' final
grid. This saving will presumably become larger for more complex flow
problems.

\subsection{Possible future directions}

The present error indicator was derived for the continuous governing
equation and as such does not directly estimate any numerical errors
(despite the similarity of the Leonard-like stress to
the truncation error).
The adaptation algorithm was still found to perform well for the 
DNS of channel flow (which has only numerical errors),
but it may be required to explicitly include the numerical
errors in the error indicator in other flows.

We also note that the error indicator
arguably measures the
sensitivity of the solution at the test-filter level
and not at the LES filter level $\lpf{\Delta}$.
This is perfectly fine 
(perhaps even desirable)
for the final grid(s),
but may not be ideal for the initial grids in the sequence that are
far from resolving the inertial subrange of the turbulence.
One possibility is to investigate approaches similar to that of
Porte-Agel~\etal~\cite{porte-agel:00}
who revised the dynamic procedure to work better on underresolved
grids.

Throughout this work we assumed a constant
scaling exponent of $\alpha(\x,\n)=2$ for all $\x$ and $\n$.
In~\ref{sec:appendix-optimal}
we show that a spatially/directionally varying exponent $\alpha$
changes
the grid selection criterion, in terms of both the
aspect ratio and the spatial distribution. 
This may become important when $\alpha$
is dramatically different in different directions
(e.g., in codes where only the spanwise direction
is handled by the Fourier expansion)
or in different locations.
A possible improvement of the method
could be achieved by either using the theoretical values of 
$\alpha$ in different directions (if they are known), 
or to add a second level of test-filtering
to estimate $\alpha(\x,\n)$ as well.

There are also several possibilities to improve
the grid selection criteria used to generate the grids.
For compressible solvers with explicit time-stepping, it may be
important to include the number of time steps in the estimated
computational cost. This would then effectively ``penalize'' very
thin cells near boundaries.
More generally, since it is well known that numerical 
and commutation
errors are
highly sensitive to the smoothness of the mesh, one should certainly
include a penalization of too rapid filter-width transitions when
solving the optimization problem.

A major improvement would be to include the adjoint of the
quantity of interest (QoI) and thus make the adaptation ``output-based''.
This has been the major advancement in steady-state grid-adaptation
over the last few decades~\cite[cf.][]{fidkowski:11}.
Inclusion of the adjoint would first require the problem of
exponential divergence of the adjoint for chaotic problems to be
solved~\cite[cf.][]{lea:00}.

\section*{Acknowledgements}

This work has been supported by NSF grants CBET-1453633 and CBET-1804825. 
Computing time has been provided by the University of Maryland 
supercomputing resources (http://hpcc.umd.edu).

\appendix

\section{Directionally and spatially varying grid scaling exponent $\alpha$
and how it changes the grid selection criterion
\label{sec:appendix-optimal}
}

The grid scaling exponent $\alpha$ in the model
$ \check{\G}(\x,\n) \approx \lpf{g}(\x,\n) \check{\Delta}^\alpha (\x,\n) $
is in general a function of both space $\x$ and direction $\n$,
i.e.,
$\alpha = \alpha(\x,\n)$.
If we follow the same approach used in Section~\ref{sec:optimal},
but allow for spatially and directionally varying $\alpha$,
the solution to the optimization problem can be obtained by minimizing
the Lagrangian 
\beq 
\begin{aligned}
\mathcal{L} 
&= 
\int_{\Omega}{
\left(
\sqrt{ \check{\G}^2(\x,\n_1) +  \check{\G}^2(\x,\n_2) +
  \check{\G}^2(\x,\n_3) } 
  -
  \frac{ \lambda }{
\check{\Delta}_{\n_1}
\check{\Delta}_{\n_2}
\check{\Delta}_{\n_3}} 
\right)}
\ 
 d\x
\\
&=
\int_{\Omega}{
\left(
\sqrt{ 
\lpf{g}_{\n_1}^2 \check{\Delta}_{\n_1}^{2\alpha_{\n_1}}
+
\lpf{g}_{\n_2}^2 \check{\Delta}_{\n_2}^{2\alpha_{\n_2}}
+
\lpf{g}_{\n_3}^2 \check{\Delta}_{\n_3}^{2\alpha_{\n_3}}
} 
  -
  \frac{ \lambda }{
\check{\Delta}_{\n_1}
\check{\Delta}_{\n_2}
\check{\Delta}_{\n_3}} 
\right)}
\ 
 d\x
 \, ,
\end{aligned}
\eeq
where $\lpf{g}_{\n_i}= \lpf{g}(\x,\n_i)$ 
and $\alpha_{\n_i} = \alpha(\x,\n_i)$.
The minimum of $\mathcal{L}$
can be found by setting its functional derivatives
with respect to each of $\check{\Delta}_{\n_i}$ to zero,
i.e., 
$\delta \mathcal{L} / \delta \check{\Delta}_{\n_i}=0$.
This leads to the solution
\beq  \label{eq:AR-variableAlpha}
\alpha(\x,\n_1) \check{\G}_{\rm opt}^2(\x,\n_1)
=
\alpha(\x,\n_2) \check{\G}_{\rm opt}^2(\x,\n_2)
=
\alpha(\x,\n_3) \check{\G}_{\rm opt}^2(\x,\n_3)
\, ,
\eeq
for the optimal anisotropy, and
\beq  \label{eq:spatial-variableAlpha}
\frac{
\check{\Delta}_{{\rm opt}} (\x,\n_1)
\check{\Delta}_{{\rm opt}} (\x,\n_2)
\check{\Delta}_{{\rm opt}}  (\x,\n_3)
\sqrt{
\check{\G}_{\rm opt}^2(\x,\n_1)
+
\check{\G}_{\rm opt}^2(\x,\n_2)
+
\check{\G}_{\rm opt}^2(\x,\n_3)
}
}
{
1/\alpha(\x,\n_1)
+
1/\alpha(\x,\n_2)
+
1/\alpha(\x,\n_3)
}
=
{\rm const.}
\, ,
\eeq
for the optimal spatial distribution. 
Note that 
(because of the spatially varying denominator)
this is different from our previous criterion 
to have the cell-integrated error indicator equidistributed in space. 

In the case of directionally varying scaling exponent,
the optimal anisotropy depends on the exact definition
of $\check{e}_{\rm tot}$ in Eqn.~\ref{eq:etot};
for example, depending on whether the integrand 
is defined as
$\sqrt{ \check{\G}^2(\x,\n_1) +  \check{\G}^2(\x,\n_2) +
  \check{\G}^2(\x,\n_3) } $
 or
 $\check{\G}(\x,\n_1) +  \check{\G}(\x,\n_2) +
  \check{\G}(\x,\n_3) $
the optimal anisotropy is obtained, respectively,
by Eqn.~\ref{eq:AR-variableAlpha}
or by the relation
$\alpha(\x,\n_1) \check{\G}_{\rm opt}(\x,\n_1)
=
\alpha(\x,\n_2) \check{\G}_{\rm opt}(\x,\n_2)
=
\alpha(\x,\n_3) \check{\G}_{\rm opt}(\x,\n_3)$. 
In the latter case the optimal spatial distribution of the filter-width
can be obtained by simply replacing for the new 
definition of the integrand in 
the numerator of Eqn.~\ref{eq:spatial-variableAlpha}.
If $\alpha$ is assumed constant in different directions,
both methods lead to the same prediction of the 
optimal aspect ratio.
Additionally, in our early tests we did not see a significant
difference in terms of the target grids between the two
definitions of the integrand 
(for the backward-facing step, using $\alpha(\x,\n) = 2$).
Therefore, these discussions were avoided in the text.

\section{Definition of the error indicator for other formulations of LES \label{sec:appendix-otherGn}}

In this section we define the error indicator for 
the cases of explicitly filtered LES and
implicit LES (ILES) of incompressible flows,
as well as the implicitly filtered LES of compressible flows.
Our formulation of the error indicator suggests a small modification 
to the standard compressible version of the 
dynamic procedure.

In explicitly filtered LES, 
the convective term of the momentum equation
is filtered at each time step 
to take the form $\lpf{ \lpf{u}_i \lpf{u}_j}$,
and the definition of the ``subfilter" stress is modified accordingly as
${\tau}_{ij} = \lpf{u_i u_j} - \lpf{ \lpf{u}_i \lpf{u}_j }$~\cite[cf.][]{bose:10}.
The definition of $\tlpf{ \lpf{ \F }}^{(\n)}_i \!\!\!\!\! (\x)$
used in Eqn.~\ref{eq:Gn}
should also be modified as
\beq
\tlpf{ \lpf{ \F }}^{(\n)}_i \!\!\!\!\! (\x)
\equiv
\pd{}{x_j} 
\left[
 	\tlpf{ \lpf{ \lpf{u}_i \lpf{u}_j } }^{(\n)} 
 - 	  \tlpf{ \lpf{
 \tlpf{ \lpf{u}}^{(\n)}_i  \tlpf{ \lpf{u}}^{(\n)}_j
 }}^{(\n)}
 \right]
 +
 \pd{}{x_j} 
 \left[
  	\tlpf{ {\tau}_{ij}^{\rm mod}( \lpf{u}_k ) }^{(\n)} 
  - 	\tau^{\rm mod}_{ij} (\tlpf{ \lpf{u}}^{(\n)}_k)
  \right]
  \, .
\eeq
This again becomes the divergence of the tensor that is used
in formulation of the dynamic procedure in explicitly filtered LES.

In ILES there is no explicit SGS model in the code, i.e.,
${\tau}_{ij}^{\rm mod} \equiv 0$
and the effect of subgrid scales is accounted for by
numerics designed to mimic an LES model.
This is usually done by modifying the convective term,
for which case
the definition of $\tlpf{ \lpf{ \F }}^{(\n)}_i $
could be modified as
\beq
\tlpf{ \lpf{ \F }}^{(\n)}_i \!\!\!\!\! (\x)
=
\frac{\delta}{\delta x_j}
\left(
 	\tlpf{ \lpf{u}_i \lpf{u}_j}^{(\n)} 
 - 	 \tlpf{ \lpf{u}}^{(\n)}_i  \tlpf{ \lpf{u}}^{(\n)}_j
 \right)
\, ,
\eeq
where $\delta/\delta x_j$ denotes the specific numerics used in the code
and it has replaced $\partial/\partial x_j$ to emphasize the need to implement 
numerics that are consistent with the goal of mimicking a SGS model.
Similarly, one could get a slightly different definition for other 
formulations of ILES (e.g., when it is implemented
by the diffusive term).
We have not tested whether or not an inconsistent implementation of numerics
(e.g., a central scheme) could produce acceptable results,
but we should probably expect a similar behavior to what 
reported in appendix~\ref{section:incons}
for ``Vr$/$DNS" or ``DSM$/$DNS" grids.

The definition of the error indicator for LES of compressible flows
becomes more complicated due to the extra equations involved
and the Favre-filtering of the primitive variables.
The governing equations for implicitly filtered LES of
compressible flows (in their conservative form) read~\cite[cf.][]{garnier:09}
\begin{gather} \nonumber
\pd{\lpf{\rho}}{t}
+
\pd{}{x_j} \lpf{\rho} \widetilde{u}_j
=0
\\ \nonumber
\pd{}{t}\lpf{\rho} \widetilde{u}_i
+
\pd{}{x_j}\lpf{\rho} \widetilde{u}_i \widetilde{u}_j
+
\pd{\lpf{p}}{x_i}
-
\pd{\widetilde{\sigma}_{ij}}{x_j}
=
\pd{ {\mathcal{T}}_{ij} }{x_j}
\\ \nonumber
\pd{}{t} \lpf{\rho} \widetilde{E}
+
\pd{}{x_j} ( \lpf{\rho} \widetilde{E} +\lpf{p} ) \widetilde{u}_j
-
\pd{}{x_j} \widetilde{\sigma}_{ij} \widetilde{u}_i
+
\pd{ \widetilde{ \mathcal{Q} }_j }{x_j}
=
\pd{ { \mathcal{S} }_j }{x_j}
\end{gather}
where $\lpf{\cdot}$ is the filtering operation that is 
implicitly applied,
$\lpf{\rho}$ and $\lpf{p}$ 
are the resolved density and pressure, respectively,
and $\widetilde{u}_i$ and $\widetilde{E}$ are the Favre-filtered
velocity and total internal energy
with Favre filtering defined as
$\tilde{\phi} = \lpf{\rho \phi}/\lpf{\rho}$.
The terms $\widetilde{\sigma}_{ij}$ and $\widetilde{ \mathcal{Q} }_j$
describe the viscous stress and conductive heat flux
and are defined as
\beq \nonumber
\begin{aligned}
\widetilde{\sigma}_{ij}
&=
\sigma_{ij} ( \widetilde{T} , \widetilde{u}_i   )
=
\mu( \widetilde{T} )
\left(
2 \widetilde{S}_{ij} 
- \frac{2}{3}
\widetilde{S}_{kk} \delta_{ij} 
\right)
\\
\widetilde{ \mathcal{Q} }_j
&= \mathcal{Q}_j (\widetilde{T})
=
-\kappa( \widetilde{T} )
\pd{\widetilde{T}}{x_j}
\end{aligned}
\eeq
where $\widetilde{T}=\lpf{p}/R\lpf{\rho}$ is the Favre-filtered temperature and 
$\mu( \widetilde{T} )$ and $\kappa( \widetilde{T} )$
are the molecular viscosity and thermal conductivity 
that are functions of $\widetilde{T}$.
The Favre-filtered strain rate is defined as
$\widetilde{S}_{ij} = ( \partial \widetilde{u}_i/\partial x_j + \partial \widetilde{u}_j/\partial x_i )/2$.
The terms ${\mathcal{T}}_{ij}$ and ${ \mathcal{S} }_j$
contain the entire effect of subgrid scales
on the momentum and energy equations
and are modeled using the LES model
(${\mathcal{T}}_{ij}$ could be slightly different from
${\tau}_{ij} = \lpf{\rho} ( \widetilde{u_i u_j} -  \widetilde{u}_i  \widetilde{u}_j )$
since there might be extra subgrid processes involved).

If we follow the approach of Section~\ref{section:Gn-formulation}
and apply a directional test-filter $\tlpf{\cdot}^{(\n)}$ to the momentum equation
at filter level $\lpf{\Delta}$ and subtract it from
the momentum equation at the test-filter level $\tlpf{ \lpf{ \cdot}}^{(\n)}$
we can identify the following as the residual forcing term,
\beq \label{eq:Fi-comp}
\begin{aligned}
\tlpf{\lpf{\F}}^{(\n)}_i \!\!\!\!\! (\x)
=
\pd{}{x_j}
\left[
\tlpf{ \lpf{\rho} \widetilde{u}_i  \widetilde{u}_j }^{(\n)}
-
\frac{
\tlpf{\lpf{\rho u_i}}^{(\n)} \tlpf{\lpf{\rho u_j}}^{(\n)}
}{\tlpf{\lpf{\rho}}^{(\n)} } 
\right]
&+
\pd{}{x_j}
\left[
\tlpf{\mathcal{T}_{ij}( \lpf{\rho},   \widetilde{u}_k  )}^{(\n)} 
-
\mathcal{T}_{ij}( \tlpf{\lpf{\rho}}^{(\n)} ,   \check{u}_k^{(\n)}  )
\right]
\\
&+
\pd{}{x_j}
\left[
\tlpf{\sigma_{ij}( \widetilde{{T}} , \widetilde{u}_k  )}^{(\n)}
-
{\sigma}_{ij}( \check{{T}}^{(\n)}  , \check{u}^{(\n)}_k  ) 
\right]
\,,
\end{aligned}
\eeq
where $\check{\cdot}^{(\n)}$ denotes Favre-filtering
at the test-filter level $\tlpf{\lpf{\Delta }}^{(\n)}$
defined as
$\check{\phi}^{(\n)} = \tlpf{\lpf{\rho \phi}}^{(\n)} / \tlpf{\lpf{\rho }}^{(\n)}$.
If we neglect the last term
(the nonlinearity of the viscous term),
the residual forcing term becomes
the divergence of the tensor used in 
the standard compressible version 
of the dynamic procedure~\cite[cf.][]{garnier:09}.
However, based on our discussions in Section~\ref{section:dynamic},
one should in principle include this term when calculating the model coefficient dynamically.
The most important application of this modification is probably in 
flows with strong heating/cooling, 
where $\mu( \widetilde{T} )$ has large variations,
especially in complex flows where one cannot specify the preferred filtering direction.

If we repeat our approach for the mass conservation equation we 
end up with
$\partial \tlpf{\lpf{\Gamma}}^{(\n)}\!\!\!\!\!/\partial t + \partial \tlpf{\lpf{e}}^{(\n)}_j \!\!\!/\partial x_j = 0$,
where $\tlpf{\lpf{\Gamma}}^{(\n)}$ and $\tlpf{\lpf{e}}^{(\n)}_j $
are the errors in the density and mass flux.
This suggests that we can exclude the mass conservation equation from 
our analysis of the source of error:
the momentum equation essentially leads to an evolution
equation for error in the mass flux $\tlpf{\lpf{e}}^{(\n)}_j $,
and thus,
by minimizing the source of error in the momentum equation
(which is the forcing term of Eqn.~\ref{eq:Fi-comp})
we automatically minimize the error in the
mass equation as well.

We can define a separate error indicator for the energy equation as
\beq
\lpf{\G}^\prime(\x,\n)
=
\sqrt{
\avg{ \tlpf{ \lpf{ \mathcal{J}}}^{(\n)}  \!\!\!\!\!\!\! (\x) \tlpf{ \lpf{ \mathcal{J}}}^{(\n)}  \!\!\!\!\!\!\! (\x)  }
}
\eeq
where $\tlpf{ \lpf{ \mathcal{J}}}^{(\n)}  \!\!\!\!\!\!\!(\x) $ is defined as
\beq \label{eq:Ji}
\begin{aligned}
\tlpf{ \lpf{ \mathcal{J}}}^{(\n)}  \!\!\!\!\!\!\! (\x)
=
&-\pd{}{x_j}
\left[
 \tlpf{\lpf{\rho}}^{(\n)} \check{{E}}^{(\n)} \check{u}^{(\n)}_j
 -
 \tlpf{\lpf{\rho} \widetilde{E} \widetilde{u}_j }^{(\n)}
 \right]
-
 \pd{}{x_j}
 \left[
 \tlpf{\lpf{p}}^{(\n)}  \check{u}^{(\n)}_j
 -
  \tlpf{ \lpf{p} \widetilde{u}_j }^{(\n)} 
  \right]
   \\
 &-
 \pd{}{x_j}
 \left[
 \frac{1-\gamma}{2}
 \check{u}^{(\n)}_j
\left(
\tlpf{\lpf{\rho}}^{(\n)}  \check{u}^{(\n)}_k  \check{u}^{(\n)}_k
-
\tlpf{\lpf{\rho}  \widetilde{u}_k  \widetilde{u}_k }^{(\n)} 
+
\tau^{\rm mod}_{kk} (\tlpf{\lpf{u}}^{(\n)}_i)
-
\tlpf{ {\tau}_{kk}^{\rm mod} (\lpf{u}_i)  }^{(\n)}
\right)
  \right]
   \\
 &+
 \pd{}{x_j}
 \left[
 \check{u}^{(\n)}_i \sigma_{ij} ( \check{{T}}^{(\n)},  \check{u}^{(\n)}_k )
 -
  \tlpf{ \widetilde{u}_i \sigma_{ij} ( \widetilde{{T}},  \widetilde{u}_k ) }^{(\n)}
  \right]
   %
 %
 %
 %
 -
 \pd{}{x_j}
 \left[
  \mathcal{Q}_j ( \check{ {T} }^{(\n)} )
  -
  \tlpf{ \mathcal{Q}_j ( \widetilde{T}  ) }^{(\n)}
  \right]
     \\
 &+
 \pd{}{x_j}
 \left[
  \mathcal{S}^{\rm mod}_j (\tlpf{\lpf{ \rho }}^{(\n)} , \tlpf{\lpf{ \rho u_i }}^{(\n)} , \tlpf{\lpf{ \rho{E} }}^{(\n)} )
  -
  \tlpf{ \mathcal{S}^{\rm mod}_j ( \lpf{ \rho } , \lpf{ \rho u_i } , \lpf{ \rho{E} } ) }^{(\n)}
  \right] \,.
\end{aligned}
\eeq
Here $\gamma$ is the ratio of the specific heats,
and ${\tau}_{ij}^{\rm mod}$ is the LES model used
for ${\tau}_{ij} = \lpf{\rho}
( \widetilde{u_i u_j} - \widetilde{u}_i \widetilde{u}_j ) $.
Note how different this error indicator is from 
an intuition-based error indicator for the energy equation,
that could be defined for instance as 
$\sqrt{ \avg{ \widetilde{T}^{*,(\n)} \widetilde{T}^{*,(\n)} } }$,
where $\widetilde{T}^{*,(\n)} = \widetilde{T} - \tlpf{\widetilde{T}}^{(\n)}$
(same as~\cite[][]{bose:12:sp} but applied in a directional sense).

One important point to keep in mind is that
we arrived at Eqns.~\ref{eq:Fi-comp} and~\ref{eq:Ji}
by excluding the error in the energy equation
from the residual term in the momentum equation and vice versa.
This is a relatively \emph{ad hoc} assumption,
driven by the appeal to make the equations simpler,
and could be suboptimal.

\section{The effect of commutation errors \label{sec:appendix-commute}}

Generally speaking, the commutation errors may pose
strict requirements on how rapidly the filter-width can vary in 
space (e.g., how large the stretching factor might be, etc.). 
While these requirements
can probably be best imposed as extra constraints when formulating
the optimization problem of finding $\check{\Delta}_{\rm opt}$
in Section~\ref{sec:optimal}
(and thus have not been the focus of this paper),
it is interesting to note that they can also be directly included
in the definition of the error indicator,
by relaxing the assumption of 
commutation between filtering and differentiation
in Eqn.~\ref{eq:ffns-u}. 
In that case, all the terms in the equation
will have a commutation error that is moved to
the right-hand side of Eqns.~\ref{eq:ffns-u}
and~\ref{eq:eeq} and is included in what 
we consider as the source of error $\tlpf{ \lpf{ \F }}^{(\n_0)}_i $.
As a simplified version of this,
where we only include the commutation error
of the convective term and the SGS term,
the definition of the forcing term is modified as
\beq
\label{eq:F-commute}
\tlpf{ \lpf{ \F }}^{(\n_0)}_i \!\!\!\!\!\!\!  (\x)
=
\tlpf{ \pd{ \lpf{u}_i \lpf{u}_j}{x_j} }^{(\n_0)} 
+
\tlpf{ \pd{  \tau_{ij}^{\rm mod}(\lpf{u}_k) }{x_j} }^{(\n_0)} 
-\pd{}{x_j}
\left[
 	 \tlpf{ \lpf{u}}^{(\n_0)}_i  \tlpf{ \lpf{u}}^{(\n_0)}_j
	 +
	 \tau_{ij}^{\rm mod}\left(\tlpf{\lpf{u}}^{(\n_0)}_k \right)
 \right]
  \,.
\eeq

\begin{figure}[t!]
	\centering	
	\includegraphics[width=38mm,clip=true,trim=0mm 0mm 10mm 3mm]{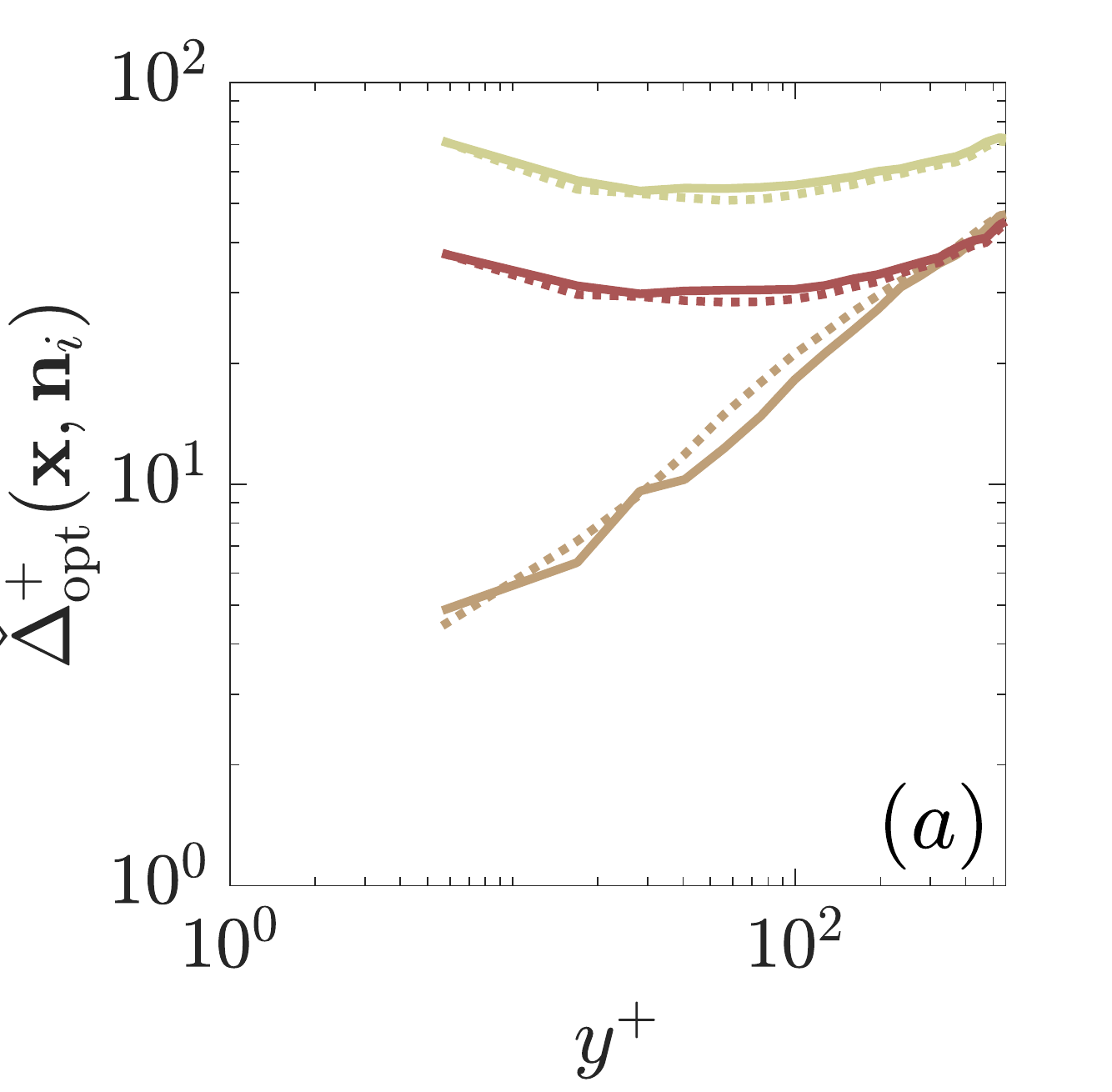}
	\includegraphics[width=38mm,clip=true,trim=0mm 0mm 10mm 3mm]{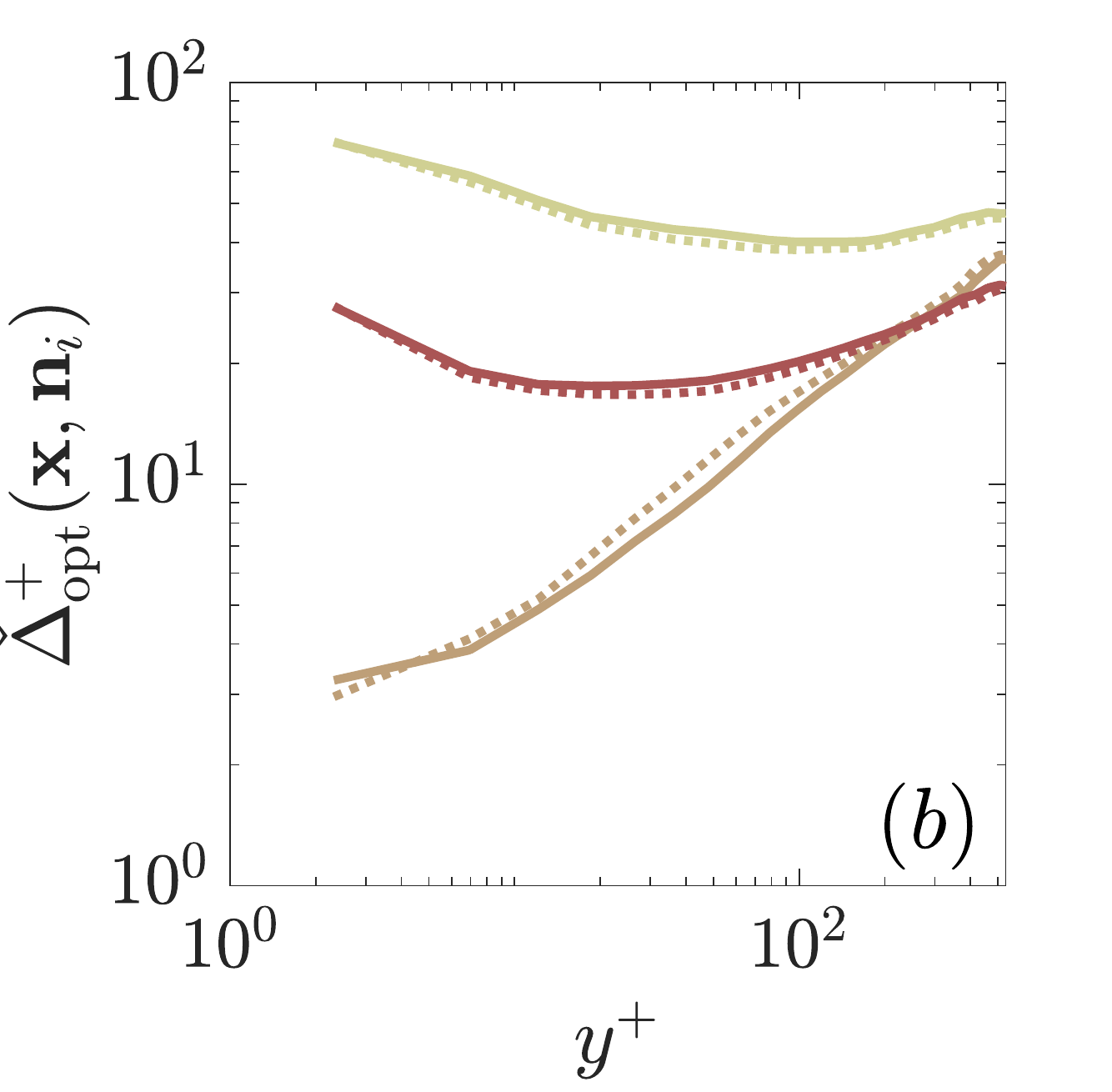}
	\includegraphics[width=38mm,clip=true,trim=0mm 0mm 10mm 3mm]{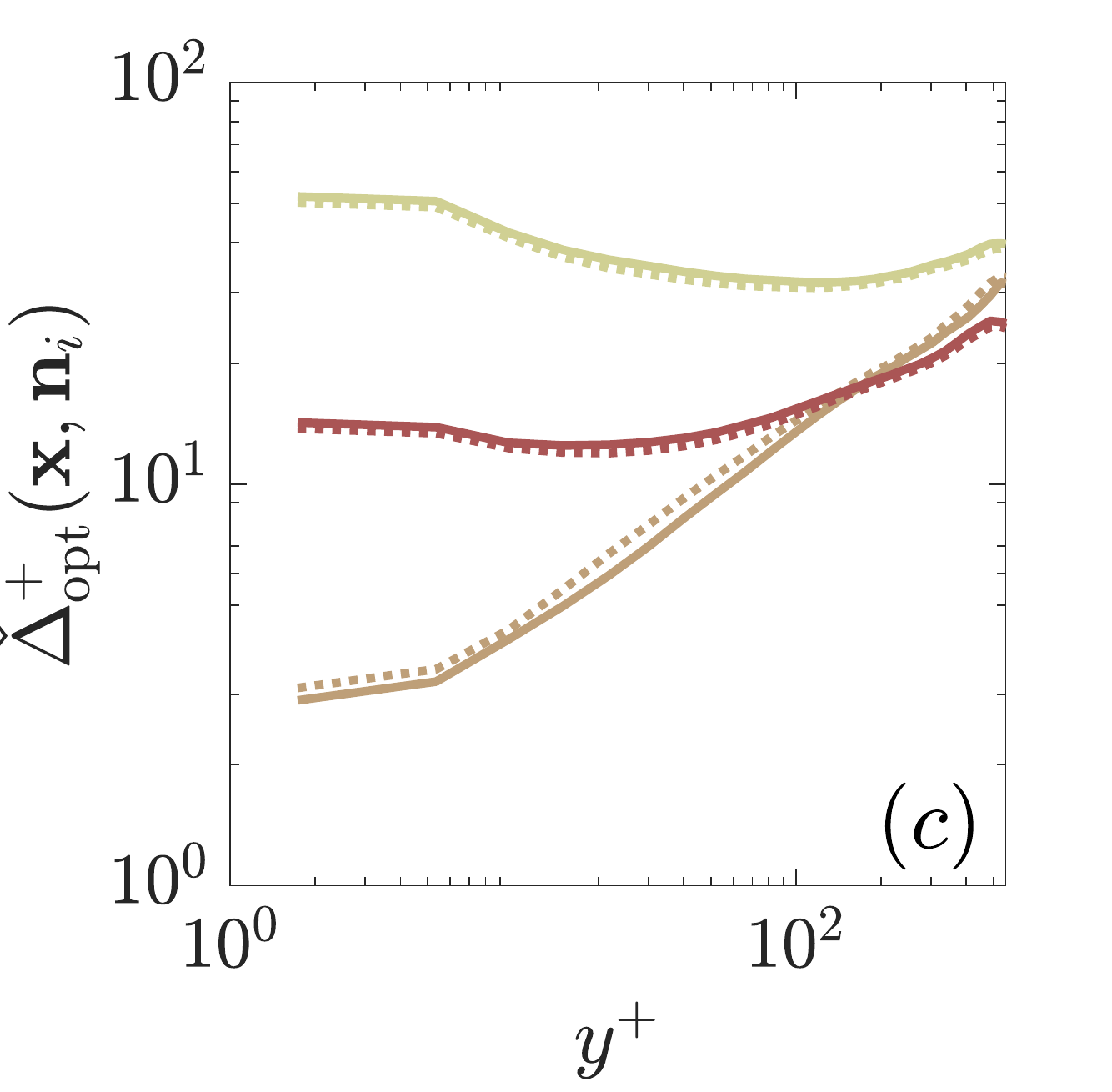}
	\caption{\label{fig:dx-profiles-commute}
	Comparison of the target resolutions 
	corresponding to grids (\textit{a}) DSM-3, 
	(\textit{b}) DSM-4 and (\textit{c}) DSM-5
	of Table~\ref{Table:channel-Gn-DSM}
	between error indicators computed using
	the original definition of Eqn.~\ref{eq:F} (dotted lines)
	and the modified definition of Eqn.~\ref{eq:F-commute} (solid lines).
	The streamwise, 
	wall-normal 
	and spanwise 
	resolutions
	are shown by the lightest to the darkest colors (in that order).
	}
\end{figure}

Figure~\ref{fig:dx-profiles-commute} shows 
how the target resolutions change in the case of
the channel flow for 
the modified
error indicator defined by Eqn.~\ref{eq:F-commute}.
The wall-normal resolution only changes slightly
to have an apparently slower stretching across the channel,
which is consistent with the smooth but slightly aggressive stretching of
the grid predicted by the original definition of the error indicator. 
Note that the streamwise and spanwise resolutions
are uniform, and thus there is no difference in the 
computed value of the error indicator in those directions.
Therefore, the small change in the resolution in $x$ and $z$
is caused by the change in the wall-normal resolution 
and the constraint on the total number of grid points.

A better test-filter for this work would be one that 
more closely resembles the effect of implicit filtering
(or cut-off) of the grid. 
For such filters
the commutation errors can become larger, and thus
more important.

\section{Comparison with the heuristic-based error indicator \label{section:An}}

In our previous work~\cite[]{toosi:17} we defined a heuristic-based 
error indicator as 
\beq \label{eq:An}
\lpf{\A}(\x,\n) = \sqrt{ \avg{ \lpf{u}_i^{*,(\n)}  \lpf{u}_i^{*,(\n)}  } }
\eeq
where $\lpf{u}_i^{*,(\n)} = \lpf{u}_i - \tlpf{ \lpf{u}}_i^{(\n)}$
is the directionally high-pass test-filtered LES velocity field
(using the same filter of Eqn.~\ref{eq:filter}).
From a physical point of view,
$\lpf{\A}(\x,\n)$ only measures the small scale content of the velocity fields
in any direction $\n$;
however, if we employ the 
classical intuitive argument that the small scale energy
controls both the numerical and modeling errors in 
LES~\cite[]{pope:04,bose:thesis},
we can assume that the errors are proportional to
$\lpf{\A}(\x,\n)$ and thus use it as an error indicator.

It is useful to compare the grids generated by the proposed error indicator 
of this study (Eqn.~\ref{eq:Gn}) with those of Eqn.~\ref{eq:An}.
In that sense, the results of this section complement 
our assessments of $\lpf{\G}(\x,\n)$
in Sections~\ref{section:DSM} and~\ref{section:BFS}.

\begin{figure}[t!]
  \centering	
  \includegraphics[width=60mm,clip=true,trim=0mm 0mm 15mm 5mm]{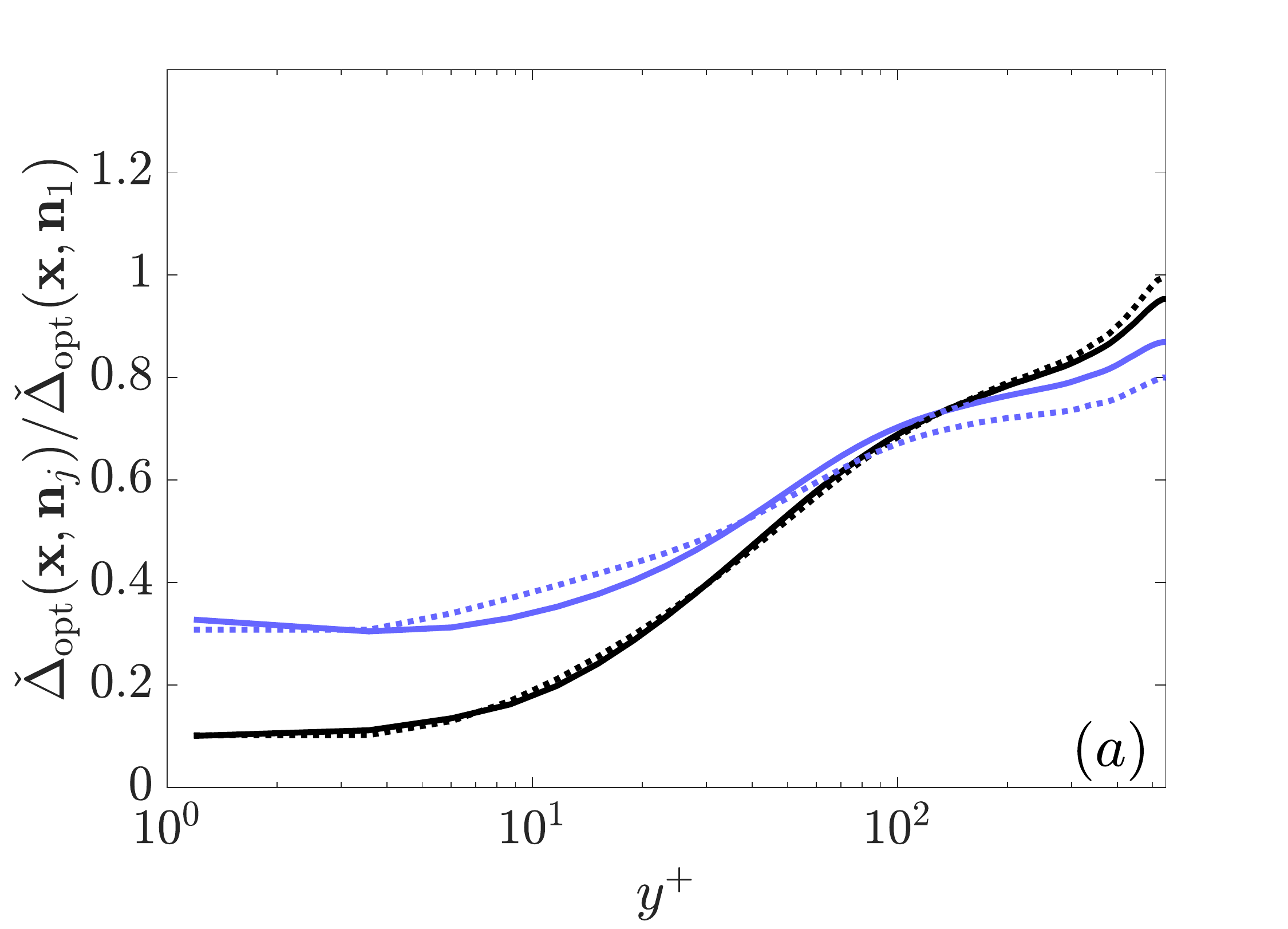}
  \includegraphics[width=60mm,clip=true,trim=0mm 0mm 15mm 5mm]{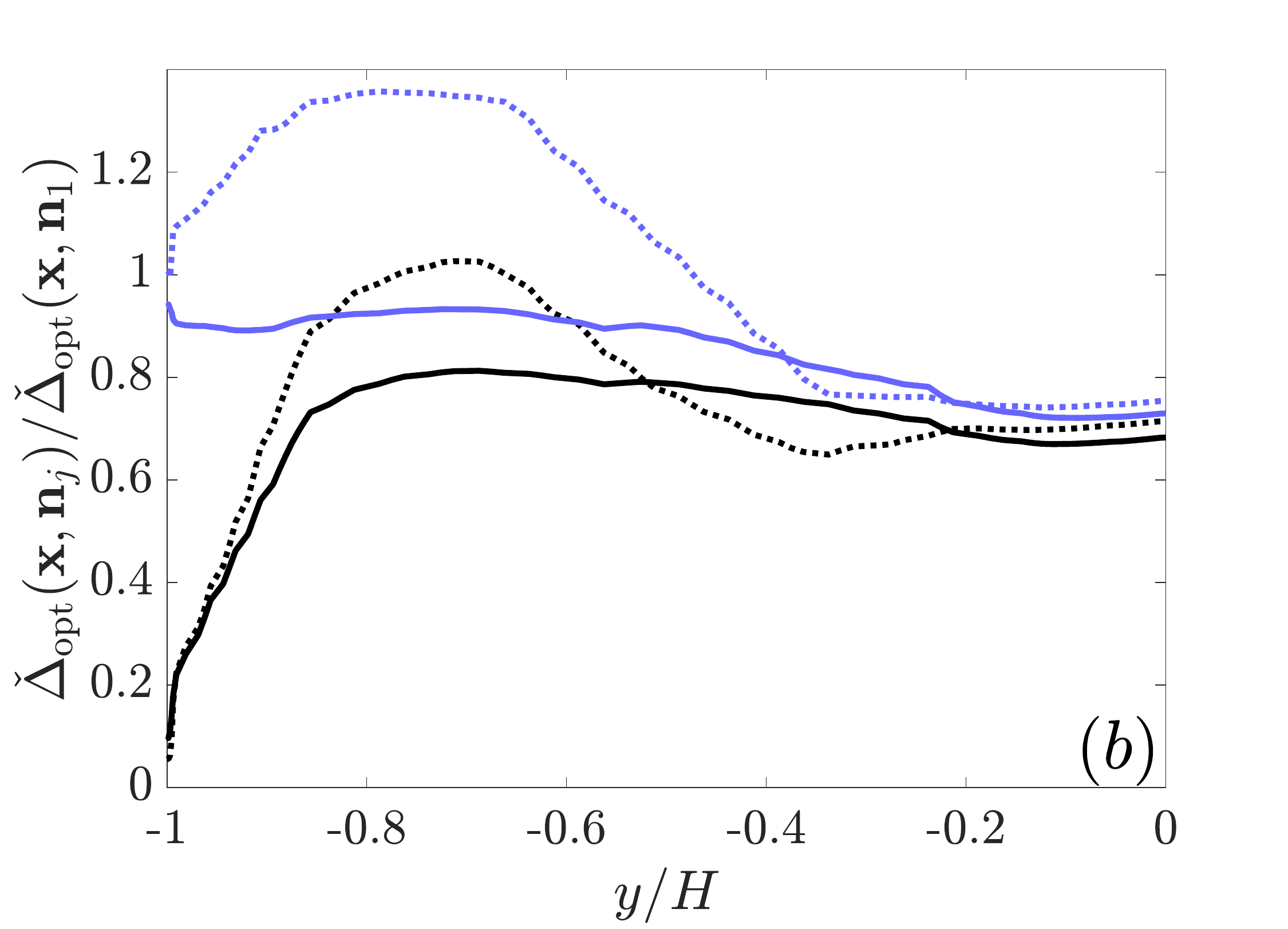}
  \caption{\label{fig:sample-ARs-compare}
    Examples of the predicted optimal cell aspect ratios
    $\check{\Delta}_{\rm opt}(\x,\n_2)/\check{\Delta}_{\rm opt}(\x,\n_1)$ (black lines)
    and
    $\check{\Delta}_{\rm opt}(\x,\n_3)/\check{\Delta}_{\rm opt}(\x,\n_1)$ (blue lines)
    using error indicators 
    $\lpf{\A}(\x,\n)$ (solid lines)
    and 
    $\lpf{\G}(\x,\n)$ (dotted lines)
    for 
    (\textit{a}) turbulent channel flow
    and 
    (\textit{b}) an $x$-normal plane inside the recirculation region
    ($x/H=2$)
    of the flow over a backward-facing step.
    Note that even in the channel flow where the predicted aspect
    ratios are similar, the difference in the spatial distribution of 
    the two error indicators (combined with the structured nature of the grids)
    causes the final aspect ratios to be different from
    each other
    (since the minimum of streamwise and spanwise
    resolutions are reached at different $y$ locations).
  }
\end{figure}

The flow setups used in~\cite[]{toosi:17} are slightly
different from those used here;
thus, all grid sequences are repeated for the exact flow setups
of this paper.
Additionally, we use the same
grid selection criteria of Eqns.~\ref{eq:optimal-direction} and~\ref{eq:optimal-space}
($\lpf{\G}(\x,\n)$ replaced by $\lpf{\A}(\x,\n)$),
which is different from the prior work.
Figure~\ref{fig:sample-ARs-compare} shows a comparison of
the optimal aspect ratios predicted by each error indicator
for the test cases of this work.

The sequence of grids generated for LES of channel flow 
is summarized in Table~\ref{Table:channel-An-new}.
An interesting observation is the qualitative difference between the two sets of grids:
although the grids generated using $\lpf{\A}(\x,\n)$
have a similar streamwise resolution 
to those generated by $\lpf{\G}(\x,\n)$,
their wall-normal resolution is coarser near the wall
and finer at the center of the channel 
(i.e., a flatter $\lpf{\Delta}_y$ profile with a smaller stretching factor).
This leads to a higher number of points across the channel
that has to be compensated by a coarser spanwise resolution to
keep $N_{\rm tot}$ constant.

\begin{table}[t!]
\begin{center}
\begin{tabular}{ r  c  c  c  c  c c  }
  Grid
  & $N_{\rm tot}$	
  & $N_y$
  & $(\fDelta_x^+, \lpf{\Delta}_{y_w}^+/2,\fDelta_z^+)$	
  & $(\fDelta_x , \fDelta_{y_c} , \fDelta_z)/H$	
  & $Re_{\tau}$
  &  $\errref$ (\%)
  \\[3pt]
    DSM-2
  & $74k$
  & 34
  & $(77,5.6,55)$
  & $(0.14,0.099,0.10)$
  & 553	
  & 11
  \\
  A-2
  & $76k$
  & 36
  & $(80,6.9,56)$
  & $(0.14,0.075,0.10)$
  & 562	
  & 12
  \\[2pt]  
  DSM-3
  & $251k$
  & 44
  & $(53,2.3,29)$
  & $(0.098,0.091,0.054)$
  & 536
  & 7.3
  \\
  A-3
  & $245k$
  & 48
  & $(57,3.6,32)$
  & $(0.10,0.061,0.058)$
  & 560
  & 8.3
 \\[2pt] 
    DSM-4
  & $514k$
  & 50
  & $(45,1.7,19)$
  & $(0.082,0.080,0.035)$
  & 544	
  & 3.3
  \\
  A-4
  & $526k$
  & 56
  & $(46,2.9,22)$
  & $(0.082,0.052,0.039)$
  & 559	
  & 5.5
  \\[2pt] 
    DSM-5
  & $1.18M$
  & 60
  & $(34,1.4,13)$
  & $(0.063,0.065,0.024)$
  & 544	
  & 1.8
  \\
  A-5
  & $1.17M$
  & 66
  & $(35,2.6,15)$
  & $(0.063,0.044,0.027)$
  & 559	
  & 4.2
  \\[2pt] 
    DSM-6
  & $2.53M$
  & 72
  & $(25,1.6,10)$
  & $(0.046,0.052,0.018)$
  & 542
  & 1.1
    \\
  A-6
  & $2.52M$
  & 80
  & $(26,2.2,11)$
  & $(0.048,0.035,0.020)$
  & 552
  & 2.4
   \\[2pt] 
      DSM-7
  & $5.80M$
  & 90
  & $(18,1.4,7.6)$
  & $(0.033,0.041,0.014)$
  & 540	
  & 1.1
      \\
  A-7
  & $5.90M$
  & 100
  & $(18,1.8,8.2)$
  & $(0.034,0.028,0.015)$
  & 543	
  & 0.8
  \\[2pt] 
  DSM-8
  & $11.1M$
  & 108
  & $(14,1.2,6.3)$
  & $(0.025,0.033,0.012)$
  & 541
  & 0.9
        \\
  A-8
  & $11.3M$
  & 118
  & $(14,1.6,6.8)$
  & $(0.025,0.024,0.013)$
  & 542
  & 1.4
\end{tabular}
\end{center}
\caption{ \label{Table:channel-An-new}
  Sequence of grids generated for LES of channel flow
  at $Re_\tau \approx 545$ using the dynamic Smagorinsky model.
  All A-$k$ grids are generated using the error indicator of Eqn.~\ref{eq:An}
  and the criteria of Eqns.~\ref{eq:optimal-direction} and~\ref{eq:optimal-space} with $\alpha=2$.
  Grids DSM-$k$ are simply copied from Table~\ref{Table:channel-Gn-DSM}.
  See caption of Table~\ref{Table:channel-Gn-DSM} for more details.
}
\end{table}

Almost all grids generated using the new error indicator $\lpf{\G}(\x,\n)$
have lower values of the error metric $\errref$.
If we accept the lower values of $\errref$ 
as a measure of optimality 
(this is not exactly true)
we can conclude that the grids generated by $\lpf{\G}(\x,\n)$
have a more optimal distribution.
This conclusion is consistent with our experience; 
in fact, grids generated by $\lpf{\A}(\x,\n)$
seem to have a slightly coarser resolution near the wall 
(especially in the last grids)
compared to what we expect for such high-resolution grids.

\begin{table}[t!]
\begin{center}
\begin{tabular}{ r  c  c  c  c  }
  Grid
  & $N_{\rm tot}$	
  & $(\fDelta_x^+ , \lpf{\Delta}_{y_w}^+/2 , \fDelta z^+)$
  & $(\fDelta_x, \fDelta_y, \fDelta_z)/\delta_{\rm shear}$
  & $\errref$ (\%)
 \\[3pt] 
  G-2
  & 297k
  & $(42,2.6,21)$
  & $(0.16,0.078,0.16)$
  & 10.5
  \\
  A-2
  & 297k
  & $(45, 5.7, 23)$
  & $(0.15,0.093,0.15)$
  & 10.5
  \\[2pt] 
    G-3
  & 611k
  & $(45,1.4,11)$
  & $(0.16,0.049,0.078)$
  & 5.6
  \\
  A-3
  & 599k
  & $(47,2.9,12)$
  & $(0.15,0.074,0.074)$
  & 6.1
  \\[2pt] 
    G-4
  & 1.32M
  & $(47,1.5,12)$
  & $(0.076,0.038,0.076)$
  & 4.9
  \\
   A-4
  & 1.35M
  & $(22,2.8,11)$
  & $(0.15,0.036,0.073)$
  & 6.6
  \\[2pt] 
    G-5
  & 2.13M
  & $(25,0.77,6.2)$
  & $(0.070,0.035,0.035)$
  & 5.4
    \\
  A-5
  & 2.17M
  & $(24,1.5,6.1)$
  & $(0.068,0.034,0.034)$
  & 4.2
  \\[2pt] 
      G-6
  & 3.41M
  & $(25,0.77,6.1)$
  & $(0.068,0.034,0.034)$
  & 3.5
    \\
  A-6
  & 3.70M
  & $(25,1.6,6.2)$
  & $(0.065,0.033,0.033)$
  & 4.4
   \\[2pt]   
  G-7
  & 6.72M
  & $(12,0.76.6.0)$
  & $(0.034,0.017,0.034)$
  & 2.5
  \\
  A-7
  & 7.26M
  & $(12,1.5,6.0)$
  & $(0.068,0.034,0.034)$
  & 2.0
\end{tabular}
\end{center}
\caption{ \label{Table:bfs-newAn}
Sequence of grids generated 
for LES of flow over a backward-facing step
using $\lpf{\A}(\x,\n)$ of Eqn.~\ref{eq:An}
and grid selection criteria of Eqns.~\ref{eq:optimal-direction} 
and~\ref{eq:optimal-space} with $\alpha=2$. 
Grids labeled by ``G" are simply copied from Table~\ref{Table:bfs-Gn}.
Refer to caption of Table~\ref{Table:bfs-Gn} for 
more details including
interpretation of each quantity.
}
\end{table}

\begin{figure}[t!]
  \centering	
  \ifIncludeFigures
  \includegraphics[width=130mm,clip=true,trim=450mm 0mm 200mm 130mm]{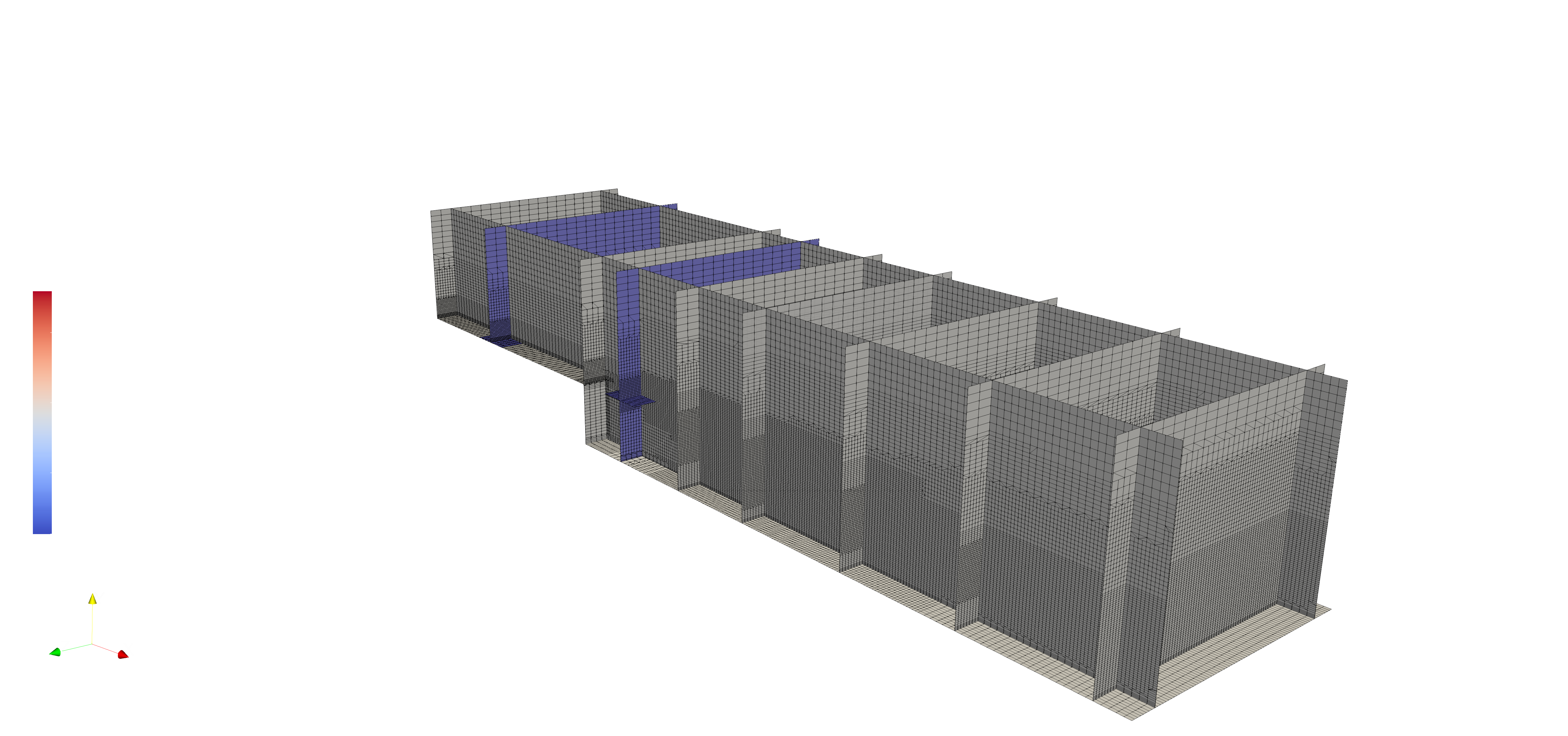}
  \fi
  \caption{\label{fig:grid-An6}
    Grid A-6 in Table~\ref{Table:bfs-newAn} with 3.70M cells.
    Intersections of the blue planes highlight locations whose resolutions are 
    reported in Table~\ref{Table:bfs-newAn}.
    Note that this grid is qualitatively different from G-6 shown in Fig.~\ref{fig:grid-Gn6}.
    See text for more details.
  }
\end{figure}

As a second comparison we consider the flow over a backward-facing
step,
with results 
summarized in Table~\ref{Table:bfs-newAn}.
The A-6 grid is shown in Fig.~\ref{fig:grid-An6}.
Interestingly, we can identify the same general grid distribution patterns as
what we saw in the channel flow; i.e.,
a general tendency to refine $\lpf{\Delta}_y$
less near the wall and more
towards the edge of the boundary layer,
as well as
refinement regions that 
are extended to a larger portion of the domain
(for all three resolution directions).

The error in (our specific) quantities of interest is generally lower for grids generated
by $\lpf{\G}(\x,\n)$.
Again, (assuming $\errref$ as a measure of optimality)
we can conclude 
that the grids generated by $\lpf{\G}(\x,\n)$ 
have a more optimal distribution.
Once more, our experience with LES confirms this conclusion,
as the reported resolutions in the boundary layer and 
shear layer of G-$k$ grids are closer to what we expect, 
especially in the last few grids
with relatively high resolutions.

\section{Sensitivity to approximate implementations of the error indicator \label{section:incons}}

In this section, we assess how sensitive the target grids are
(in terms of error in QoIs and predicted distribution of filter-width)
to approximations made when computing
$\check{\G}(\x,\n)$.
This is of special interest to us 
because such approximations 
are almost unavoidable in practice,
the most common of which
happens when computing 
$\tau_{ij}^{\rm mod}(\tlpf{\lpf{u}}_k^{(\n)})$ 
in Eqn.~\ref{eq:F}.
Examples of such approximations in the present study include: 
our assumption 
that the model coefficients remain unchanged
between filter levels $\lpf{\Delta}$ and $\tlpf{\lpf{\Delta}}^{(\n)}$
to avoid performing the full dynamic procedure
(used in Sections~\ref{section:DSM} and~\ref{section:BFS}),
estimating the change in $\lpf{k}_{\rm sgs}$ 
by an approximate formula to avoid solving
an extra transport equation (Section~\ref{section:BFS}),
different numerics used in computing $\lpf{\G}(\x,\n)$ 
to avoid
reimplementation of the exhaustively elaborate numerical
schemes used in the LES solver,
and so on.

As a relatively extreme test
we use different LES models in the code
and in computation of the error indicator.
To this end, we generate five new sequences of grids,
all starting from a grid with resolution of
$(\lpf{\Delta}_x,\lpf{\Delta}_y,\lpf{\Delta}_z)/H = (0.20,0.10,0.20)$.
In three of these sequences the constant Vreman model is
used in the LES solver with $c_v=0.07$ and $\lpf{\Delta}/\lpf{h}=1$
(slightly different setup from what was used in Section~\ref{section:vreman}
to test it in a regime of comparable magnitude for modeling and numerical errors),
while $\tau_{ij}^{\rm mod}(\tlpf{\lpf{u}}_k^{(\n)})$ in
the error indicator is computed once by the Vreman model
(as it should),
once by using the (dynamic) Smagorinsky model 
(Eqn.~\ref{eq:DSM-approximate} of the paper),
and once by setting the SGS terms to zero 
(${\tau}_{ij}^{\rm mod} \equiv 0$, corresponding to the DNS case).
These three sets of grids are labeled
``Vr$/$Vr", ``Vr$/$DSM" and ``Vr$/$DNS", respectively.
Similarly, the other three sequences are generated by using the dynamic Smagorinsky
model in the LES solver and using the Smagorinsky model in $\lpf{\G}(\x,\n)$
(labeled ``DSM$/$DSM"),
DSM in the solver and the Vreman model in the error indicator
(labeled ``DSM$/$Vr") 
or setting the SGS terms to zero in the error indicator. 
The last sequence (corresponding to ``DSM$/$DNS" grids)
is discontinued after the fourth grid,
since the target grids had identical resolutions 
(within two significant digits)
to ``DSM$/$Vr" grids.

\begin{figure}[t!]
	\centering	
	\includegraphics[width=80mm,clip=true,trim=00mm 00mm 10mm 0mm]{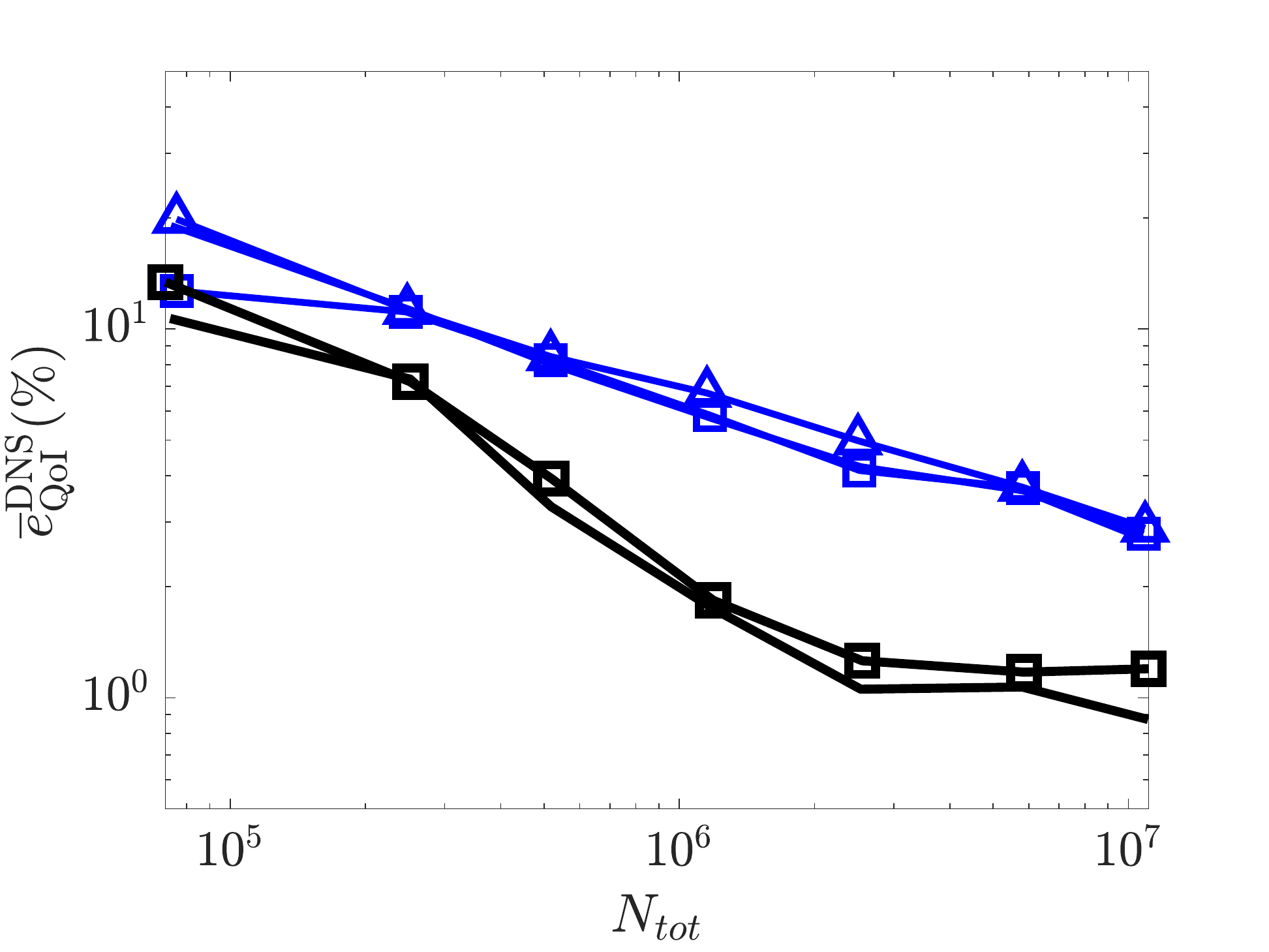}
	\caption{\label{fig:error-incons} 
	Sensitivity of error in the QoIs to 
	approximations in computation of $\lpf{\G}(\x,\n)$.
	Figure summarizes results of Tables~\ref{Table:channel-Gn-inconsVreman} 
	and~\ref{Table:channel-Gn-inconsDSM} for 
	convergence of grids labeled as
	``DSM$/$DSM" (plain black line),
	``DSM$/$Vr" (black with squares),
	``Vr$/$Vr" (plain blue),
	``Vr$/$DSM" (blue with squares)
	and 
	``Vr$/$DNS" (blue with triangles).
	See text and Tables for more details. 
	}
\end{figure}

The generated grids are summarized in 
Tables~\ref{Table:channel-Gn-inconsVreman} 
(for sequences with the Vreman model in the LES solver)
and~\ref{Table:channel-Gn-inconsDSM} (for DSM).
Convergence of $\errref$ with total number of cells $N_{\rm tot}$
is further illustrated in Fig.~\ref{fig:error-incons}.

\begin{table}[t!]
\begin{center}
\begin{tabular}{ l c  c  c  c  c  c   }
  Grid
  & $N_{\rm tot}$	
  & $N_y$
  & $(\fDelta_x^+, \lpf{\Delta}_{y_w}^+/2,\fDelta_z^+)$	
  & $(\fDelta_x , \fDelta_{y_c} , \fDelta_z)/H$	
  & $Re_{\tau}$
  &  $\errref$ (\%)
  \\[3pt] 
  Vr$/$Vr-2
  & $74k$
  & 34
  & $(70,4.0,47)$
  & $(0.14,0.10,0.097)$
  & 488
  & 19
  \\
    Vr$/$DSM-2
  & $76k$
  & 32
  & $(64,7.0,55)$
  & $(0.12,0.11,0.10)$
  & 528
  & 13
    \\
    Vr$/$DNS-2
  & $76k$
  & 36
  & $(70,3.3,46)$
  & $(0.15,0.096,0.097)$
  & 476
  & 20
   \\[2pt] 
    Vr$/$Vr-3
  & $253k$
  & 48
  & $(58,1.9,25)$
  & $(0.11,0.081,0.050)$
  & 507
  & 11
  \\
   Vr$/$DSM-3
  & $246k$
  & 40
  & $(41,3.8,33)$
  & $(0.078,0.10,0.063)$
  & 526
  & 11
    \\
   Vr$/$DNS-3
  & $245k$
  & 48
  & $(59,1.5,25)$
  & $(0.12,0.085,0.050)$
  & 504
  & 11
   \\[2pt] 
    Vr$/$Vr-4
  & $513k$
  & 52
  & $(46,2.0,18)$
  & $(0.089,0.076,0.034)$
  & 516
  & 8.1
  \\
   Vr$/$DSM-4
  & $517k$
  & 48
  & $(34,2.5,22)$
  & $(0.065,0.083,0.043)$
  & 522
  & 8.2
    \\
   Vr$/$DNS-4
  & $512k$
  & 54
  & $(46,1.1,18)$
  & $(0.091,0.075,0.035)$
  & 511
  & 8.4
  \\[2pt] 
    Vr$/$Vr-5
  & $1.18M$
  & 64
  & $(34,1.6,13)$
  & $(0.065,0.059,0.025)$
  & 521
  & 5.7
  \\
   Vr$/$DSM-5
  & $1.17M$
  & 60
  & $(29,2.0,15)$
  & $(0.054,0.065,0.028)$
  & 525
  & 5.8
    \\
   Vr$/$DNS-5
  & $1.15M$
  & 64
  & $(34,0.80,13)$
  & $(0.067,0.063,0.025)$
  & 517
  & 6.7
   \\[2pt] 
    Vr$/$Vr-6
  & $2.54M$
  & 76
  & $(25,1.4,10)$
  & $(0.047,0.049,0.019)$
  & 527
  & 4.2
    \\
   Vr$/$DSM-6
  & $2.52M$
  & 74
  & $(23,1.7,11)$
  & $(0.044,0.051,0.020)$
  & 530
  & 4.1
    \\
   Vr$/$DNS-6
  & $2.50M$
  & 78
  & $(25,0.61,10)$
  & $(0.048,0.050,0.020$
  & 521
  & 5.0
  \\[2pt] 
    Vr$/$Vr-7
  & $5.85M$
  & 96
  & $(18,1.1,7.8)$
  & $(0.033,0.037,0.019)$
  & 531
  & 3.6
    \\
   Vr$/$DSM-7
  & $5.83M$
  & 94
  & $(17,1.4,8.0)$
  & $(0.032,0.039,0.015)$
  & 534
  & 3.7
      \\
   Vr$/$DNS-7
  & $5.80M$
  & 98
  & $(18,0.47,7.9)$
  & $(0.034,0.038,0.015)$
  & 527
  & 3.7
   \\[2pt] 
    Vr$/$Vr-8
  & $10.7M$
  & 112
  & $(14,0.97,6.5)$
  & $(0.026,0.031,0.012)$
  & 533
  & 2.7
  \\
   Vr$/$DSM-8
  & $10.8M$
  & 110
  & $(14,1.2,6.5)$
  & $(0.025,0.032,0.012)$
  & 535
  & 2.8
      \\
   Vr$/$DNS-8
  & $10.9M$
  & 114
  & $(14,0.38,6.5)$
  & $(0.026,0.032,0.012)$
  & 530
  & 2.9
\end{tabular}
\end{center}
\caption{ \label{Table:channel-Gn-inconsVreman}
Sensitivity of target grids to 
approximations in computation of 
$\tau_{ij}^{\rm mod}(\tlpf{\lpf{u}}_k^{(\n)})$ in Eqn.~\ref{eq:F}.
All simulations use the constant Vreman model in the solver
with $c_v=0.07$ and $\lpf{\Delta}/\lpf{h}=1$.
Refer to caption of Table~\ref{Table:channel-Gn-DSM} for more details and
interpretation of what each quantity means.
See text for how grids ``Vr$/$Vr", ``Vr$/$DSM" and ``Vr$/$DNS" are generated.
}
\end{table}

\begin{table}[t!]
\begin{center}
\begin{tabular}{ l  c  c  c  c  c  c   }
  Grid
  & $N_{\rm tot}$	
  & $N_y$
  & $(\fDelta_x^+, \lpf{\Delta}_{y_w}^+/2,\fDelta_z^+)$	
  & $(\fDelta_x , \fDelta_{y_c} , \fDelta_z)/H$	
  & $Re_{\tau}$
  &  $\errref$ (\%)
  \\[3pt] 
  DSM$/$DSM-2
  & $74k$
  & 34
  & $(77,5.6,55)$
  & $(0.14,0.099,0.10)$
  & 553	
  & 11
  \\
    DSM$/$Vr-2
  & $72k$
  & 34
  & $(78,4.1,51)$
  & $(0.15,0.10,0.097)$
  & 529
  & 13
  \\[2pt] 
    DSM$/$DSM-3
  & $251k$
  & 44
  & $(53,2.3,29)$
  & $(0.098,0.091,0.054)$
  & 536
  & 7.3
  \\
   DSM$/$Vr-3
  & $252k$
  & 46
  & $(55,1.4,28)$
  & $(0.10,0.091,0.053)$
  & 527
  & 7.2
  \\[2pt] 
    DSM$/$DSM-4
  & $514k$
  & 50
  & $(45,1.7,19)$
  & $(0.082,0.080,0.035)$
  & 544	
  & 3.3
  \\
   DSM$/$Vr-4
  & $519k$
  & 52
  & $(46,1.0,19)$
  & $(0.086,0.079,0.035)$
  & 538
  & 3.9
  \\[2pt] 
    DSM$/$DSM-5
  & $1.18M$
  & 60
  & $(34,1.4,13)$
  & $(0.063,0.065,0.024)$
  & 544	
  & 1.8
  \\
   DSM$/$Vr-5
  & $1.19M$
  & 62
  & $(34,0.79,13)$
  & $(0.064,0.065,0.024)$
  & 537
  & 1.8
  \\[2pt] 
    DSM$/$DSM-6
  & $2.53M$
  & 72
  & $(25,1.6,10)$
  & $(0.046,0.052,0.018)$
  & 542
  & 1.1
    \\
   DSM$/$Vr-6
  & $2.54M$
  & 76
  & $(25,0.62,10)$
  & $(0.048,0.051,0.019)$
  & 535
  & 1.3
    \\[2pt] 
   DSM$/$DSM-7
  & $5.80M$
  & 90
  & $(18,1.4,7.6)$
  & $(0.033,0.041,0.014)$
  & 540	
  & 1.1
      \\
  DSM$/$Vr-7
  & $5.83M$
  & 94
  & $(18,0.51,7.7)$
  & $(0.034,0.040,0.014)$
  & 535
  & 1.2
  \\[2pt] 
   DSM$/$DSM-8
  & $11.1M$
  & 108
  & $(14,1.2,6.3)$
  & $(0.025,0.033,0.012)$
  & 541
  & 0.9
  \\
   DSM$/$Vr-8
  & $11.1M$
  & 112
  & $(14,0.43,6.4)$
  & $(0.025,0.032,0.012)$
  & 537
  & 1.2
\end{tabular}
\end{center}
\caption{ \label{Table:channel-Gn-inconsDSM}
Sensitivity of target grids to 
approximations in computation of 
$\tau_{ij}^{\rm mod}(\tlpf{\lpf{u}}_k^{(\n)})$ in Eqn.~\ref{eq:F}.
Refer to caption of Table~\ref{Table:channel-Gn-DSM} for more details.
All results are for LES using the dynamic Smagorinsky model.
Grids labeled ``DSM$/$DSM" are the same as those reported in Table~\ref{Table:channel-Gn-DSM}
and are simply copied from there.
See text for grids ``DSM$/$Vr".
}
\end{table}

Note that (see Fig.~\ref{fig:error-incons})
the error in the QoIs 
is not significantly affected by these relatively extreme 
inconsistencies in
our implementation of the error indicator.
The change in the target resolution is slightly more noticeable
(Tables~\ref{Table:channel-Gn-inconsVreman} 
and~\ref{Table:channel-Gn-inconsDSM}).
For instance, the grid Vr$/$DSM-4 has a friction resolution
of $(34,2.5,22)$ which is somewhat different from 
grid Vr-4 with resolution of $(46,2.0,18)$.
Interestingly,
this change in the target resolutions has a general trend 
that is present for almost all grids in the sequence
(e.g., ``DSM$/$Vr" grids
have similar streamwise and spanwise
resolutions compared to ``DSM$/$DSM" grids, 
while their wall-normal resolution is finer 
adjacent to the wall).
However, we should emphasize that
despite the relative change in the resolution
of the target grids
these are still suitable grids for 
LES of wall bounded turbulence.
In other words, the aspect ratio of the cells 
may be slightly affected and suboptimal,
but the spanwise resolutions of the cells are
still significantly finer than their streamwise resolution,
and their wall-normal resolution is such that
it resolves all the scales in the $y$ direction.
The small change in $\errref$ is 
in fact another proof of
suitability of generated grids for channel flow:
although this little effect on error in the QoIs 
is more related to the ability of the
solver and its LES model in handling different grids,
we can still conclude that the change in the target resolution
is within some acceptable value 
to not deteriorate LES results significantly.

The presented results are specific to
the LES code (numerics and models) 
and implementation of the error indicator
used in this study,
while
the conclusion that the target grids 
are still close to what we would get 
by accurate implementation of the LES model 
is probably more general.

\section{Sensitivity of the target grids to the initial grid \label{section:BFS-initialGrid}}

In Sections~\ref{sec:channel}
and~\ref{section:BFS}
we tested the robustness of the
error indicator and its target grids
to coarse initial grids and underresolved turbulence. 
While in all cases the proposed methodology led to adapted grids
that were quite close to what is considered ``best-practice'',
there are still two questions that remain to be answered:
(i) how repeatable the generated grids are
(i.e., if we can generate the same target grids by starting from 
a different initial grid),
and
(ii) how sensitive the final results are
to the skeletal grid 
(i.e., if the proposed method can sufficiently refine the grid
in all important regions, even if the maximum resolution 
is not limited by the skeletal grid).
The goal of this section is to answer these two questions.

Only the flow over a backward-facing step is studied in this section.
The sequence is started from an initial grid G-1$'$
with a resolution of
$(\fDelta_x,\fDelta_y,\fDelta_z)/H = (0.8,0.1,0.1)$.
The reason for the finer spanwise resolution
is to keep the flow turbulent. 
The skeletal grid in this section has a
resolution of
$(\Delta_{0,x},\Delta_{0,y},\Delta_{0,z})/H = (0.8,0.2,0.2)$ 
beyond which coarsening is not possible. 
At each iteration of the process $N_{\rm tot}$
is matched between grids G-$k'$ of this section and
G-$k$ (the original sequence in Section~\ref{section:BFS}).
The sequence is terminated once the next
target grid is deemed sufficiently similar to 
its equivalent in the original sequence. 
The results are summarized in Table~\ref{Table:BFS-Gn-prime}.

\begin{table}[t!]
\begin{center}
\begin{tabular}{ l c  c  c  c   c }
  Grid
  & $N_{\rm tot}$	
  & $(\lpf{\Delta}_x^+ , \lpf{\Delta}_{y_w}^+/2 , \lpf{\Delta}_z^+)$
  & $(\lpf{\Delta}_x, \lpf{\Delta}_y, \lpf{\Delta}_z)/\delta_{\rm shear}$
  & $\errref$ (\%)
  \\[3pt]
  G-1$'$
  & 79.6k
  & (173,11,22)
  & (1.2,0.14,0.14)
    & 41
  \\
    G-1
  & 149k
  & $(42,10,42)$
  & $(0.21, 0.17, 0.33)$
    & 11.1
  \\[2pt]
    G-2$'$
  & 285k
  & (46,2.9,12)
  & (0.17,0.083,0.083)
    & 15.5
  \\
  G-2
  & 297k
  & $(42,2.6,21)$
  & $(0.16,0.078,0.16)$
  & 10.5
  \\[2pt]
    G-3$'$
  & 570k
  & (45,1.4,11)
  & (0.15,0.038,0.076)
    & 6.9
  \\
  G-3
  & 611k
  & $(45,1.4,11)$
  & $(0.16,0.049,0.078)$
  & 5.6
  \\[2pt]
    G-4$'$
  & 1.18M
  & (46,0.73,12)
  & (0.075,0.037,0.037)
    & 5.2
  \\
  G-4
  & 1.32M
  & $(47,1.5,12)$
  & $(0.076,0.038,0.076)$
  & 4.9
\end{tabular}
\end{center}
\caption{ \label{Table:BFS-Gn-prime}
Sequence of grids generated for LES of the flow
over a backward-facing step starting from a 
different initial grid and how it compares 
with results of Section~\ref{section:BFS}.
See the caption of Table~\ref{Table:bfs-Gn}
for more details. 
Sequence is terminated at grid G-4$'$
since from this point forward the target grids
become nearly identical. 
Note that grid G-5$'$ in Fig.~\ref{fig:eR-convergence-GnPrime}
is generated from grid G-4$'$, 
without the need for running the LES on it.
}
\end{table}

\begin{figure}[t!]
  \centering	
  \includegraphics[width=70mm,clip=true,trim=0mm 0mm 10mm 10mm]{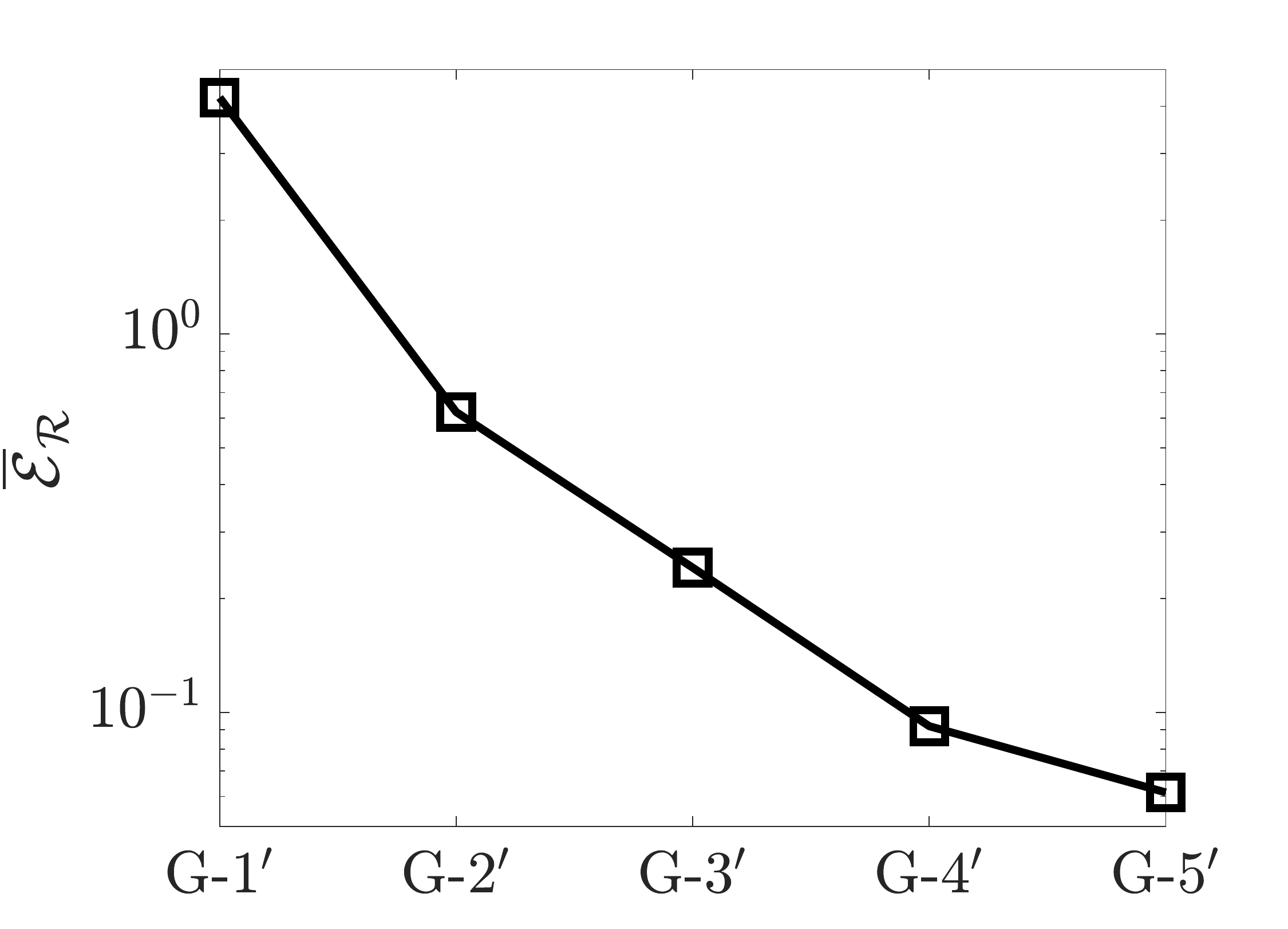}
  \caption{ \label{fig:eR-convergence-GnPrime}
  Decrease in the difference between grids G-$k'$ and  G-$k$
  as the sequence progresses.
  The difference is characterized by $\lpf{\mathcal{E}}_\mathcal{R}$ defined in Eqn.~\ref{eq:eM}
  with some modifications (see text for more details).
  Depending on the acceptable value of $\lpf{\mathcal{E}}_\mathcal{R}$
  either of grids G-4$'$ ($\lpf{\mathcal{E}}_\mathcal{R} \approx 0.09$)
  or G-5$'$ ($\lpf{\mathcal{E}}_\mathcal{R} \approx 0.06$) 
  can be considered sufficiently close
  to grids G-4 and G-5 of Table~\ref{Table:BFS-Gn-prime}.
  }
\end{figure}

The difference between grids the G-$k'$ and G-$k$
is quantified 
by the error in the refinement regions,
$\lpf{\mathcal{E}}_\mathcal{R}$, defined in Eqn.~\ref{eq:eM}.
In this section, 
$\lpf{\mathcal{R}}_{\rm ref}$ is taken from grid G-$k$
and the integration domain is limited to $\Omega: \x=(x,y) \in [-20H,25H] \times [-H,0.5H]$ 
to eliminate the effect of coarser streamwise resolution of grids G-$k'$
further away from the wall.
The convergence of $\lpf{\mathcal{E}}_\mathcal{R}$ with iteration number
is shown in Fig.~\ref{fig:eR-convergence-GnPrime}.

\begin{figure}[t!]
  \centering	
  \includegraphics[width=130mm,clip=true,trim=50mm 10mm 10mm 11.6mm]{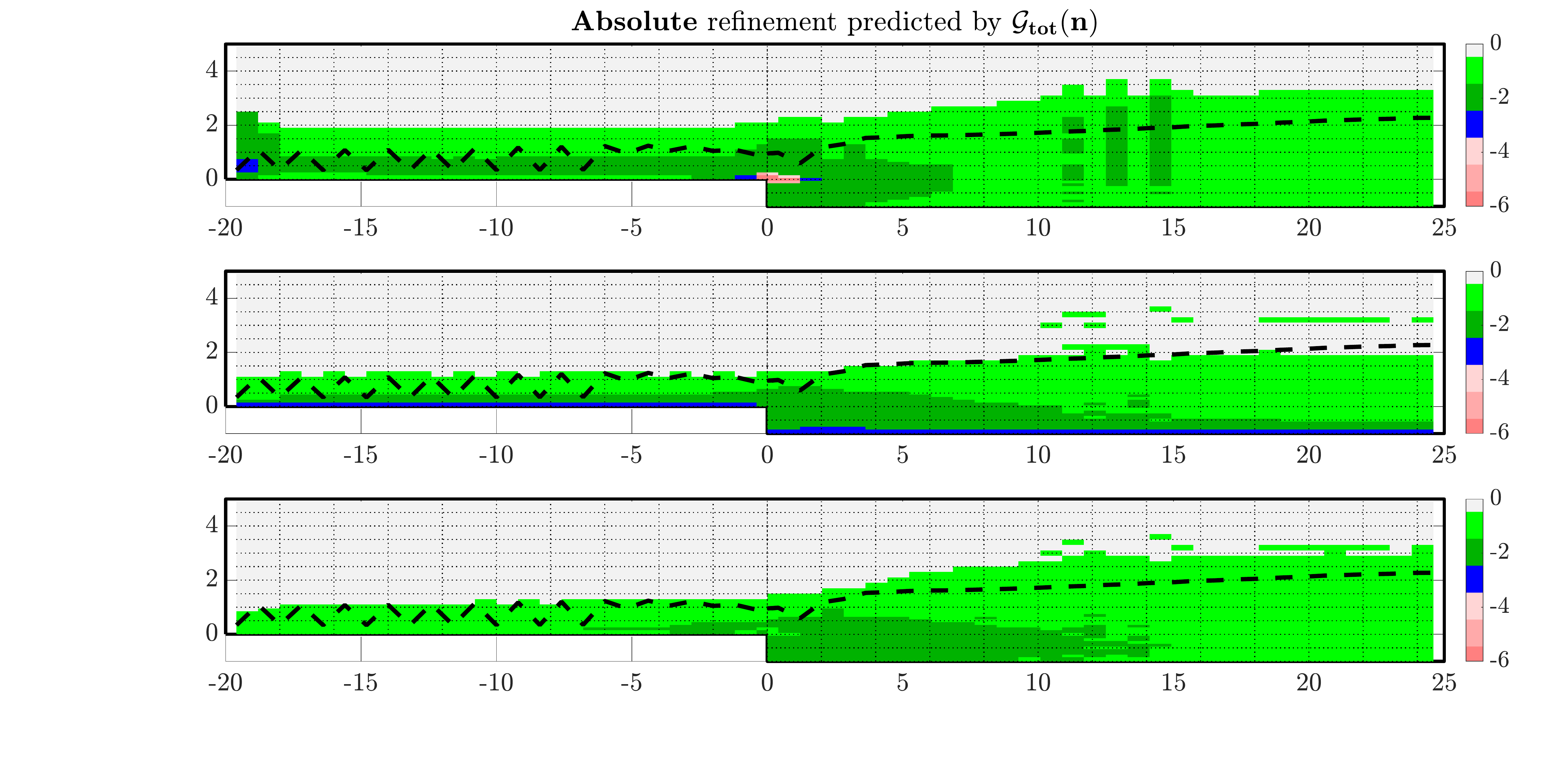}
  \caption{\label{fig:grid-Gn2-prime}
    The grid G-2$'$ from Table~\ref{Table:BFS-Gn-prime}
    illustrated by its refinement levels 
    in $x$ (top),
    $y$ (middle), and $z$ (bottom).
    Refinement levels are computed based on a skeletal grid
    with $(\Delta_{0,x},\Delta_{0,y},\Delta_{0,z})/H = (0.8,0.2,0.2)$ 
    for all $\x$ and $\n$.
    The light green, dark green and blue colors illustrate
    regions with one ($\lpf{\Delta}_{\n_x}/H=0.4$ or $\lpf{\Delta}_{\n_{y/z}}/H=0.1$), 
    two ($\lpf{\Delta}_{\n_x}/H=0.2$ or $\lpf{\Delta}_{\n_{y/z}}/H=0.05$), 
    and three ($\lpf{\Delta}_{\n_x}/H=0.1$ or $\lpf{\Delta}_{\n_{y/z}}/H=0.025$) 
    refinement levels, respectively.
    The white regions 
     correspond to areas of the domain
     that are left untouched 
    (i.e., $\lpf{\Delta}_{\n_x}/H=0.8$ or $\lpf{\Delta}_{\n_{y/z}}/H=0.2$).
    The dashed line highlights the $\delta_{95}$ boundary layer thickness.
    Compare with Fig.~\ref{fig:grid-Gn2}.
  }
\end{figure}

Grid G-2$'$ from Table~\ref{Table:BFS-Gn-prime} 
is illustrated by its refinement regions
in Fig.~\ref{fig:grid-Gn2-prime}.
Note how different this first adapted grid is compared 
to grid G-2 illustrated in Fig.~\ref{fig:grid-Gn2}. 
Namely, the extremely coarse streamwise resolution of G-1$'$
does not allow the solution, and thus the error indicator,
to capture the recirculation bubble after the step
and to resolve it on G-2$'$. 
On the other hand, the coarser resolution of the skeletal grid
gives the algorithm more flexibility to optimally distribute the filter-width,
such that even on this coarse grid the wall-normal and
spanwise resolutions are quite reasonable
(both resolutions are finer than grid G-2).

Note that the results of this section are particular to 
this test case, and may not be necessarily true for other cases.
For instance, if we had coarsened the grid enough 
such that it relaminarized, most probably the error indicator 
would have failed to predict the correct target grid.
Besides, in transitional flows
(including some separated flows where
turbulence is triggered by the Kelvin-Helmholtz instability)
the error indicator itself is probably not adequate 
and the adjoint fields may become necessary to 
capture the extreme sensitivity of the entire flowfield
to those specific regions and 
to predict sufficient resolutions in those regions.

\section{More details on the test-filter and a more general class of suitable filters \label{sec:appendix-filter}}

\begin{figure}[t!]
	\centering	
	\includegraphics[width=38mm,clip=true,trim=0mm 0mm 10mm 3mm]{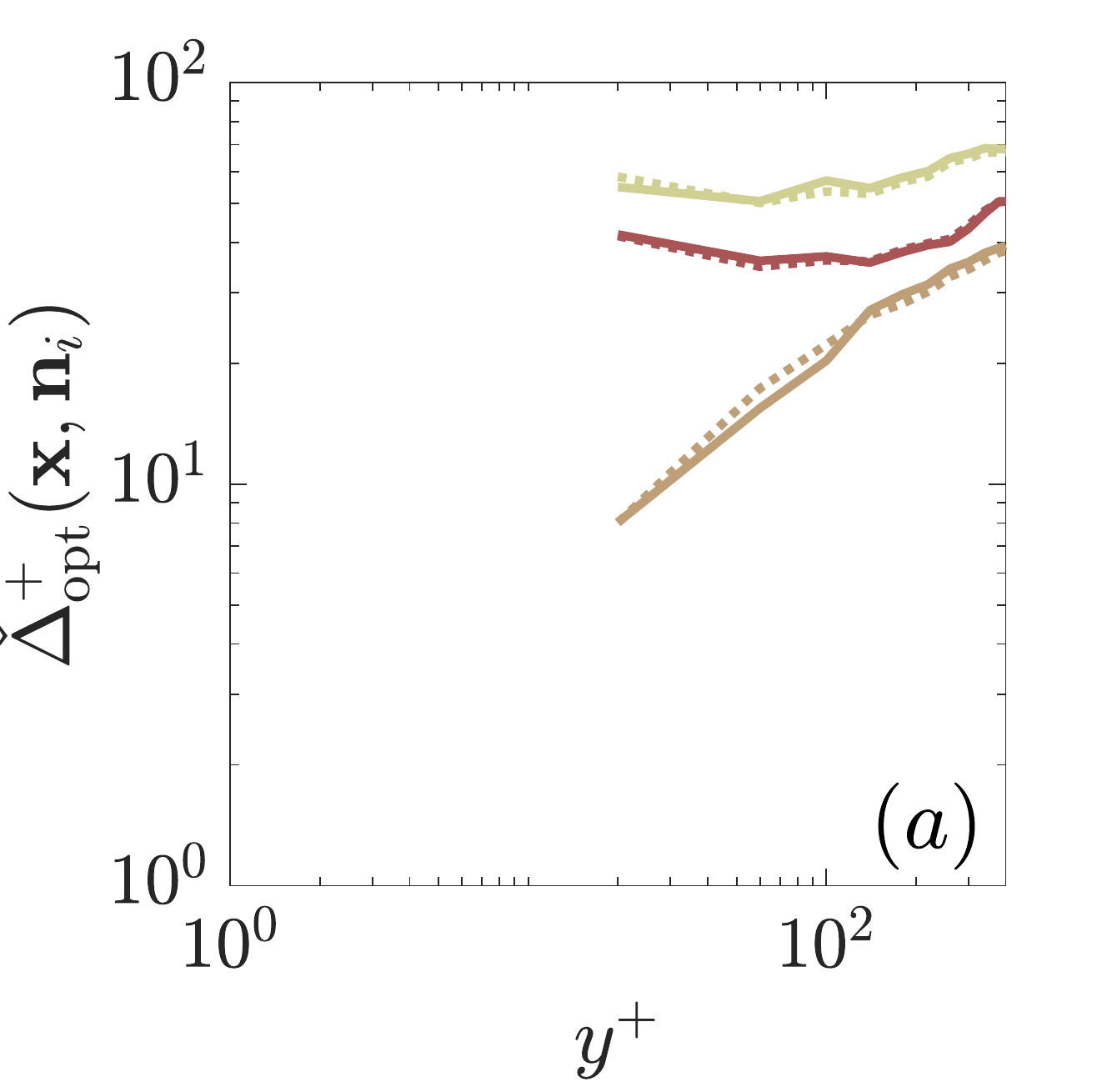}
	\includegraphics[width=38mm,clip=true,trim=0mm 0mm 10mm 3mm]{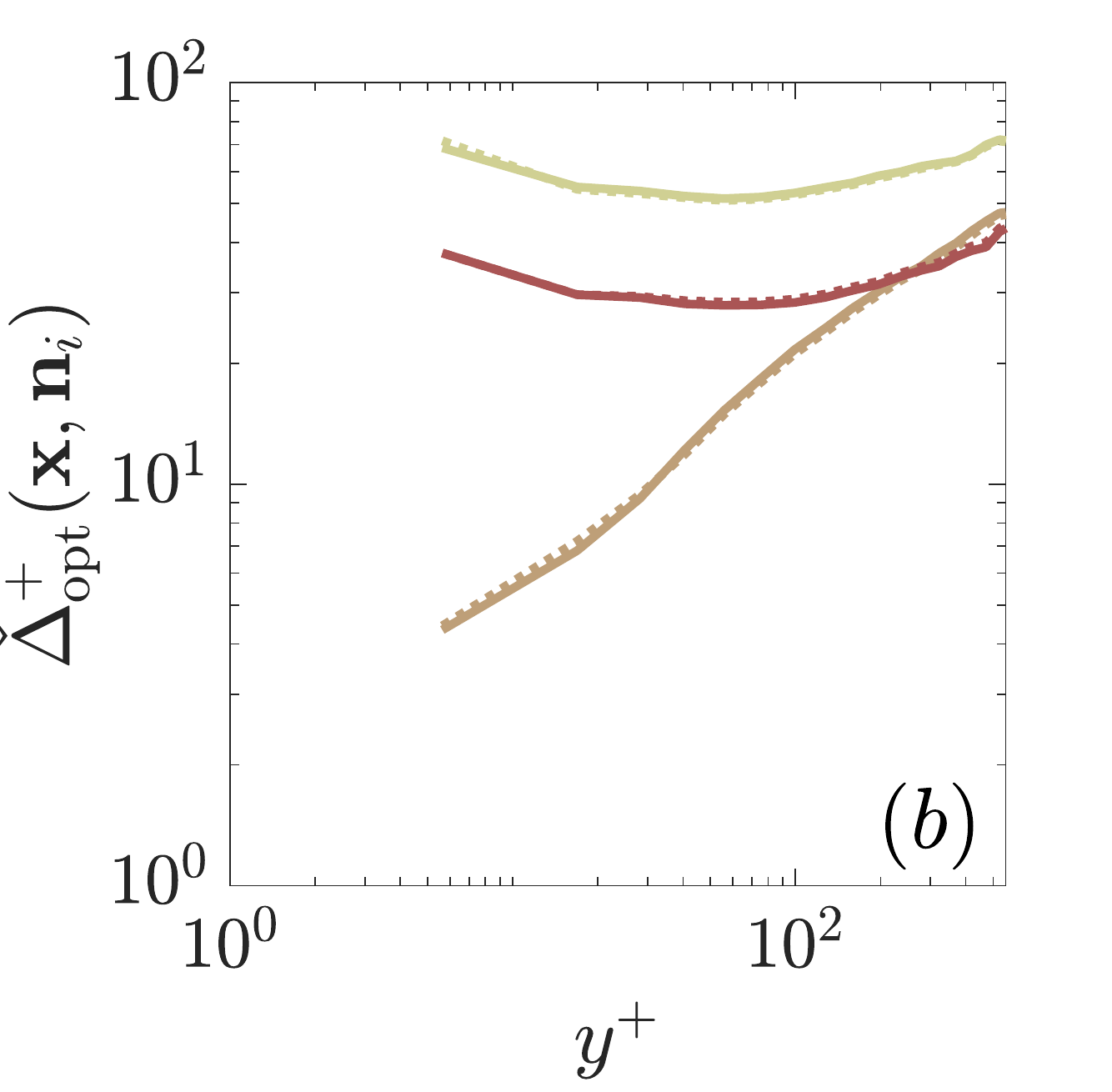}
	\includegraphics[width=38mm,clip=true,trim=0mm 0mm 10mm 3mm]{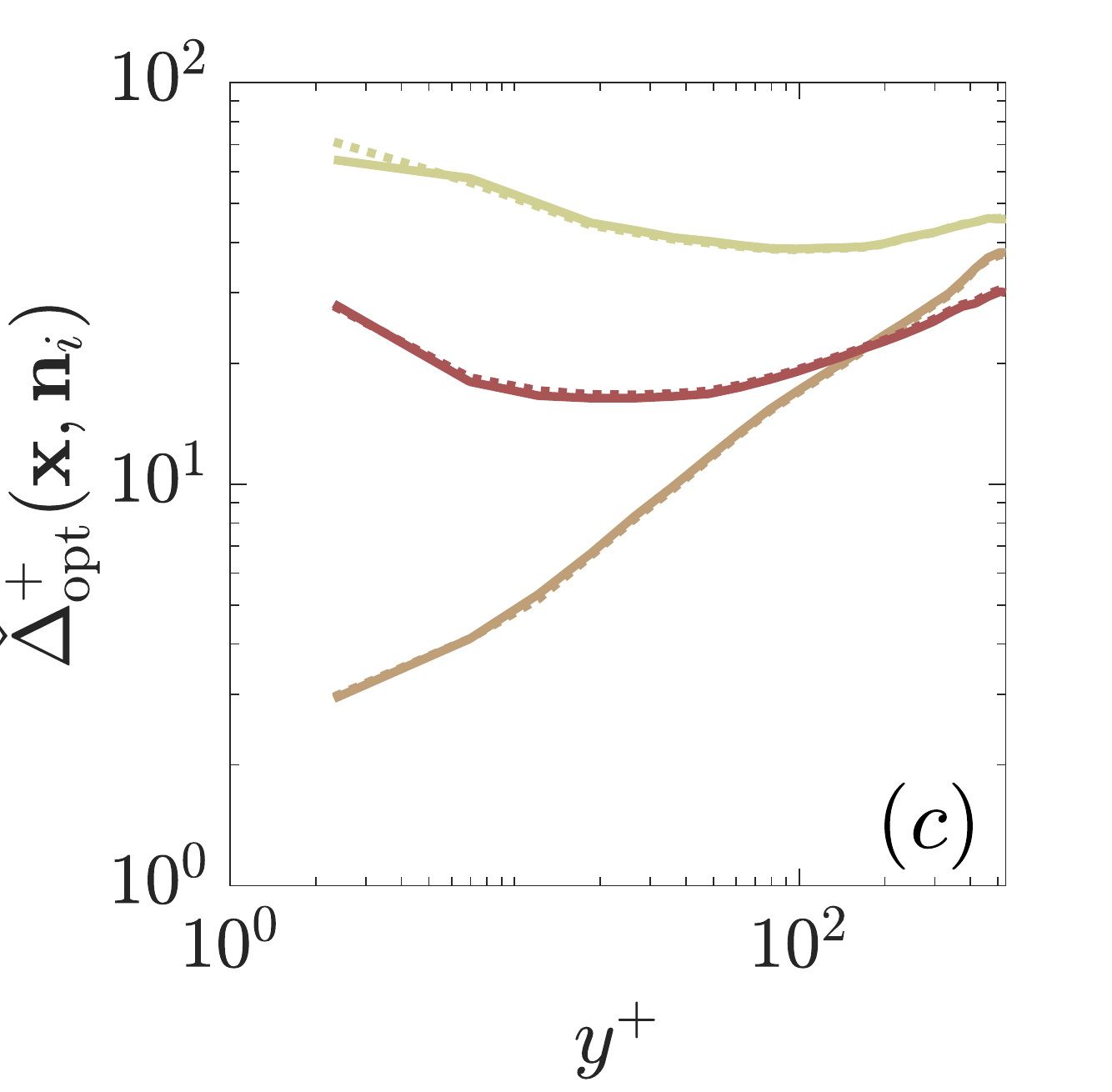}
	\caption{\label{fig:dx-profiles}
	Comparison of the target resolutions 
	corresponding to grids (\textit{a}) DSM-2, 
	(\textit{b}) DSM-3 and (\textit{c}) DSM-4
	of Table~\ref{Table:channel-Gn-DSM}
	between error indicators computed using
	the approximate relation of Eqn.~\ref{eq:filter} (dotted lines)
	and the full deconvolution of Eqn.~\ref{eq:filter-fullDeconv} (solid lines).
	The streamwise, 
	wall-normal 
	and spanwise 
	resolutions
	are shown by the lightest to the darkest colors (in that order).
	}
\end{figure}

The test-filter used throughout this work
(Eqn.~\ref{eq:filter})
is obtained by the van Cittert approximation
to the implicit differential filter
\beq \label{eq:filter-fullDeconv}
\lpf{\phi}
= 
\left( 
I - \frac{ \lpf{ \Delta}_{\n_0}^2}{4} \n_0^T \nabla \nabla^T \n_0
\right)
\tlpf{\lpf{\phi}}^{(\n_0)} 
\, ,
\eeq
which is a modified version of the filter
originally proposed by Germano~\cite[]{germano:86:lesfilter}
to make it directionally dependent
\cite[see][]{toosi:17}.
The van Cittert approximation is truncated after two
terms to reduce the computational cost and complexity.
Since this filter is used only to compute the error indicator,
which is then only used to find the optimal filter-width,
the effect of this truncation is assumed 
(and assessed)
to be negligible in the final predictions of $\check{\Delta}_{\rm opt}(\x,\n)$
(even though the computed values of the error indicator change).
This is demonstrated in Fig.~\ref{fig:dx-profiles}
for the example of LES of the channel flow
using the dynamic Smagorinsky model.

The current implementation of the filter (Eqn.~\ref{eq:filter})
falls into a more general class of differential filters
defined as 
\cite[cf.][]{sagaut:06, vasilyev:98}
\beq
\tlpf{\lpf{\phi}}^{(\n_0)} 
=
\sum_{k=0}^\infty
{
\frac{(-1)^k}{k!}
\fDelta_{\n_0}^k
M_k^{(\n_0)}
\frac
{\partial^k \lpf{\phi}}
{\partial x_{\n_0}^k}
}
\eeq
where $M_k^{(\n_0)}$ is the $k$th moment of the filter kernel
(in direction $\n_0$).
If only $k=0$ and $k \geq K$
are kept in the expansion~\cite[cf.][]{vasilyev:98},
the commutation error of this general class of filters
is of order $K$ (with some extra assumptions).
In other words, by replacing the second derivative of
Eqn.~\ref{eq:filter}
with higher order derivates the commutation error
of the filter and differentiation 
(which was assumed to be negligible in our 
derivations in Section~\ref{section:Gn-formulation})
can be made sufficiently small. 
On structured grids with several neighbors 
available for each point higher derivatives are trivial to implement;
however, in complex geometries 
(especially in the finite volume approach)
computation of derivatives higher than $\partial^2/\partial x_{(\n_0)}^2$
are generally impractical
(at least in the finite-volume framework). 
As a result, the filter used in this work
was defined based on the second derivative
with a commutation error of order $\Delta_{\n_0}^2$.
The reason for the use of a differential filter is that
it is easily applicable to complex geometries with fully
unstructured grids, where the second derivative
can be computed using the Taylor expansion and 
solving a least-square problem
(see~\cite[][]{frey:05,fidkowski:07,toosi:17} for more details).

\bibliographystyle{elsarticle-num}
\bibliography{./references.bib}

\end{document}